%% file: paper1.tex
\documentclass [12pt]{article}
\usepackage{amsmath,amssymb,cite}
\usepackage{appendix}
\usepackage{feynmp}
\usepackage[vcentermath,enableskew]{youngtab}
\usepackage{tabularx}
\usepackage[english]{babel}
\usepackage[dvips]{graphicx}
\usepackage{indentfirst}
\usepackage{epsfig}
\usepackage{physics}
\usepackage{pdfpages}
\usepackage{cases}
\usepackage{appendix}
\usepackage{tabularx}
\usepackage[english]{babel}
\usepackage[dvips]{graphicx}
\usepackage{indentfirst}
\usepackage{epsfig}
\usepackage{slashed}
\usepackage{fancyhdr}
\usepackage{cases}
\usepackage{tcolorbox}
\usepackage{amsmath,amssymb,cite}
\usepackage{amsfonts, geometry, hyperref}
\usepackage{booktabs,longtable}
\usepackage{appendix}
\usepackage{feynmp}
\usepackage{float}
\usepackage{subfigure}
\usepackage[vcentermath,enableskew]{youngtab}
\usepackage{tabularx}
\usepackage[english]{babel}
\usepackage{graphicx}
\usepackage{indentfirst}
\usepackage{epsfig}
\usepackage{color}
\usepackage{epstopdf}
\usepackage[]{hyperref}
\usepackage[section]{placeins}
\usepackage[stable]{footmisc}
\usepackage{appendix}
\usepackage{adjustbox}
\usepackage{tabularx}
\usepackage[english]{babel}
\usepackage{graphicx}
\usepackage{indentfirst}
\usepackage{epsfig}
\usepackage{slashed}
\usepackage{fancyhdr}
\usepackage{url}
\usepackage{pgfplots}
\usepackage{amsmath}

\pgfplotsset{compat=1.18}
\numberwithin{equation}{section}

\makeatletter
\patchcmd{\beamer@sectionintoc}
  {\vfill}
  {\vskip\itemsep}
  {}
  {}

\long\def\beamer@section[#1]#2{%
  \beamer@savemode%
  \mode<all>%
  \ifbeamer@inlecture
    \refstepcounter{section}%
    \beamer@ifempty{#2}%
    {\long\def\secname{#1}\long\def\lastsection{#1}}%
    {\global\advance\beamer@tocsectionnumber by 1\relax%
      \long\def\secname{#2}%
      \long\def\lastsection{#1}%
      \addtocontents{toc}{\protect\beamer@sectionintoc{\the\c@section}{#2\hfill\the\c@page}{\the\c@page}{\the\c@part}%
        {\the\beamer@tocsectionnumber}}}%
    {\let\\=\relax\xdef\sectionlink{{Navigation\the\c@page}{\noexpand\secname}}}%
    \beamer@tempcount=\c@page\advance\beamer@tempcount by -1%
    \beamer@ifempty{#1}{}{%
      \addtocontents{nav}{\protect\headcommand{\protect\sectionentry{\the\c@section}{#1}{\the\c@page}{\secname}{\the\c@part}}}%
      \addtocontents{nav}{\protect\headcommand{\protect\beamer@sectionpages{\the\beamer@sectionstartpage}{\the\beamer@tempcount}}}%
      \addtocontents{nav}{\protect\headcommand{\protect\beamer@subsectionpages{\the\beamer@subsectionstartpage}{\the\beamer@tempcount}}}%
    }%
    \beamer@sectionstartpage=\c@page%
    \beamer@subsectionstartpage=\c@page%
    \def\insertsection{\expandafter\hyperlink\sectionlink}%
    \def\insertsubsection{}%
    \def\insertsubsubsection{}%
    \def\insertsectionhead{\hyperlink{Navigation\the\c@page}{#1}}%
    \def\insertsubsectionhead{}%
    \def\insertsubsubsectionhead{}%
    \def\lastsubsection{}%
    \Hy@writebookmark{\the\c@section}{\secname}{Outline\the\c@part.\the\c@section}{2}{toc}%
    \hyper@anchorstart{Outline\the\c@part.\the\c@section}\hyper@anchorend%
    \beamer@ifempty{#2}{\beamer@atbeginsections}{\beamer@atbeginsection}%
  \fi%
  \beamer@resumemode}%

\def\beamer@subsection[#1]#2{%
  \beamer@savemode%
  \mode<all>%
  \ifbeamer@inlecture%
    \refstepcounter{subsection}%
    \beamer@ifempty{#2}{\long\def\subsecname{#1}\long\def\lastsubsection{#1}}
    {%
      \long\def\subsecname{#2}%
      \long\def\lastsubsection{#1}%
      \addtocontents{toc}{\protect\beamer@subsectionintoc{\the\c@section}{\the\c@subsection}{#2\hfill\the\c@page}{\the\c@page}{\the\c@part}{\the\beamer@tocsectionnumber}}%
    }%
    \beamer@tempcount=\c@page\advance\beamer@tempcount by -1%
    \addtocontents{nav}{%
      \protect\headcommand{\protect\beamer@subsectionentry{\the\c@part}{\the\c@section}{\the\c@subsection}{\the\c@page}{\lastsubsection}}%
      \protect\headcommand{\protect\beamer@subsectionpages{\the\beamer@subsectionstartpage}{\the\beamer@tempcount}}%
    }%
    \beamer@subsectionstartpage=\c@page%
    \edef\subsectionlink{{Navigation\the\c@page}{\noexpand\subsecname}}%
    \def\insertsubsection{\expandafter\hyperlink\subsectionlink}%
    \def\insertsubsubsection{}%
    \def\insertsubsectionhead{\hyperlink{Navigation\the\c@page}{#1}}%
    \def\insertsubsubsectionhead{}%
    \Hy@writebookmark{\the\c@subsection}{#2}{Outline\the\c@part.\the\c@section.\the\c@subsection.\the\c@page}{3}{toc}%
    \hyper@anchorstart{Outline\the\c@part.\the\c@section.\the\c@subsection.\the\c@page}\hyper@anchorend%
    \beamer@ifempty{#2}{\beamer@atbeginsubsections}{\beamer@atbeginsubsection}%
  \fi%
  \beamer@resumemode}

\makeatother

\begin{document}

\title{Endpoint formulation and Molien--Weyl structure for the \(N=2\), large--\(d\) BFSS/BMN models\\
\large\emph{BFSS/BMN Matrix Quantum Mechanics IV}
}

\author{
Badis Ydri\\[2mm]
Department of Physics, Badji Mokhtar Annaba University, Algeria\\
}

\maketitle

\begin{abstract}

We study the \(N=2\), large--\(d\) sector of BFSS/BMN-type matrix quantum mechanics on the lattice in the Gaussian regime. We develop a radial endpoint formulation in which the bulk, gauge, and longitudinal degrees of freedom are integrated out, leaving transverse endpoint variables governed by an effective holonomy potential. We show that this planar endpoint formulation is equivalent to the angular Molien--Weyl description of the gauge-projected partition function, up to a universal spectator factor. This relation allows the low-temperature expansion of the endpoint partition function to be obtained from the Molien--Weyl result, whose quadratic coefficient \(d(d+1)/2\) counts Gaussian singlet states above the vacuum.

We then analyze the continuum limit of the quadratic coefficient and show that it separates into a Gaussian contribution, a \(D\)-channel, and a \(\beta\)-channel. The naive Gaussian term becomes trivial, while the exact holonomy kernel generates finite continuum contributions through singular dependence on the endpoint Gaussian width and anisotropic coupling.

We then study the geometry of the holonomy potential and show that its relevant saddle is a constrained boundary saddle on the aligned branch, rather than an unconstrained critical point. The associated transverse expansion captures the local saddle geometry, but any finite polynomial truncation has a trivial continuum limit. Finally, we introduce a non-polynomial toy model based on \(V_{\rm toy}(B)=-\log\cosh B\), which provides a completion of the transverse expansion and reproduces exactly the continuum \(D\)-channel contribution \(-2d\). This prepares the geometric interpretation of the \(D\)-channel as a Wishart--Stiefel entropy associated with an emergent four-dimensional geometry embedded \(\mathbb R^d\) in the endpoint formulation.

\end{abstract}


\newpage
\tableofcontents

\section{Introduction, Goal, and Summary}
\subsection{Generalities}
\subsubsection{BFSS/BMN systems}
\medskip
\noindent 
A fundamental class of matrix quantum mechanics (MQM) consists of one-dimensional supersymmetric gauge theories with \(d\) Higgs/scalar fields \(X_a\) in the adjoint representation of \(U(N)\). These models are obtained by dimensional reduction of ten-dimensional \(\mathcal N = 1\) super Yang--Mills theory to one time dimension~\cite{Brink:1976bc}, leading to the celebrated BFSS\(_{d+1}\) matrix quantum mechanical models with Euclidean bosonic action
\begin{eqnarray}
S_{\rm BFSS,B}^{\rm E}
=
\frac{1}{g^2}
\int_{0}^{\beta}\! dt~ 
\Tr\!\left[
\frac{1}{2}(D_t X_a)^2
-
\frac{1}{4}[X_a,X_b]^2
\right]
+
\text{fermionic terms}.
\label{BFSSIntro}
\end{eqnarray}
where
\begin{eqnarray}
D_t = \partial_t - i[A_t,\,\cdot\,].
\end{eqnarray}
\medskip
\noindent The allowed dimensions for these supersymmetric matrix quantum mechanical models are constrained by the Fierz identity analysis of Baake, Reinicke and Rittenberg~\cite{Baake:1984ie}. Equivalently, these are precisely the dimensions in which the corresponding superstring theories and super Yang-Mills theories exist:
\begin{eqnarray}
D_{\rm YM}=d_s=d+1=10,6,4,3,2.
\end{eqnarray}
Here \(d\) is the number of bosonic matrix coordinates appearing after dimensional reduction to matrix quantum mechanics, while the associated supermembrane or M-theory/supergravity dimension is shifted by one further dimension,
\begin{eqnarray}
D_{\rm M}=d+2=11,7,5,4,3.
\end{eqnarray}

\medskip
\noindent The holographic, strongly coupled regime of these models is obtained in the ’t~Hooft planar limit~\cite{tHooft1974}
\begin{eqnarray}
N\to\infty,
\qquad
g^2\to 0,
\qquad
\lambda \equiv g^2N=\text{fixed}.
\end{eqnarray}
This limit produces an effectively infinite number of degrees of freedom and is the natural setting for holography~\cite{tHooft1993,Susskind1995}. At large \(\lambda\), the supersymmetric gauge theory is expected to admit a dual description in terms of classical, weakly coupled and weakly curved supergravity.

The seminal example is the \(d=9\) case, which corresponds to the Banks--Fischler--Shenker--Susskind BFSS\(_{10}\) model, also known as M-(atrix) theory~\cite{BanksFischlerShenkerSusskind1997}. This model provides, in the large \(N\) limit, a non-perturbative formulation of M-theory in the infinite–momentum frame. More precisely, the BFSS\(_{10}\) model describes the low–energy worldvolume dynamics of \(N\) coincident D\(0\)-branes~\cite{Witten1996}. Its dynamics and thermodynamics exhibit a strong/weak duality with type~IIA supergravity~\cite{Itzhaki1998}, which is the low-energy limit of type~IIA superstring theory where D\(0\)-branes arise~\cite{Polchinski1995}. In particular, the near-horizon geometry of the black 0-brane—viewed as a bound state of \(N\) D\(0\)-branes—is described at low energy by the discrete light-cone quantization (DLCQ) of D\(0\)-branes as realized by the above maximally supersymmetric Yang--Mills BFSS\(_{10}\) model.

Type IIA supergravity can also be obtained by dimensional reduction of eleven-dimensional supergravity~\cite{Cremmer1978} on a circle \(S^1\)~\cite{Witten1995}. Correspondingly, the dimensional reduction of the eleven-dimensional M-wave solution along the compact direction \(x_{10}\) yields the ten-dimensional non-extremal black 0-brane solution of type IIA supergravity~\cite{Hyakutake:2014maa,Hyakutake:2006aq}. This further establishes the relevance of the BFSS\(_{10}\) model for the description of quantized supermembranes in the light-cone gauge \cite{Hoppe1982,Hoppe1988,deWitHoppeNicolai1988} and for the light-cone quantization of superparticles in maximally supersymmetric pp-wave backgrounds~\cite{Kowalski-Glikman:1984qtj,Blau:2001ne}.

Indeed, the bosonic matrices \(X_a\) play the role of noncommutative coordinate matrices of the D\(0\)-branes:
\begin{itemize}
  \item Diagonal elements $\big(X^{ii}_1, X^{ii}_2, \ldots, X^{ii}_{d}\big)$ encode the position of the \(i\)-th D\(0\)-brane along the transverse directions \(x^{1},\ldots,x^{d}\).
  \item Off-diagonal elements $\big(X^{ij}_1, X^{ij}_2, \ldots, X^{ij}_{d}\big)$ correspond to open strings stretched between the \(i\)-th and \(j\)-th branes, mediating interactions and gauge couplings. When the branes coincide, these off-diagonal modes become massless.
\end{itemize}
See, for example,~\cite{Azeyanagi2009} and the pedagogical presentations~\cite{Zwiebach2009,Becker2006}.

\medskip
\noindent Maldacena's conjecture~\cite{Maldacena1999,Gubser1998,Witten1998} states that, in the decoupling limit
\begin{eqnarray}
\alpha' \longrightarrow 0, 
\qquad 
g_s \longrightarrow 0, 
\qquad
\lambda \equiv g^2 N \ \text{fixed and large}, 
\qquad 
N \gg 1,
\end{eqnarray}
the strongly coupled maximally supersymmetric one-dimensional \(U(N)\) gauge theory is equivalent to weakly coupled type II string theory on the black 0-brane background. Since the gauge theory admits a nonperturbative lattice definition~\cite{Wilson:1974sk}, this duality provides a concrete nonperturbative definition of quantum gravity in this setting.

This correspondence has been extensively tested through Monte Carlo simulations~\cite{Catterall2008,Anagnostopoulos2008,Hanada2014,Hanada2016b,Filev:2015hiaF} and analytic techniques~\cite{Kabat2001,Hanada2009,Hyakutake2014}. In particular, the BFSS\(_{10}\)/black 0-brane duality can be used to derive thermodynamic properties of quantum black holes and to relate quantum-gravity corrections to finite-\(N\) effects in the matrix model. See~\cite{Hanada2016} for a pedagogical review.

The black 0-brane therefore provides a concrete and testable realization of gauge/gravity duality, linking matrix quantum mechanics and gauge theory to quantum gravity and black-hole thermodynamics.

\medskip
\noindent More generally, one may consider the lower-dimensional BFSS\(_{d+1}\) matrix quantum mechanical models obtained from the allowed supersymmetric Yang--Mills dimensions
\begin{eqnarray}
D_{\rm YM}=d+1=6,4,3,2,
\end{eqnarray}
in addition to the maximally supersymmetric BFSS\(_{10}\) case.  These models provide:
\begin{itemize}
  \item A description of the DLCQ dynamics of D\(0\)-branes in the allowed Yang--Mills/string dimensions
  \[
  D_{\rm YM}=d_s=d+1=10,6,4,3,2.
  \]
  This includes, as a consequence, a formulation of the gauge/gravity duality.

  \item A matrix regularization of quantized supermembranes in the light-cone gauge, whose corresponding spacetime dimensions are
  \[
  D_{\rm M}=d+2=11,7,5,4,3.
  \]

  \item A description of the light-cone quantization of superparticles in maximally supersymmetric pp-wave backgrounds of M-theory and supergravity in the same dimensions \(D_{\rm M}=d+2\).
\end{itemize}
Hence these models may be viewed as lower-dimensional analogues of BFSS\(_{10}\), describing the corresponding reduced supersymmetric gauge dynamics, light-cone supermembrane truncations, and noncritical holographic sectors.

\medskip
\noindent 
These lower-dimensional models inherit, as a consequence,  many essential dynamical features of the BFSS\(_{10}\) theory---including confinement--deconfinement transitions, eigenvalue condensation, gauge/gravity-inspired matrix dynamics and emergent geometric phases---while having a simpler structure that makes them more tractable analytically and numerically.

\medskip
\noindent
One may also generalize the BFSS\(_{d+1}\) models by considering maximally supersymmetric mass deformations. Beginning with the BMN plane-wave matrix model~\cite{BerensteinMaldacenaNastase2002}, such deformations were systematically classified in~\cite{Kim:2006,Park:2005}.

\medskip
\noindent
This second class of models is obtained by introducing maximally supersymmetric mass deformations, including quadratic matrix mass terms and a Myers-type cubic interaction~\cite{Myers}, together with fermionic terms that preserve all supercharges. The Euclidean bosonic action is schematically of the form
\begin{eqnarray}
S_{\rm BMN,B}^{\rm E}
=
S_{\rm BFSS,B}^{\rm E}
+
\frac{1}{g^2}
\int_{0}^{\beta}\! dt~
\Tr\!\left[
\mu_1 X_a^2 + \mu_2 \epsilon_{ijk} X_i X_j X_k
\right]
+
\text{fermionic terms}.
\label{BMNIntro}
\end{eqnarray}
The allowed dimensions are \(d=2,3,5,9\), corresponding to BMN\(_{3,4,6,10}\), while the special case \(d=1\), namely BMN\(_2\), is discussed in~\cite{Park:2005} and also in~\cite{Ydri2025}. These deformations need not preserve the full \(SO(d)\) symmetry of the undeformed model: the quadratic mass terms may split the transverse \(SO(d)\) symmetry into smaller rotational factors, while the Myers-type cubic interaction typically selects an \(SO(3)\subset SO(d)\) sector.

The quadratic mass term is sometimes referred to as a cosmological term, since it introduces the curvature scale of the background: it lifts the flat directions of the BFSS model and replaces the flat target-space interpretation by a curved pp-wave one.

From the dual gravity viewpoint, this is precisely the distinction between the undeformed and deformed models. The undeformed model corresponds to type IIA superstring theory on black 0-brane backgrounds, with an uplift to M-theory, whereas the maximally supersymmetric mass deformations correspond to pp-wave backgrounds. In the maximally supersymmetric case, their half-BPS sectors are related to LLM bubbling geometries~\cite{Lin:2004nb}.

Monte Carlo simulations of the BMN model can be found, for example, in~\cite{Asano:2018nol,Asano:2020yry} and references therein.

In summary, each BFSS\(_{d+1}\) model admits a corresponding BMN deformation preserving maximal supersymmetry. These deformations describe supermembranes and superparticles in maximally supersymmetric pp-wave backgrounds. The corresponding classification is summarized in Table~\ref{so3}.

\begin{table}[h]
\centering
\begin{tabular}{@{}lllll@{}}
\toprule
\textbf{Model} & \(D_{\rm YM}\) & Splitting of \(SO(D_{\rm YM}-1)\) & Superalgebra & Deformation parameter \\
\midrule
\(\mathcal N=16\) & 10 & \(SO(6)\times SO(3)\) & \(\mathfrak{su}(2|4)\) & \(\mu\) \\
\(\mathcal N=8\) type I & 6 & \(SO(3)\times SO(2)\) & \(\mathfrak{su}(2|2)\) & \(\mu\) \\
\(\mathcal N=8\) type II & 6 & \(SO(4)\) & \(\mathfrak{su}(2|1)\oplus\mathfrak{su}(2|1)\) & \(\mu\) \\
\(\mathcal N=4\) type I & 4 & \(SO(3)\) & \(\mathfrak{su}(2|1)\) & \(\mu_1,\mu_2\) \\
\(\mathcal N=4\) type II & 4 & \(SO(2)\) & \(\mathrm{Clifford}_{4}(\mathbb R)\) & \(\mu\) \\
\(\mathcal N=2\) & 3 & \(SO(2)\) & \(\mathrm{Clifford}_{2}(\mathbb R)\) & \(\mu\) \\
\(\mathcal N=1+1\) & 2 & \(SO(1,2)\) & \(\mathfrak{osp}(1|2,\mathbb R)\) & \(\Lambda(t),\rho(t)\) \\
\bottomrule
\end{tabular}
\caption{Classification of massive supersymmetric Yang--Mills quantum mechanics models and their deformation parameters.}
\label{so3}
\end{table}

\subsubsection{Large--\(d\) saddle and Molien--Weyl integrals}
\medskip
\noindent 
A third important class of models consists of Gaussian, or large-mass, approximations to BFSS/BMN in various dimensions. In this regime, the theory reduces to supersymmetric gauged matrix harmonic oscillators, which are exactly solvable and admit a formulation in terms of Molien--Weyl integrals~\cite{OConnor:2023mss,OConnor:2024udv}. From a mathematical viewpoint, these integrals compute the Hilbert series of invariant operators, see for example~\cite{CoxLittleOShea2005}.

\medskip
\noindent 
In fact, the large--\(d\) dynamics of the BFSS systems \eqref{BFSSIntro} also yields supersymmetric gauged matrix harmonic oscillators, in which the Yang--Mills interaction is effectively replaced by a mass term with a parameter scaling as \(s\sim d^{1/3}\). See~\cite{Mandal:2009vzN,Mandal:2011hbN} and also~\cite{Kabat:2000zv,Kabat:2001ve}. For the BMN systems \eqref{BMNIntro}, a correlated double--scaling limit is identified in~\cite{Ydri2025}, in which both the deformation mass parameter \(m\) and the number of matrices \(d\) are taken large, while the combination
\begin{eqnarray}
\hat m=\kappa^{2/3}
\;\equiv\;
\frac{m}{d^{2/3}}
\end{eqnarray}
is held fixed.

\medskip
\noindent 
In this limit, the large--$d$ expansion renders the original commutator--squared interaction
self--consistently Gaussian, effectively replacing it by a dynamical mass term.
The resulting effective action is that of a gauged matrix harmonic oscillator,
\begin{eqnarray}
S_{\rm MHO}[X;\theta]
&=&
N\int_{0}^{\beta}\! dt\ \mathrm{Tr}\bigg[
\frac{1}{2}(D_tX_a)^2
+\frac{s^2}{2}X_a^2
\bigg]+
\text{fermionic terms},\label{MHO_action_for_Ward}
\end{eqnarray}
where the effective mass is
\begin{eqnarray}
s^2 = m + k_0,
\end{eqnarray}
with $k_0$ determined self--consistently by the gap equation:
\begin{eqnarray}
s^3-m\,s=d,
\qquad d>0.
\end{eqnarray}
\medskip
\noindent 
The corresponding confinement/deconfinement transition  \cite{Kawahara:2007fnF,Aharony:2003sxF,Aharony:2004igF,AlvarezGaume:2005fvF, Gross:1980heF,Wadia:1980cpF} is governed by the holonomy effective action, yielding a critical temperature
\begin{eqnarray}
T_c(\kappa)
=
\frac{s(\kappa)}{\log d}
=
\frac{d^{1/3}}{\log d}
\Big(\kappa^{1/3}+\frac{1}{2\kappa^{2/3}}+\cdots\Big),
\end{eqnarray}
which is parametrically pushed to higher values as $d\to\infty$.

\begin{figure}[htbp]
\begin{center}
   \includegraphics[width=10cm,angle=-0,page=3]{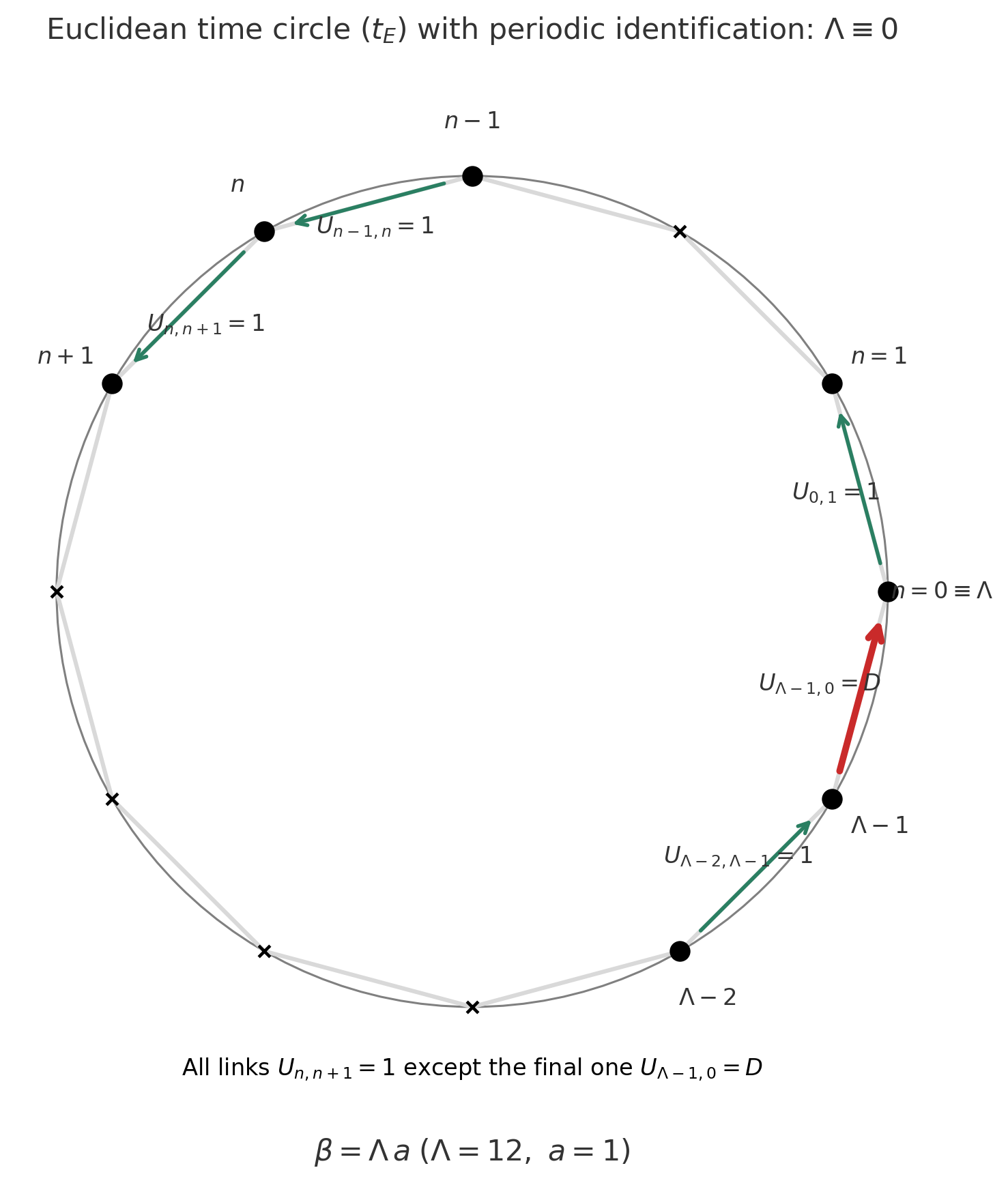}
\end{center}
\caption{The static diagonal (Polyakov) gauge.}\label{GF}
\end{figure}

\medskip
\noindent 
Discretizing the thermal circle into \(\Lambda\) sites with spacing \(a=\beta/\Lambda\), we denote
\(X_{a,n}\equiv X_a(t_n)\), with \(t_n=na\), and introduce link variables
\(U_{n,n+1}\in U(N)\) implementing parallel transport between adjacent time slices.
We impose periodic boundary conditions for bosons and anti-periodic boundary conditions
for fermions, and employ the static Polyakov gauge~\cite{Filev:2015hiaF}. More explicitly,
using gauge invariance, together with Haar invariance of the \(U_{n,n+1}\) measures, one can
gauge-transform all links to unity except the closing link, so that the entire gauge dependence is
captured by a single holonomy linking the \(n=\Lambda-1\) site to the \(n=\Lambda\equiv 0\) site.
See Fig.~\ref{GF}.

\medskip
\noindent
Thus, gauge fixing on the thermal circle reduces the gauge field \(A_t\) to a constant holonomy,
\begin{eqnarray}
g=\mathcal{P}\exp\Big(i\int_0^\beta dt\,A_t\Big)\in U(N).
\end{eqnarray}
After integrating out the coordinate matrices \(X_a\), the path integral becomes a single group integral over \(g\). Since the \(d\) matrices factorize in the Gaussian approximation, the normal-ordered partition function is the \(d\)-th power of the BFSS\(_2\) Molien--Weyl integrand:
\begin{eqnarray}
Z_{N,d}(x)
=
\int d\mu(g)\;
\frac{1}{\Big[{\bf det}\big(1-x\,g\otimes g^{-1}\big)\Big]^d},
\qquad x=e^{-\beta s}.
\label{eq:MW_BFSSd}
\end{eqnarray}
\medskip
\noindent 
Diagonalizing $g=\mathrm{diag}(z_1,\ldots,z_N)$ gives the explicit Molien--Weyl form
\begin{eqnarray}
Z_{N,d}(x)
=
\frac{1}{N!}\oint\prod_{i=1}^{N}\frac{dz_i}{2\pi i z_i}\;
\Delta(z)\Delta(z^{-1})
\prod_{i,j=1}^{N}\frac{1}{\big(1-x\,z_i z_j^{-1}\big)^d},
\label{eq:MW_BFSSd_explicit}
\end{eqnarray}
which can equivalently be written as

\begin{eqnarray}
  Z_{N,d}(x)&=&\frac{1}{N!}\frac{1}{(1-x_b)^{n_bN}}\oint \prod_{i=1}^N\frac{dz_i}{2\pi i z_i}\Delta_A(-1,z)\frac{1}{\Delta_B^{n_b}(-x_b,z)}\nonumber\\
  &&x_b=e^{-\beta m_b}\equiv x, \qquad m_b\equiv s,\qquad n_b\equiv d.\label{eq:MW_BFSSd_explicit1}
\end{eqnarray}
The Faddeev--Popov--Vandermonde determinant $\Delta_A(1,z)$ and the bosonic determinant
$\Delta_B^{\,n_b}(x_b,z)$ are defined in terms of

 \begin{eqnarray}
   \Delta(x,z)=\prod_{i<j}(1+x\frac{z_i}{z_j})\prod_{i<j}(1+x\frac{z_j}{z_i}).
 \end{eqnarray}
 The factor  $1/|W|\equiv 1/N!$ in the partition function (\ref{eq:MW_BFSSd_explicit}) or (\ref{eq:MW_BFSSd_explicit1}) is part of the Haar measure normalization. It is the Weyl group volume, corresponding to division by permutations of the eigenvalues:
diagonal matrices related by permutations represent the same group element.

\medskip
It is clear that in the Molien--Weyl integral (\ref{eq:MW_BFSSd_explicit}), the terms with $i=j$ contribute an overall factor
\begin{equation}
\prod_{i=1}^{N}\frac{1}{(1-x_b)^d}=\frac{1}{(1-x_b)^{dN}},
\end{equation}
which corresponds to the $N$ zero--weights of the $U(N)$ adjoint.  Since
\(
\mathfrak{u}(N)=\mathfrak{su}(N)\oplus\mathfrak{u}(1),
\)
one of these zero--weights is associated with the decoupled $U(1)$ sector.
Projecting to the $SU(N)$ adjoint therefore amounts to removing a single
zero--weight factor, yielding
\begin{equation}
Z^{SU(N)}_{N,d}(x)
=
(1-x_b)^d\,Z^{U(N)}_{N,d}(x).
\end{equation}
We should then work with the formula
\begin{eqnarray}
  Z_{N,d}(x)&=&\frac{1}{N!}\frac{1}{(1-x_b)^{n_b(N-1)}}\oint \prod_{i=1}^N\frac{dz_i}{2\pi i z_i}\Delta_A(-1,z)\frac{1}{\Delta_b^{n_b}(-x_b,z)}.
\end{eqnarray}
\medskip
\noindent 
The explicit supersymmetric completion of the above gauged matrix harmonic oscillator models, together with their corresponding Molien--Weyl integrals, can be found in~\cite{Ydri2025}.

\subsection{Goal: towards matrix quantum mechanics and matrix quantum gravity}

\medskip
\noindent
This work belongs to a broader program devoted to one-dimensional gauge theories
with an arbitrary number \(d\) of noncommuting matrix coordinates. The central
objects are the BFSS\(_{d+1}\) matrix quantum mechanics models and their
mass--deformed BMN\(_{d+1}\) extensions. Our approach combines large--\(d\)
Gaussian reduction, Molien--Weyl singlet projection, endpoint formulations, and
Monte Carlo methods, with the aim of elucidating the structure of matrix quantum
mechanics and its possible interpretation as matrix quantum gravity.

\medskip
\noindent
Here, quantum gravity may refer either to the gauge/gravity-duality approach and
its quantum black-hole dynamics, or to the noncommutative-geometry/matrix-model
approach to emergent geometry and gravity
\cite{Ydri:2022ueu,Ydri:2021cam,Ydri:2020fry}.

\medskip
\noindent
The program is guided by two working assumptions. The first is that matrix
quantum mechanics provides the more fundamental dynamical framework, while
zero-dimensional matrix models, including the IKKT model~\cite{Ishibashi:1996xs},
may be understood as reductions, limits, or approximations of an underlying
one-dimensional matrix quantum theory. The second is the ``unreasonable
effectiveness'' of Gaussian structures in matrix quantum mechanics: even before
the full interacting theory is restored, the Gaussian reduction already captures
a substantial part of the nontrivial quantum dynamics.

\medskip
\noindent
The broader project is organized around several connected themes~\cite{Ydri2025}:
the large-\(d\) limit of BFSS/BMN systems; Molien--Weyl singlet counting and
BFSS\(_2\) factorization; endpoint formulations of Gaussian matrix quantum
mechanics  and their Wishart/Stiefel geometry; Monte Carlo studies of supersymmetric BFSS\(_3\)/BMN\(_3\); and the
relation of BFSS\(_2\)/BMN\(_2\) to noncommutative AdS\(_2\)/dS\(_2\) geometry
and emergent or latent geometry.

\medskip
\noindent
The present paper focuses on the endpoint formulation and Molien--Weyl structure
of the \(N=2\), large-\(d\) BFSS/BMN models in the Gaussian regime. Its main
purpose is to derive the radial planar endpoint theory, relate it to the angular
Molien--Weyl description of the gauge-projected partition function, and use this
relation to analyze the low-temperature singlet expansion. In particular, we
study the continuum decomposition of the quadratic coefficient into Gaussian,
\(D\)-channel, and \(\beta\)-channel contributions, the constrained aligned
saddle of the holonomy potential, and the non-polynomial toy completion which
reproduces the universal \(-2d\) \(D\)-channel contribution.

\input{paper1_p1.tex}

\end{document}

%% file: paper1_p1.tex
\subsection{Summary of results}
\subsubsection{Model}
\medskip
\noindent 
The starting point is the endpoint form of the path integral for the
\emph{\(N=2\), large--\(d\) BFSS/BMN matrix quantum mechanics on the lattice}.

\medskip
\noindent
After integrating out the bulk fluctuations along the thermal circle,
the path integral reduces to a boundary theory involving only the two endpoints
\(n=0\) and \(n=\Lambda\). By gauge fixing to the static Polyakov gauge, the two endpoints are still
connected by the holonomy. The gauge field is then integrated out, followed by
the longitudinal endpoint variables.

\medskip
\noindent 
The remaining degrees of freedom are therefore transverse two-dimensional vectors. More precisely, for each matrix
direction \(a=1,\ldots,d\), one obtains two endpoint vectors
\begin{eqnarray}
V_a=(V_a^1,V_a^2),
\qquad
W_a=(W_a^1,W_a^2),
\qquad
a=1,\ldots,d,
\end{eqnarray}
where \(V_a\) and \(W_a\) encode the initial and final transverse endpoint
configurations of the \(a\)-th coordinate matrix. Thus the endpoint theory is
described by two sets of \(d\) two-dimensional vectors,
\begin{eqnarray}
\{V_a\}_{a=1}^{d},
\qquad
\{W_a\}_{a=1}^{d}.
\end{eqnarray}

\medskip
\noindent
In terms of these variables, the planar endpoint action takes the form
\begin{eqnarray}
S_{N=2,\rm eff}^{\rm BFSS_{d+1}}=
m\sum_{a=1}^{d}
\Big(
V_a^2+W_a^2
\Big)
+
V'_{\rm hol}(A,R),
\label{Send_intro}
\end{eqnarray}
where
\begin{eqnarray}
V_a^2=(V_a^1)^2+(V_a^2)^2,
\qquad
W_a^2=(W_a^1)^2+(W_a^2)^2,
\end{eqnarray}
and
\begin{eqnarray}
V_a\cdot W_a=V_a^1W_a^1+V_a^2W_a^2,\qquad V_a\times W_a =V_a^1W_a^2-V_a^2W_a^1.
\end{eqnarray}
\medskip
\noindent
The planar collective variables are
\begin{eqnarray}
A
=\lambda\sum_{a=1}^{d}V_a\cdot W_a,
\qquad
B=\lambda
\sum_{a=1}^{d}V_a\times W_a,
\qquad
R=\sqrt{A^2+B^2},
\end{eqnarray}
where
\begin{eqnarray}
\lambda=\frac{2N}{a},
\end{eqnarray}
and \(a\) is the lattice spacing.

\medskip
\noindent
The holonomy-induced potential is then given by 
\begin{eqnarray}
V'_{\rm hol}(A,R)
=
-2\beta_\Lambda A
-
\log\Big(
I_0(R)-\frac{A}{R}I_1(R)
\Big),
\qquad
R=\sqrt{A^2+B^2}.
\label{Vhol_intro}
\end{eqnarray}
\medskip
\noindent In the continuum limit
\begin{eqnarray}
a\to 0,
\qquad
\Lambda\to\infty,
\qquad
\beta=a\Lambda
\ \ \text{fixed},
\end{eqnarray}
the endpoint coefficients \(\alpha_\Lambda\) and \(\beta_\Lambda\), which define the mass scale
\begin{eqnarray}
m=\frac{2N}{a}\alpha_\Lambda
\end{eqnarray}
and the anisotropic endpoint coupling, are given explicitly in terms of the oscillator mass \(s\), the lattice spacing \(a\), and the fugacity \(x=e^{-\beta s}\) by
\begin{eqnarray}
\alpha_\Lambda
=
\frac12+\frac{as}{2}\frac{1+x^2}{1-x^2}+O(a^2),
\qquad 
\beta_\Lambda
=
as\,\frac{x}{1-x^2}+O(a^2),
\qquad
x=e^{-\beta s}.
\label{alphabeta_x_intro}
\end{eqnarray}

\subsubsection{Radial endpoint formulation versus angular Molien--Weyl description}

\medskip
\noindent
The first result of the paper is the identification of two equivalent ways of organizing the planar sector of the \(N=2\) Gaussian matrix quantum mechanics. The first is the \emph{angular}, or Molien--Weyl, description, which emphasizes the gauge-symmetry side of the problem: one integrates out the coordinate matrices \(X_a\), keeps the gauge-field holonomy, diagonalizes it by an angle \(\Delta\), and performs the Haar integral at the end. The second is the \emph{radial} endpoint description, which emphasizes the spacetime or coordinate-matrix side: one instead integrates out the holonomy angle and obtains an effective Bessel kernel for the endpoint variables.

\medskip
\noindent
In the holonomy-last route, the fixed-\(\Delta\) planar endpoint integral is Gaussian. Its fixed-holonomy integrand is
\begin{eqnarray}
Z_{\perp}^{\rm fixed\text{-}\Delta}
\propto
\Big(
\alpha_\Lambda^2-\beta_\Lambda^2-\beta_\Lambda\cos\Delta-\frac14
\Big)^{-d}.
\end{eqnarray}
Taking the continuum limit and removing the \(x\)-independent normalization gives
\begin{eqnarray}
\widetilde Z_\perp(\Delta)
=
\left(
\frac{1-x^2}{1-2x\cos\Delta+x^2}
\right)^d.
\label{Ztilde_fixedDelta_intro}
\end{eqnarray}

\medskip
\noindent
The corresponding planar part of the \(SU(2)\) Molien--Weyl integrand is
\begin{eqnarray}
Z_{\rm MW}^{(\perp)}(\Delta)
\propto
\frac{1}{(1-x)^d(1-2x\cos\Delta+x^2)^d}.
\label{MW_fixedDelta_intro}
\end{eqnarray}
Comparing \eqref{Ztilde_fixedDelta_intro} with \eqref{MW_fixedDelta_intro} gives the fixed-holonomy dictionary
\begin{eqnarray}
\widetilde Z_\perp(\Delta)
=
(1-x)^d(1-x^2)^d\,
Z_{\rm MW}^{(\perp)}(\Delta).
\label{dictionary_fixedDelta_intro}
\end{eqnarray}

\medskip
\noindent
Since
\begin{eqnarray}
\mathcal X(x)=(1-x)^d(1-x^2)^d
\end{eqnarray}
is independent of the holonomy angle, it factors through the \(SU(2)\) Haar integral. Hence the integrated radial endpoint partition function and the Molien--Weyl partition function are related by
\begin{eqnarray}
\widetilde Z_\perp(x)
=
\mathcal X(x)\,Z_{\rm MW}(x),
\qquad
Z_{\rm MW}(x)
=
\frac{\widetilde Z_\perp(x)}{(1-x)^d(1-x^2)^d}.
\end{eqnarray}
Thus the radial endpoint and angular Molien--Weyl descriptions are two organizations of the same planar gauge-projected partition function. At the planar level, their only difference is the universal spectator factor \(\mathcal X(x)\), independent of both the holonomy angle and the endpoint variables.

\medskip
\noindent
The Molien--Weyl side can be treated directly. For the \(SU(2)\) Gaussian matrix quantum mechanics, the exact Molien--Weyl partition function may be written in closed form as
\begin{eqnarray}
Z_{\mathrm{MW}}^{SU(2)}(x)
=
(1-x)^{-3d}\,
{}_2F_1\!\left(
d,\frac32;2;-\frac{4x}{(1-x)^2}
\right).
\label{MW_SU2_intro_hyper}
\end{eqnarray}
Expanding this expression at low temperature, \(x=e^{-\beta s}\ll1\), gives
\begin{eqnarray}
Z_{\mathrm{MW}}^{SU(2)}(x)
=
1+\frac{d(d+1)}{2}x^2
+\frac{d(d-1)(d-2)}{6}x^3
+O(x^4).
\label{MW_SU2_intro_series}
\end{eqnarray}
The absence of a linear term reflects the absence of one-particle singlet excitations. The first nontrivial coefficient,
\begin{eqnarray}
k=\frac{d(d+1)}{2},
\end{eqnarray}
counts the quadratic Gaussian singlet states above the vacuum, while the cubic coefficient counts the independent cubic singlet states.

\medskip
\noindent
Using the dictionary \eqref{dictionary_fixedDelta_intro}, the low-temperature expansion of the Molien--Weyl partition function can be re-expressed in the radial endpoint formulation. This yields
\begin{eqnarray}
\widetilde Z_\perp(x)
=
1-dx+d(d-1)x^2
-\frac12 d^2(d-1)x^3
+O(x^4).
\end{eqnarray}
This provides a useful consistency check: the angular Molien--Weyl singlet-counting expansion and the radial planar endpoint expansion encode the same gauge-projected physics, with their difference accounted for by the universal spectator factor \(\mathcal X(x)\).

\subsubsection{Direct endpoint computation from resummed moments}

\medskip
\noindent
The same planar endpoint expansion can also be obtained directly, without using the Molien--Weyl dictionary. In the endpoint route, one first integrates the holonomy angle. This produces the exact Bessel kernel
\begin{eqnarray}
\Phi(A,B)
=
I_0(R)-\frac{A}{R}I_1(R),
\qquad
R^2=A^2+B^2.
\end{eqnarray}
The reduced planar partition function is then written as a Gaussian average,
\begin{eqnarray}
\widetilde Z_\perp(x)
=
Z_{\perp,0}(x)\,
\left\langle
\Phi(A,B)
\right\rangle_0,
\end{eqnarray}
where \(Z_{\perp,0}(x)\) is the free transverse Gaussian partition function, viz.
\begin{eqnarray}
Z_{\perp,0}(x)
&=&
\int \prod_{a=1}^{d} d^2V_a\,d^2W_a\;
\exp\!\left[
-\alpha_\Lambda\sum_{a=1}^{d}(V_a^2+W_a^2)
+
2\beta_\Lambda\sum_{a=1}^{d}V_a\cdot W_a
\right]\nonumber\\
&=&
\left(
\frac{\pi^2}{\lambda^2 D_\Lambda}
\right)^d,
\qquad
D_\Lambda=\alpha_\Lambda^2-\beta_\Lambda^2.
\end{eqnarray}
Thus \(\langle\cdots\rangle_0\) denotes expectation value with respect to this normalized Gaussian measure.

\medskip
\noindent
Expanding the Bessel kernel gives
\begin{eqnarray}
\Phi(A,B)
=
\sum_{n=0}^{\infty}c_nR^{2n}
-
A\sum_{n=0}^{\infty}d_nR^{2n},
\qquad
c_n=\frac{1}{4^n(n!)^2},
\qquad
d_n=\frac{1}{2\cdot4^n n!(n+1)!}.
\end{eqnarray}
This is not a finite expansion in the thermal parameter \(x\). Rather, it is an expansion in the collective endpoint variables \(A\) and \(R\), and each order in \(x\) receives contributions from an infinite tower of Gaussian moments. Thus the low-temperature expansion is naturally organized as a resummed moment expansion:
\begin{eqnarray}
G_{\rm ex}(x)=\frac{\widetilde Z_\perp(x)}{Z_{\perp,0}(x)}
=
\sum_{n=0}^{\infty}c_n\langle R^{2n}\rangle_0
-
\sum_{n=0}^{\infty}d_n\langle A\,R^{2n}\rangle_0.
\end{eqnarray}

\medskip
\noindent
The required moments are generated by
\begin{eqnarray}
{\cal M}(t,u)
=
\left\langle e^{tA+uB}\right\rangle_0
=
\left[
\frac{D_\Lambda}
{D_\Lambda-\beta_\Lambda t-\frac{t^2+u^2}{4}}
\right]^d.
\end{eqnarray}
The even and odd towers are extracted by
\begin{eqnarray}
\langle R^{2n}\rangle_0
=
\left.
(\partial_t^2+\partial_u^2)^n{\cal M}(t,u)
\right|_{t=u=0},
\qquad
\langle A\,R^{2n}\rangle_0
=
\left.
\partial_t(\partial_t^2+\partial_u^2)^n{\cal M}(t,u)
\right|_{t=u=0}.
\end{eqnarray}

\medskip
\noindent
Expanding \(D_\Lambda\) and \(\beta_\Lambda\) in powers of \(x\), the cubic endpoint expansion is controlled by four infinite families of moments,
\begin{eqnarray}
E_{n,0},\qquad E_{n,2},\qquad O_{n,1},\qquad O_{n,3},
\end{eqnarray}
corresponding respectively to the constant and quadratic parts of the even tower \(\langle R^{2n}\rangle_0\) and the linear and cubic parts of the odd tower \(\langle A\,R^{2n}\rangle_0\). Resumming these four families gives the pre-normalized endpoint expansion
\begin{eqnarray}
G_{\rm ex}(x)
&=&
(1-z)^{-d}
-
\frac{d\beta_1}{2D_0}(1-z)^{-d-1}x+
\left[
-\frac{dD_2}{4D_0^2}(1-z)^{-d-1}
+
\frac{d(d+1)\beta_1^2}{4D_0^2}(1-z)^{-d-2}
\right]x^2
\nonumber\\
&-&
\left[\frac{d\beta_3}{2D_0}(1-z)^{-d-1}-\frac{dD_2\beta_1}{2D_0^2}(1+dz)(1-z)^{-d-2}+\frac{d(d+1)(d+2)\beta_1^3}{16D_0^3}(1-z)^{-d-3}\right]x^3+O(x^4).\nonumber\\\label{Ztilde_before_cont_intro}
\end{eqnarray}
Here
\begin{eqnarray}
\beta_\Lambda=\beta_1x+\beta_3x^3+O(x^5),
\qquad
D_\Lambda=D_0+D_2x^2+O(x^4),
\end{eqnarray}
with
\begin{eqnarray}
\beta_1=\beta_3=\mu,
\qquad
D_0=\frac{1+2\mu}{4},
\qquad
D_2=\mu,
\qquad
\mu=as,
\end{eqnarray}
and therefore
\begin{eqnarray}
z=\frac{1}{4D_0}=\frac{1}{1+2\mu},
\qquad
1-z=\frac{2\mu}{1+2\mu}.
\end{eqnarray}

\medskip
\noindent
The overall factor \((1-z)^{-d}\) is the residual \(x=0\) massless-mode normalization of the planar Gaussian. It diverges as \(\mu^{-d}\) in the strict continuum limit and should not be retained as part of the normalized thermal answer. We therefore define the fully normalized partition function by the formula 
\begin{eqnarray}
\widehat Z_\perp
:=\frac{G_{\rm ex}(x)}{G_{\rm ex}(0)}=
(1-z)^d\frac{\widetilde Z_\perp(x)}{Z_{\perp,0}(x)}.
\label{Zhat_def_intro}
\end{eqnarray}
This gives
\begin{eqnarray}
\widehat Z_\perp
&=&
1-dx
+
\left[
-\frac{2d}{1+2\mu}
+
d(d+1)
\right]x^2
\nonumber\\
&&
-
\left[
d
-
\frac{2d(1+2\mu+d)}{1+2\mu}
+
d(d+1)(d+2)
\right]x^3
+
O(x^4).
\label{Zhat_mu_finite_intro}
\end{eqnarray}
Finally, in the strict continuum limit \(\mu\to0\),
\begin{eqnarray}
\widehat Z_\perp(x)
=
1-dx+d(d-1)x^2-\frac12 d^2(d-1)x^3+O(x^4),
\end{eqnarray}
in agreement with the expansion obtained from the Molien--Weyl dictionary.

\subsubsection{Constrained boundary saddle and transverse expansion}

\medskip
\noindent
The endpoint formulation makes the geometry of the holonomy potential particularly transparent. The exact holonomy potential is
\begin{eqnarray}
V_{\rm hol}(A,R)
=
-\log\!\left(
I_0(R)-\frac{A}{R}I_1(R)
\right),
\qquad
-R\le A\le R,
\qquad
R\ge0.
\end{eqnarray}
Equivalently, writing \(X=R^2\), the physical domain is constrained by
\begin{eqnarray}
A^2\le X.
\end{eqnarray}
Thus the relevant saddle is not an unconstrained saddle in the ambient \((A,X)\)-plane. Since \(\partial_A V_{\rm hol}\neq0\) throughout the physical domain, the full ambient gradient cannot vanish. Instead, the stationary point lies on the aligned component of the physical boundary,
\begin{eqnarray}
X=A^2,
\qquad A\ge0,
\end{eqnarray}
or equivalently 
\begin{eqnarray}
A=R.
\end{eqnarray}

\medskip
\noindent
Restricting the potential to this boundary gives
\begin{eqnarray}
V_{\rm b}(A)
=
-\log\!\Big(I_0(A)-I_1(A)\Big).
\end{eqnarray}
The boundary saddle \(A=A_*\) is determined by
\begin{eqnarray}
I_0(A_*)
=
\left(1+\frac{1}{A_*}\right)I_1(A_*),
\end{eqnarray}
with
\begin{eqnarray}
A_*=R_*\simeq1.545,
\qquad
R_*^2\simeq2.387.
\end{eqnarray}
It is worth emphasizing, especially because it explains the form of the expansion below, that this saddle is constrained. The ambient gradient does not vanish; rather, only its projection along the tangent direction to the physical boundary vanishes. Equivalently,
\begin{eqnarray}
\partial_A V_{\rm hol}(A_*,R_*^2)
+
2A_*\,\partial_X V_{\rm hol}(A_*,R_*^2)
=
0,
\end{eqnarray}
whereas
\begin{eqnarray}
\partial_A V_{\rm hol}(A_*,R_*^2)\neq0,
\qquad
\partial_X V_{\rm hol}(A_*,R_*^2)\neq0.
\end{eqnarray}
This is why linear terms survive in the ambient Taylor expansion: they encode the fact that the saddle is stationary only after imposing the boundary constraint.

\medskip
\noindent
This constrained saddle is in fact a local maximum along the aligned boundary branch, since the second derivative of the boundary potential is negative at the saddle.

\medskip
\noindent
To make this geometry explicit, one introduces boundary-adapted variables
\begin{eqnarray}
\Sigma:=R^2-R_*^2,
\qquad
\Delta:=R^2-A^2.
\end{eqnarray}
Here \(\Sigma\) measures displacement along the aligned boundary direction, while \(\Delta\) measures departure from the boundary itself. In these variables, the expansion of the holonomy potential around the constrained saddle takes the form
\begin{eqnarray}
V_{\rm hol}(A,R)
&=&
V_*
-\frac{1}{2R_*}\,\Delta
+\frac{2-R_*}{4R_*^3}\,\Delta\,\Sigma
-\frac{R_*-1}{8R_*^3}\,\Sigma^2
+\frac{1}{8R_*^2}\,\Delta^2
+O(3),
\label{Vhol_SigmaDelta_intro}
\end{eqnarray}
where
\begin{eqnarray}
V_*
=
-\log\!\Big(I_0(R_*)-I_1(R_*)\Big).
\end{eqnarray}
The absence of a linear term in \(\Sigma\) expresses stationarity along the boundary branch, while the linear term in \(\Delta\) expresses the constrained nature of the saddle.

\medskip
\noindent
It is then natural to rewrite the expansion in terms of the transverse variable \(B\), defined by
\begin{eqnarray}
B^2=R^2-A^2,
\qquad
R^2=A^2+B^2.
\end{eqnarray}
The aligned branch corresponds to \(B=0\), while nonzero \(B\) measures transverse departure from that branch. Substituting \(R^2=A^2+B^2\) into \eqref{Vhol_SigmaDelta_intro}, one obtains
\begin{eqnarray}
V_{\rm hol}(A,B)
&=&
V(A)
+
c_2(A)B^2
+
c_4B^4
+
O_3\!\left(A^2-R_*^2,B^2\right),
\label{Vhol_AB_intro}
\end{eqnarray}
where
\begin{eqnarray}
V(A)
&=&
V_*
-\frac{R_*-1}{8R_*^3}\left(A^2-R_*^2\right)^2,
\\
c_2(A)
&=&
-\frac{1}{2R_*}
+
\frac{3-2R_*}{8R_*^3}\left(A^2-R_*^2\right),
\\
c_4
&=&
\frac{3-R_*}{8R_*^3}.
\end{eqnarray}
Thus the local expansion is organized by the longitudinal displacement \(A^2-R_*^2\) and the transverse variable \(B^2\).

\medskip
\noindent
Fixing the longitudinal variable at its saddle value,
\begin{eqnarray}
A=R_*,
\end{eqnarray}
gives the transverse expansion
\begin{eqnarray}
V_{\rm hol}(A=R_*,B)
=
V_*
-\frac{1}{2R_*}\,B^2
+
\frac{3-R_*}{8R_*^3}\,B^4
+
O(B^6).
\label{Vhol_fixedAstar_intro}
\end{eqnarray}
Thus the constrained boundary saddle is a maximum along the aligned branch but is unstable in the transverse direction at quadratic order. The quartic term is positive and provides the leading stabilizing correction. This transverse Landau-type expansion is the starting point for the quartic approximation and for the toy model used to isolate the \(D\)-channel contribution.

\subsubsection{Continuum limit, channels, and completion by a toy model}

\medskip
\noindent
The continuum-limit analysis is most transparent at the level of the quadratic coefficient in the low-\(x\) expansion of the normalized full endpoint ratio
\begin{eqnarray}
\frac{\widetilde Z_\perp(x)}{\widetilde Z_\perp(0)}.
\end{eqnarray}
Let us recall first that, in the exact endpoint theory, all \(x\)-dependence enters through the two continuum-sensitive quantities
\begin{eqnarray}
D_\Lambda(x)=D_0+D_2x^2+O(x^4),
\qquad
\beta_\Lambda(x)=\beta_1x+\beta_3x^3+O(x^5).
\end{eqnarray}
The exact holonomy kernel is
\begin{eqnarray}
\Phi(A,B)
=
I_0(R)-\frac{A}{R}I_1(R),
\qquad
R^2=A^2+B^2,
\end{eqnarray}
and its Gaussian average defines
\begin{eqnarray}
G_{\rm ex}(D_\Lambda,\beta_\Lambda)
=
\frac{\widetilde Z_\perp(x)}{Z_{\perp,0}(x)}
=
\langle \Phi(A,B)\rangle_0.
\end{eqnarray}
Thus the kernel-normalized exact contribution is

\begin{eqnarray}
\widehat Z_\perp(x)
=
\frac{G_{\rm ex}(D_\Lambda(x),\beta_\Lambda(x))}
{G_{\rm ex}(D_0,0)}.
\label{Zhat_exact_intro}
\end{eqnarray}

\medskip
\noindent
Expanding \eqref{Zhat_exact_intro} to quadratic order gives
\begin{eqnarray}
\widehat Z_\perp(x)
=
1
+
\beta_1
\frac{\partial_\beta G_{\rm ex}(D_0,0)}
{G_{\rm ex}(D_0,0)}\,x
+
\left[
D_2
\frac{\partial_D G_{\rm ex}(D_0,0)}
{G_{\rm ex}(D_0,0)}
+
\frac{\beta_1^2}{2}
\frac{\partial_\beta^2G_{\rm ex}(D_0,0)}
{G_{\rm ex}(D_0,0)}
\right]x^2
+
O(x^3).
\label{Taylor_Gex_intro}
\end{eqnarray}
Thus the quadratic coefficient of the normalized full endpoint partition function 
\begin{eqnarray}
\frac{\widetilde Z_\perp(x)}{\widetilde Z_\perp(0)}
=
\frac{Z_{\perp,0}(x)}{Z_{\perp,0}(0)}
\widehat Z_\perp(x)
\end{eqnarray}
is

\begin{eqnarray}
{\cal C}_2^{\rm ex}
=
-\frac{dD_2}{D_0}
+
D_2
\frac{\partial_D G_{\rm ex}(D_0,0)}
{G_{\rm ex}(D_0,0)}
+
\frac{\beta_1^2}{2}
\frac{\partial_\beta^2G_{\rm ex}(D_0,0)}
{G_{\rm ex}(D_0,0)}.
\label{C2_exact_intro}
\end{eqnarray}
The three terms have distinct meanings. The first is the universal Gaussian contribution from the free endpoint determinant. The second is the \(D\)-channel, controlled by the dependence of the exact kernel on the isotropic Gaussian width \(D_\Lambda\). The third is the \(\beta\)-channel, controlled by the explicit anisotropic endpoint coupling.

\medskip
\noindent
We now apply this decomposition to the continuum limit of the exact theory. In the continuum low-temperature regime one has
\begin{eqnarray}
D_0=\frac{1+2\mu}{4},
\qquad
D_2=\mu,
\qquad
\beta_1=\mu,
\qquad
z=\frac{1}{4D_0},
\qquad
1-z=\frac{2\mu}{1+2\mu},
\qquad
\mu=as.
\label{continuum_data_exact_intro}
\end{eqnarray}
The exact resummation gives
\begin{eqnarray}
G_{\rm ex}(D_0,0)
=
(1-z)^{-d},
\end{eqnarray}
and hence
\begin{eqnarray}
D_2
\frac{\partial_D G_{\rm ex}(D_0,0)}
{G_{\rm ex}(D_0,0)}
=
-\frac{dD_2}{4D_0^2(1-z)},
\qquad
\frac{\beta_1^2}{2}
\frac{\partial_\beta^2 G_{\rm ex}(D_0,0)}
{G_{\rm ex}(D_0,0)}
=
\frac{d(d+1)\beta_1^2}{4D_0^2(1-z)^2}.
\label{D_beta_exact_intro}
\end{eqnarray}
Therefore the quadratic coefficient of the normalized full endpoint ratio becomes
\begin{eqnarray}
{\cal C}_2^{\rm ex}
=
-\frac{dD_2}{D_0}
-\frac{dD_2}{4D_0^2(1-z)}
+
\frac{d(d+1)\beta_1^2}{4D_0^2(1-z)^2}.
\label{C2_exact_threepiece_intro}
\end{eqnarray}

\medskip
\noindent
Taking \(\mu\to0\), the three contributions behave as
\begin{eqnarray}
-\frac{dD_2}{D_0}
\longrightarrow 0,
\qquad
-\frac{dD_2}{4D_0^2(1-z)}
\longrightarrow -2d,
\qquad
\frac{d(d+1)\beta_1^2}{4D_0^2(1-z)^2}
\longrightarrow d(d+1).
\label{C2_exact_limits_intro}
\end{eqnarray}
Thus the Gaussian prefactor becomes trivial in the continuum limit, while the two exact holonomy channels survive. The \(D\)-channel gives the finite contribution \(-2d\), whereas the \(\beta\)-channel gives \(d(d+1)\). Together they reconstruct
\begin{eqnarray}
{\cal C}_2^{\rm ex}
\longrightarrow
d(d-1),
\end{eqnarray}
which is precisely the quadratic coefficient of the planar endpoint expansion. This shows that the nontrivial continuum physics is carried not by the naive Gaussian determinant, but by the singular \(D\)- and \(\beta\)-dependence of the exact holonomy kernel.

\medskip
\noindent
The same criterion immediately explains why the transverse expansion, and more generally any pure \(B\)-theory obtained from a finite polynomial truncation, fails to reproduce the exact continuum limit. In such a theory the holonomy kernel depends only on the transverse variable \(B\). After Gaussian averaging, the corresponding function depends only on \(D_\Lambda\):
\begin{eqnarray}
G_B(D_\Lambda)
=
\Big\langle e^{-V_B(B)}\Big\rangle_0,
\qquad
\widehat Z_\perp^{(B)}(x)
=
\frac{G_B(D_\Lambda(x))}{G_B(D_0)}.
\label{GB_intro_def}
\end{eqnarray}
There is no independent dependence on \(\beta_\Lambda\), because the explicit longitudinal variable \(A\) has been frozen. Thus the \(\beta\)-channel is absent, and the quadratic coefficient reduces to the single \(D\)-channel contribution
\begin{eqnarray}
{\cal C}_2^{(B)}
=
D_2\frac{G_B'(D_0)}{G_B(D_0)}.
\label{C2_B_intro}
\end{eqnarray}

\medskip
\noindent
For the quartic transverse approximation, and in fact for any finite polynomial potential \(V_B(B)\), the function \(G_B(D)\) is regular at the continuum point \(D=1/4\). Therefore
\begin{eqnarray}
\frac{G_B'(D_0)}{G_B(D_0)}=O(1),
\qquad
D_2=\mu\to0,
\end{eqnarray}
and hence
\begin{eqnarray}
{\cal C}_2^{(B)}\to0.
\end{eqnarray}
This is the precise sense in which the transverse expansion has a trivial continuum limit: it captures the local geometry of the constrained saddle, but its Gaussian average does not contain the singular factor needed to compensate the explicit \(D_2\sim\mu\) suppression. Consequently, the \(D\)-channel disappears, while the \(\beta\)-channel is absent from the outset.

\medskip
\noindent
The failure of finite transverse truncations suggests that the correct replacement should not be another finite polynomial in \(B\), but a non-polynomial transverse kernel whose Gaussian average has the required singular behavior at \(D_\Lambda=1/4\). This motivates the following simple toy model:
\begin{eqnarray}
e^{-V_{\rm toy}(B)}=\cosh B,
\qquad
V_{\rm toy}(B)=-\log\cosh B.\label{toyintro}
\end{eqnarray}
This choice should be viewed as a completion of the transverse expansion rather than as a term-by-term fit to the exact local saddle potential.

\medskip
\noindent
Locally, the toy potential has the same qualitative Landau structure as the transverse expansion around the constrained boundary saddle:
\begin{eqnarray}
V_{\rm toy}(B)
=
-\frac12B^2+\frac1{12}B^4+O(B^6).
\end{eqnarray}
It therefore captures the basic transverse instability: a negative quadratic term stabilized  by a positive quartic correction. However, its global geometry is more faithful to the exact holonomy potential than the finite quartic truncation. The quartic transverse potential eventually develops artificial minima away from \(B=0\), whereas \eqref{toyintro} has only the central maximum at \(B=0\) and then decreases monotonically as \(|B|\) increases. Thus the toy model preserves the qualitative global shape of the exact transverse holonomy profile better than the polynomial truncation.

\medskip
\noindent
Its main advantage is analytic: it partly restores the correct continuum limit and, in this precise sense, provides a completion of the transverse expansion.

\medskip
\noindent
Indeed, the Gaussian average of the toy kernel is exactly

\begin{eqnarray}
G_{\rm toy}(D_\Lambda)
=
\big\langle \cosh B\big\rangle_0
=
\left(
1-\frac{1}{4D_\Lambda}
\right)^{-d}.
\end{eqnarray}
This is precisely the singular \(D_\Lambda\)-dependence needed for the \(D\)-channel to survive the continuum limit. Consequently,
\begin{eqnarray}
{\cal C}_2^{\rm toy}
=
D_2\frac{G_{\rm toy}'(D_0)}{G_{\rm toy}(D_0)}
=
-\frac{dD_2}{4D_0^2(1-z)}
\longrightarrow
-2d,
\qquad
z=\frac{1}{4D_0}.
\end{eqnarray}
Thus the toy model reproduces the full continuum \(D\)-channel exactly.

\medskip
\noindent
The reason this works is that the non-polynomial kernel \(\cosh B\) resums the transverse moments into the singular factor \((1-1/(4D_\Lambda))^{-d}\). In this sense, the toy model is a global completion of the transverse expansion: it is not an exact local Taylor approximation coefficient by coefficient, but it captures both the correct qualitative global geometry and the correct continuum analytic structure. It still does not reproduce the \(\beta\)-channel, because it depends only on \(B\) and not on the longitudinal variable \(A\).

\subsection{Organization of the paper}

\medskip
\noindent
The paper is organized as follows. Section~2 develops the preliminary lattice endpoint formulation, including the bulk integration, holonomy reduction, longitudinal sector, planar sector, and the resulting low-temperature effective planar action. Section~3 establishes the relation to the Molien--Weyl partition function, compares the angular and radial descriptions, and derives the cubic low-temperature expansion from the Molien--Weyl side. Section~4 derives the same expansion directly as a resummed Gaussian moment expansion in the endpoint formulation. Section~5 analyzes the physics and geometry of the exact holonomy potential, including the constrained boundary saddle and the transverse expansion. Section~6 studies the continuum limit, explains the \(D\)- and \(\beta\)-channels, and introduces the non-polynomial toy model that completes the transverse expansion. Section~7 gives the conclusion, and Section~8 contains the acknowledgements. The appendix collects the covariance-matrix formulas used for the Gaussian moment computations.

\section{Preliminary considerations}
\subsection{Setup}

\medskip
\noindent
We reconsider the bosonic matrix quantum mechanics
 \begin{eqnarray}
   S=N\int_0^{\beta} dt \mathrm{Tr}\bigg[\frac{1}{2}(D_tX_a)^2+\frac{s^2}{2}X_a^2\bigg].
 \end{eqnarray}
 discretized on a Euclidean time lattice with $\Lambda$ sites and spacing $a=\beta/\Lambda$.
We fix the local $U(N)$ symmetry non–perturbatively by putting the holonomy
\begin{eqnarray}
D=\mathrm{diag}(e^{i\theta_1},\ldots,e^{i\theta_N})
\end{eqnarray}
on the last temporal link. The resulting lattice action is

\begin{eqnarray}
  S_{\rm lat}^{\rm BFSS_{d+1}}&=&\frac{N}{a}\bigg[(1+\frac{a^2s^2}{2})\sum_{n=0}^{\Lambda-1}{\rm Tr}{X}_a^{ 2}(n)-\sum_{n=0}^{\Lambda-2}{\rm Tr}{X}_a^{}(n){X}_a^{}(n+1)\bigg]\nonumber\\
  &-&\frac{N}{a}\sum_{i,j}e^{-i(\theta_i-\theta_j)}({X}_a^{}(\Lambda-1))_{ij}({X}_a^{}(0))_{ji}-\frac{1}{2}\sum_{i\ne j}\ln\sin^2\frac{\theta_i-\theta_j}{2}.
\end{eqnarray}
We consider $2\times 2$ traceless matrices and we write
\begin{eqnarray}
(X_a(n))_{ij}=\sum_{\mu=1}^3x_a^{\mu}(n)(\sigma_{\mu})_{ij}.
\end{eqnarray}
The action reduces then to
\begin{eqnarray}
  S_{N=2,\rm lat}^{\rm BFSS_{d+1}}
  &=&2\frac{N}{a}\bigg[(1+\frac{a^2s^2}{2})\sum_{n=0}^{\Lambda-1}{x}_a^{\mu}(n)x_a^{\mu}(n)-\sum_{n=0}^{\Lambda-2}{x}_a^{\mu}(n){x}_a^{\mu}(n+1)-{x}_a^{3}(\Lambda-1){x}_a^{3}(0)\bigg]\nonumber\\
  &-&2\frac{N}{a}\bigg[\cos\Delta.(\vec{x}_{\Lambda-1}.\vec{x}_0)+\sin\Delta.(\vec{x}_{\Lambda-1}\times \vec{x}_0)\bigg]-\ln\sin^2\frac{\Delta}{2},\qquad \Delta=\theta_1-\theta_2,\nonumber\\\label{S_lat_start}
\end{eqnarray}
where the scalar and vector products are defined explicitly by 
\begin{eqnarray}
  \vec{x}_{\Lambda-1}.\vec{x}_0=\sum_{a=1}^d\Big(x_a^{1}(\Lambda-1)x_a^{1}(0)+x_a^{2}(\Lambda-1)x_a^{2}(0)\Big),
\end{eqnarray}
and
\begin{eqnarray}
 \vec{x}_{\Lambda-1}\times \vec{x}_0=\sum_{a=1}^d\Big(x_a^{1}(\Lambda-1)x_a^{2}(0)-x_a^{2}(\Lambda-1)x_a^{1}(0)\Big).
\end{eqnarray}

\subsection{Bulk integration and exact boundary reduction}

\medskip
\noindent

For each matrix index $a=1,\ldots,d$, define the three-dimensional vectors 
\begin{eqnarray}
v_a := \vec x_a(0),
\qquad
w_a := \vec x_a(\Lambda-1),
\qquad
y_a(n) := \vec x_a(n),
\quad 1\le n\le \Lambda-2.
\end{eqnarray}
The purely Gaussian part of the action reads
\begin{eqnarray}
S_{\rm gauss}
=
2\frac{N}{a}
\sum_{a=1}^{d}
\Bigg[
r\Big(|v_a|^2+|w_a|^2
+\sum_{n=1}^{\Lambda-2}|y_a(n)|^2\Big)
-
\sum_{n=0}^{\Lambda-2}
\vec x_a(n)\!\cdot\!\vec x_a(n+1)
\Bigg],
\qquad
r=1+\frac{a^2s^2}{2}.\nonumber\\
\end{eqnarray}
This action is quadratic and diagonal in $a$ and in the adjoint components,
so the bulk integration factorizes over $a$ and over the three components
of each vector.

\medskip
\noindent
Introduce for each $a$ the $(\Lambda-2)$–component column vector
\begin{eqnarray}
Y_a :=
\begin{pmatrix}
y_a(1)\\
\vdots\\
y_a(\Lambda-2)
\end{pmatrix}.
\end{eqnarray}
Then the action can be written as
\begin{eqnarray}
S_{\rm gauss}
=
2\frac{N}{a}
\sum_{a=1}^{d}
\Big[
r(|v_a|^2+|w_a|^2)
+
\frac{1}{2} Y_a^T M Y_a
-
Y_a^T B_a
\Big],
\end{eqnarray}
where
\begin{eqnarray}
M = 2r\,\mathbf 1 - \mathbf A,
\qquad
B_a = v_a\, e_1 + w_a\, e_L,
\end{eqnarray}
where $L:=\Lambda-2$, $e_1=(1,0,\ldots,0)^T$ and $e_L=(0,\ldots,0,1)^T$. Here, $\mathbf A$ is the adjacency matrix of the open chain:
$(\mathbf A)_{i,i+1}=(\mathbf A)_{i+1,i}=1$.

\medskip
\noindent Completing the square and performing the Gaussian integral over all bulk vectors
$Y_a$ gives the result 
\begin{eqnarray}
\int \prod_{a=1}^{d} d^{3L}Y_a \;
e^{-S_{\rm gauss}}
=
\left[
\left(\frac{\pi a}{N}\right)^{\!L/2}
(\det M)^{-1/2}
\right]^{3d}
\exp\!\left[
- S_{\rm eff}(v,w)
\right],
\end{eqnarray}
with effective boundary action
\begin{eqnarray}
S_{\rm eff}(v,w)
=
2\frac{N}{a}
\sum_{a=1}^{d}
\Big[
r(|v_a|^2+|w_a|^2)
-
\frac{1}{2}
B_a^T M^{-1} B_a
\Big].
\end{eqnarray}

\medskip
\noindent The bulk integration involves the inverse of the open--chain tridiagonal matrix
\begin{eqnarray}
M = 2r\,\mathbf 1-\mathbf A,
\qquad
(\mathbf A)_{i,i+1}=(\mathbf A)_{i+1,i}=1,
\qquad
r=\cosh\gamma=1+\frac{a^2s^2}{2},
\end{eqnarray}
which is the lattice Helmholtz operator (massive 1D Laplacian) on an interval with
Dirichlet boundary conditions. Its inverse is the corresponding discrete Green's function,
\begin{eqnarray}
(M^{-1})_{ij}
=
\frac{\sinh(i\gamma)\,\sinh((L+1-j)\gamma)}
{\sinh\gamma\;\sinh((L+1)\gamma)},
\qquad 1\le i\le j\le L,
\label{Minv_ij_ordered}
\end{eqnarray}
and the full matrix is obtained by symmetry $(M^{-1})_{ij}=(M^{-1})_{ji}$.
In particular,
\begin{eqnarray}
(M^{-1})_{11}=(M^{-1})_{LL}
=
\frac{\sinh(L\gamma)}{\sinh((L+1)\gamma)},
\qquad
(M^{-1})_{1L}
=
\frac{\sinh\gamma}{\sinh((L+1)\gamma)}.
\label{Minv_endpoints_clean}
\end{eqnarray}
Therefore
\begin{eqnarray}
B_a^T M^{-1} B_a
=
(M^{-1})_{11}\,(|v_a|^2+|w_a|^2)
+
2(M^{-1})_{1L}\,v_a\!\cdot\! w_a,
\end{eqnarray}
and substituting into $S_{\rm eff}$ yields the exact boundary quadratic form
\begin{eqnarray}
S_{\rm eff}(v,w)
=
2\frac{N}{a}
\sum_{a=1}^{d}
\Big(
\alpha_\Lambda (|v_a|^2+|w_a|^2)
-
2\beta_\Lambda\, v_a\!\cdot\! w_a
\Big),
\label{Seff_alphabeta}
\end{eqnarray}
with
\begin{eqnarray}
\alpha_\Lambda
=
r-\frac{1}{2}(M^{-1})_{11}
=
\frac{\sinh(\Lambda\gamma)}{2\sinh((\Lambda-1)\gamma)},
\qquad
\beta_\Lambda
=
\frac{1}{2}(M^{-1})_{1L}
=
\frac{\sinh\gamma}{2\sinh((\Lambda-1)\gamma)}.
\label{alphabeta_final}
\end{eqnarray}

\subsection{Determinant piece, continuum limit and extent of space}

The bulk Gaussian integration produces the factor
\begin{eqnarray}
Z_{\rm det}
=
\left[
\left(\frac{\pi a}{N}\right)^{L/2}
(\det M)^{-1/2}
\right]^{3d},
\qquad
L=\Lambda-2,
\label{Zdet_def}
\end{eqnarray}
where $M=2r\,\mathbf 1-\mathbf A$ acts only on lattice-site indices.
The exponent $3d$ comes from the $d$ matrices and the
$3$ adjoint generators; for general $N$ this becomes
$(N^2-1)d$.

\medskip
\noindent
To compute $\det M$, define $D_n:=\det(2r\,\mathbf 1_n-\mathbf A_n)$ for the $n\times n$
open-chain matrix. Expanding along the last row gives the standard recursion
\begin{eqnarray}
D_n = 2r\,D_{n-1}-D_{n-2},
\qquad
D_0=1,
\qquad
D_1=2r.
\label{Dn_recursion}
\end{eqnarray}
The solution of this recursion relation is
\begin{eqnarray}
D_n=\frac{\sinh\!\big((n+1)\gamma\big)}{\sinh\gamma},\qquad \cosh\gamma=r=1+\frac{a^2 s^2}{2}.
\label{Dn_closed}
\end{eqnarray}
Therefore, for $n=L=\Lambda-2$,
\begin{eqnarray}
\det M
=
D_L
=
\frac{\sinh\!\big((L+1)\gamma\big)}{\sinh\gamma}
=
\frac{\sinh\!\big((\Lambda-1)\gamma\big)}{\sinh\gamma}.
\label{detM_closed}
\end{eqnarray}
\medskip
\noindent The continuum limit is defined by $a\to 0$ and $\Lambda\to\infty$ with
\begin{eqnarray}
\beta := a\Lambda
\label{beta_def}
\end{eqnarray}
fixed. Then
\begin{eqnarray}
\gamma = as + O(a^3),
\qquad
(\Lambda-1)\gamma = \beta s + O(a),
\qquad
\sinh\gamma = as + O(a^3),
\end{eqnarray}
so that
\begin{eqnarray}
\det M
\;\longrightarrow\;
\frac{\sinh(\beta s)}{a s}.
\label{detM_cont}
\end{eqnarray}
Up to an overall $a$--dependent normalization (independent of $s$),
the determinant contribution becomes
\begin{eqnarray}
Z_{\rm det}
\propto
\Big(\frac{s}{\sinh(\beta s)}\Big)^{\frac{d(N^2-1)}{2}}.
\label{Zdet_cont}
\end{eqnarray}
This is precisely the continuum functional determinant of the
one--dimensional operator $-\partial_\tau^2+s^2$
on an interval of length $\beta$.

\medskip
\noindent Define the free energy
\begin{eqnarray}
F
=
-\frac{1}{\beta}\log Z.
\end{eqnarray}
Keeping only the $s$--dependent part from \eqref{Zdet_cont},
\begin{eqnarray}
F
=
\frac{d(N^2-1)}{2\beta}
\Big[
\log\sinh(\beta s)
-
\log s
\Big].
\label{F_det}
\end{eqnarray}
The extent of space $R^2$ is obtained from the standard matrix-model identity
\begin{eqnarray}
\frac{\partial F}{\partial s^2}
=
\frac{N^2}{2}\, R^2.
\label{extent_def}
\end{eqnarray}
Using
\(
\frac{\partial}{\partial s^2}
=
\frac{1}{2s}\frac{\partial}{\partial s}
\),
one finds
\begin{eqnarray}
R^2
=
\frac{d}{2}
\left(1-\frac{1}{N^2}\right)
\left[
\frac{\coth(\beta s)}{s}
-
\frac{1}{\beta s^2}
\right].
\label{R2_final}
\end{eqnarray}

\medskip
\noindent
In particular, at low temperature $(\beta s\gg 1)$,
\begin{eqnarray}
R^2
\longrightarrow
\left(1-\frac{1}{N^2}\right)
\frac{d}{2s},
\end{eqnarray}
while at high temperature $(\beta s\ll 1)$,
\begin{eqnarray}
R^2
\sim
\left(1-\frac{1}{N^2}\right)
\frac{d\beta}{6}.
\end{eqnarray}
Thus the determinant piece alone reproduces the standard Gaussian
adjoint-matrix behavior.

\subsection{Holonomy reduction}

\medskip
\noindent
At this stage we simply add the remaining $\Delta$--dependent boundary couplings already present
in \eqref{S_lat_start} to the effective action \eqref{Seff_alphabeta}. The full boundary action after bulk integration, including the determinant piece \eqref{Zdet_def}, is therefore
\begin{eqnarray}
S_{\rm bdry}(v,w;\Delta)
&=&\frac{3d}{2}\log\det M+
2\frac{N}{a}\sum_{a=1}^{d}
\Big(
\alpha_\Lambda (|v_a|^2+|w_a|^2)
-2\beta_\Lambda\, v_a\!\cdot\! w_a
\Big)
\nonumber\\
&-&2\frac{N}{a}\sum_{a=1}^{d}
\Big[
\cos\Delta\;(W_a\!\cdot\! V_a)
+\sin\Delta\;(W_a\!\times\! V_a)
\Big]
-2\frac{N}{a}\sum_{a=1}^{d} w_a^{3} v_a^{3}
-\ln\sin^2\frac{\Delta}{2}.\nonumber\\
\label{Sbdry_full}
\end{eqnarray}
Here, $V_a$ and $W_a$ denote the two-dimensional projections of the
three-dimensional adjoint vectors $v_a$ and $w_a$, obtained by
restricting to the $(\sigma_1,\sigma_2)$ components which couple
to the holonomy. 

Equation \eqref{Sbdry_full} is the exact bulk--integrated
representation of the theory: all dependence on the lattice depth
$\Lambda$ and on the mass parameter $s$ is entirely encoded in the
boundary coefficients $(\alpha_\Lambda,\beta_\Lambda)$ and in the
determinant factor $\det M$.

\medskip
\noindent
We need to compute the integral over the holonomy angle $\theta$ which is defined by

\begin{eqnarray}
Z(W_a,V_a)
&=&
\int_0^{2\pi} d\Delta~\sin^2\frac{\Delta}{2}\;
\exp\!\Big(A\cos\Delta+B\sin\Delta\Big)
\nonumber\\
&=&
\frac12\int_0^{2\pi} d\Delta~(1-\cos\Delta)\;
\exp\!\Big(A\cos\Delta+B\sin\Delta\Big),
\end{eqnarray}
where
\begin{eqnarray}
A:=\frac{2N\Lambda}{\beta}(W_a\cdot V_a),\qquad B:=\frac{2N\Lambda}{\beta}(W_a\times V_a).
\end{eqnarray}
Introduce
\begin{eqnarray}
  R:=\sqrt{A^2+B^2}=\frac{2N\Lambda}{\beta}\sqrt{(W_a \cdot V_a)^2+(W_a\times V_a)^2},
  \end{eqnarray}
and
\begin{eqnarray}
A=R\cos\delta,\quad B=R\sin\delta,
\qquad
A\cos\Delta+B\sin\Delta = R\cos(\Delta-\delta).
\end{eqnarray}
Then (using standard Bessel identities)
\begin{eqnarray}
\int_0^{2\pi} d\Delta~e^{A\cos\Delta+B\sin\Delta}
&=&
\int_0^{2\pi} d\Delta~e^{R\cos(\Delta-\delta)}
=
2\pi I_0(R),
\\
\int_0^{2\pi} d\Delta~\cos\Delta~e^{A\cos\Delta+B\sin\Delta}
&=&
\frac{\partial}{\partial A}\int_0^{2\pi} d\Delta~e^{A\cos\Delta+B\sin\Delta}
=
\frac{\partial}{\partial A}\big(2\pi I_0(R)\big)
=
2\pi\,\frac{A}{R}\,I_1(R).\nonumber\\
\end{eqnarray}
Therefore
\begin{eqnarray}
Z(w_a,v_a)
&=&
\frac12\Big[\,2\pi I_0(R)-2\pi\frac{A}{R}I_1(R)\Big]
\nonumber\\
&=&
\pi\Big[I_0(R)-\frac{A}{R}I_1(R)\Big],
\qquad
R=\sqrt{A^2+B^2}.
\end{eqnarray}

\medskip
\noindent
Thus, up to an irrelevant additive constant, the effective action after integrating
out the holonomy from \eqref{Sbdry_full} is

\begin{eqnarray}
   S_{N=2,\rm eff}^{\rm BFSS_{d+1}}
&=&\frac{3d}{2}\log\det M+
2\frac{N}{a}\sum_{a=1}^{d}
\Big(
\alpha_\Lambda (|v_a|^2+|w_a|^2)
-2\beta_\Lambda\, v_a\!\cdot\! w_a
\Big)-2\frac{N}{a}\sum_{a=1}^{d} w_a^{3} v_a^{3}
\nonumber\\
&-&\log\Big(I_0(R)-\frac{A}{R}I_1(R)\Big).\label{EFF}
\end{eqnarray}
\medskip
\noindent
Recall that
\begin{eqnarray}
\alpha_\Lambda
=
\frac{\sinh(\Lambda\gamma)}{2\sinh((\Lambda-1)\gamma)},
\qquad
\beta_\Lambda
=
\frac{\sinh\gamma}{2\sinh((\Lambda-1)\gamma)},
\qquad
\cosh\gamma
=
1+\frac{a^2s^2}{2},
\end{eqnarray}
and taking the continuum limit
\begin{eqnarray}
a\to 0,
\qquad
\Lambda\to\infty,
\qquad
\beta=a\Lambda
\ \ \text{fixed},
\end{eqnarray}
one finds
\begin{eqnarray}
\gamma=as+O(a^3),
\end{eqnarray}
and therefore
\begin{eqnarray}
\alpha_\Lambda
=
\frac12+\frac{as}{2}\coth(\beta s)+O(a^2),\qquad 
\beta_\Lambda
=
\frac{as}{2\sinh(\beta s)}+O(a^2).
\end{eqnarray}

\medskip
\noindent
Introducing
\begin{eqnarray}
x=e^{-\beta s},
\end{eqnarray}
these continuum expressions become
\begin{eqnarray}
\alpha_\Lambda
=
\frac12+\frac{as}{2}\frac{1+x^2}{1-x^2}+O(a^2),\qquad 
\beta_\Lambda
=
as\,\frac{x}{1-x^2}+O(a^2).
\end{eqnarray}

\medskip
\noindent
Thus the zero-temperature and thermal pieces split naturally as
\begin{eqnarray}
\alpha_\Lambda
&=&
\underbrace{\left(\frac12+\frac{as}{2}\right)}_{\text{zero temperature}}
+
\underbrace{\frac{as\,x^2}{1-x^2}}_{\text{thermal excitations}}
+O(a^2),\\
\beta_\Lambda
&=&
\underbrace{0}_{\text{zero temperature}}
+
\underbrace{\frac{as\,x}{1-x^2}}_{\text{thermal excitations}}
+O(a^2).
\end{eqnarray}

\subsection{Longitudinal sector in the continuum limit}

\medskip
\noindent
Using
\begin{eqnarray}
|v_a|^2=(V_a)^2+(v_a^3)^2,
\qquad
|w_a|^2=(W_a)^2+(w_a^3)^2,
\qquad
v_a\!\cdot\! w_a=A+v_a^3w_a^3,
\end{eqnarray}
the longitudinal contribution is
\begin{eqnarray}
S_{\parallel}
&=&
2\frac{N}{a}\sum_{a=1}^{d}
\left[
\alpha_\Lambda\Big((v_a^3)^2+(w_a^3)^2\Big)
-2\beta_\Lambda\, v_a^3 w_a^3
\right]
-2\frac{N}{a}\sum_{a=1}^{d} v_a^3 w_a^3
\nonumber\\
&=&
2\frac{N}{a}\sum_{a=1}^{d}
\left[
\alpha_\Lambda\Big((v_a^3)^2+(w_a^3)^2\Big)
-\Big(2\beta_\Lambda+1\Big)v_a^3 w_a^3
\right].
\label{Sparallel_start}
\end{eqnarray}

\medskip
\noindent
It is convenient to diagonalize this quadratic form by introducing
\begin{eqnarray}
u_a=\frac{v_a^3+w_a^3}{\sqrt2},
\qquad
z_a=\frac{v_a^3-w_a^3}{\sqrt2}.
\end{eqnarray}
Then
\begin{eqnarray}
S_{\parallel}
&=&
2\frac{N}{a}\sum_{a=1}^{d}
\left[
\left(\alpha_\Lambda-\beta_\Lambda-\frac12\right)u_a^2
+
\left(\alpha_\Lambda+\beta_\Lambda+\frac12\right)z_a^2
\right].
\label{Sparallel_diag}
\end{eqnarray}

\medskip
\noindent
The Gaussian integral is now immediate:
\begin{eqnarray}
Z_{\parallel}
&=&
\prod_{a=1}^{d}
\int du_a\,dz_a\,
\exp\!\left[
-2\frac{N}{a}
\left(
\left(\alpha_\Lambda-\beta_\Lambda-\frac12\right)u_a^2
+
\left(\alpha_\Lambda+\beta_\Lambda+\frac12\right)z_a^2
\right)
\right]
\nonumber\\
&=&
\left(\frac{\pi a}{2N}\right)^d
\left[
\left(\alpha_\Lambda-\beta_\Lambda-\frac12\right)
\left(\alpha_\Lambda+\beta_\Lambda+\frac12\right)
\right]^{-d/2}.
\end{eqnarray}
Hence the longitudinal contribution to the effective action is
\begin{eqnarray}
S_{\parallel}^{\rm eff}
&=&
-\log Z_{\parallel}
\nonumber\\
&=&
-d\log\!\left(\frac{\pi a}{2N}\right)
+\frac{d}{2}\log\!\left[
\alpha_\Lambda^2-\left(\beta_\Lambda+\frac12\right)^2
\right].
\label{Spar_check_start}
\end{eqnarray}

\medskip
\noindent
Now insert the continuum-limit expressions
\begin{eqnarray}
\alpha_\Lambda
&=&
\frac12+\frac{as}{2}\frac{1+x^2}{1-x^2}+O(a^2),
\\
\beta_\Lambda
&=&
as\,\frac{x}{1-x^2}+O(a^2),
\qquad
x=e^{-\beta s}.
\end{eqnarray}

\medskip
\noindent
We first compute the two squares separately. For $\alpha_\Lambda^2$,
\begin{eqnarray}
\alpha_\Lambda^2
=
\left(
\frac12+\frac{as}{2}\frac{1+x^2}{1-x^2}+O(a^2)
\right)^2=
\frac14
+\frac{as}{2}\frac{1+x^2}{1-x^2}
+O(a^2).
\label{alpha_sq_check}
\end{eqnarray}
For $\left(\beta_\Lambda+\frac12\right)^2$,
\begin{eqnarray}
\left(\beta_\Lambda+\frac12\right)^2
=
\left(
\frac12+as\,\frac{x}{1-x^2}+O(a^2)
\right)^2=
\frac14
+as\,\frac{x}{1-x^2}
+O(a^2).
\label{beta_sq_check}
\end{eqnarray}
Subtracting \eqref{beta_sq_check} from \eqref{alpha_sq_check}, we get
\begin{eqnarray}
\alpha_\Lambda^2-\left(\beta_\Lambda+\frac12\right)^2=
\frac{as}{2}\,\frac{1-x}{1+x}
+O(a^2).
\label{alpha_beta_diff_check}
\end{eqnarray}

\medskip
\noindent
Thus
\begin{eqnarray}
S_{\parallel}^{\rm eff}
&=&
-d\log\!\left(\frac{\pi a}{2N}\right)
+\frac{d}{2}\log\!\left[
\frac{as}{2}\frac{1-x}{1+x}
+O(a^2)
\right]\nonumber\\
&=&
-d\log\!\left(\frac{\pi a}{2N}\right)
+\frac{d}{2}\log\!\left(\frac{as}{2}\right)
-d\,x
-\frac{d}{3}x^3
+O(x^5)+O(a).\label{Z_long_factor}
\end{eqnarray}

\subsection{Planar sector}

\medskip
\noindent

\medskip
\noindent
The final effective action \eqref{EFF} reduces to the planar theory
\begin{eqnarray}
   S_{N=2,\rm eff}^{\rm BFSS_{d+1}}
&=&\frac{3d}{2}\log\det M
+S_{\parallel}^{\rm eff}
+m\sum_{a=1}^{d}\Big((V_a)^2+(W_a)^2\Big)
+V'_{\rm hol}(A,R),\label{EFF0}
\end{eqnarray}
where the holonomy potential is now defined by
\begin{eqnarray}
V'_{\rm hol}(A,R)
=
-2\beta_\Lambda\,A
-\log\Big(I_0(R)-\frac{A}{R}I_1(R)\Big),
\qquad
R=\sqrt{A^2+B^2},
\label{HI}
\end{eqnarray}
and the mass scale is
\begin{eqnarray}
m=2\frac{N}{a}\,\alpha_\Lambda.
\end{eqnarray}

\medskip
\noindent
The first term is the bulk vacuum energy, the second term is the longitudinal contribution, while all non-Gaussian singlet dynamics is encoded in the holonomy potential, which also incorporates the $\beta_\Lambda$ coupling. The quadratic term sets the overall scale of the planar fluctuations and carries the thermal dependence through $\alpha_\Lambda$.

\subsection{Large-$R$ expansion of the holonomy-induced potential}

\medskip
\noindent
We now analyze the large--$R$ regime of the holonomy factor
\begin{eqnarray}
\Phi(A,B)
:=
I_0(R)-\frac{A}{R}I_1(R),
\qquad
R=\sqrt{A^2+B^2},
\qquad
c:=\frac{A}{R}\in[-1,1],\label{exact0}
\end{eqnarray}
and the corresponding holonomy-induced potential
\begin{eqnarray}
V_{\rm hol}(A,B):=-\log \Phi(A,B).\label{exact}
\end{eqnarray}
Our purpose is twofold: first, to keep not only the leading linear term but also the constant and the first $1/R$ correction; second, to determine the stationary point structure of the resulting approximate potential and its quadratic expansion around the maximum.

\medskip
\noindent
We start from the standard large--$R$ asymptotics of modified Bessel functions,
\begin{eqnarray}
I_0(R)
&\sim&
\frac{e^R}{\sqrt{2\pi R}}
\left(
1+\frac{1}{8R}+\frac{9}{128R^2}+O(R^{-3})
\right),\label{I0}
\\
I_1(R)
&\sim&
\frac{e^R}{\sqrt{2\pi R}}
\left(
1-\frac{3}{8R}-\frac{15}{128R^2}+O(R^{-3})
\right).\label{I1}
\end{eqnarray}
Substituting into $\Phi(A,B)$ gives
\begin{eqnarray}
\Phi(A,B)
&\sim&
\frac{e^R}{\sqrt{2\pi R}}
\left[
(1-c)
+\frac{1+3c}{8R}
+\frac{9+15c}{128R^2}
+O(R^{-3})
\right].
\label{Phi_largeR_master}
\end{eqnarray}
Hence
\begin{eqnarray}
V_{\rm hol}(A,B)
&=&
-\log\Phi(A,B)
\nonumber\\
&\sim&
-R+\frac12\log(2\pi R)
-\log\!\left[
(1-c)
+\frac{1+3c}{8R}
+\frac{9+15c}{128R^2}
+O(R^{-3})
\right].
\label{Vhol_largeR_master}
\end{eqnarray}

\medskip
\noindent
As long as $c\neq1$, the leading term inside the logarithm is nonzero. Factoring out $(1-c)$, one finds
\begin{eqnarray}
V_{\rm hol}(R,c)
&=&
-R+\frac12\log(2\pi R)-\log(1-c)
\nonumber\\
&&
-\log\!\left[
1+\frac{1+3c}{8(1-c)}\frac{1}{R}
+\frac{9+15c}{128(1-c)}\frac{1}{R^2}
+O(R^{-3})
\right].\label{approximate}
\end{eqnarray}
Expanding the last logarithm to first subleading order,
\begin{eqnarray}
V_{\rm hol}(R,c)
&=&
-R+\frac12\log(2\pi R)-\log(1-c)
-\frac{1+3c}{8(1-c)}\frac{1}{R}
+O(R^{-2}),
\qquad c\neq1.
\label{Vapp_generic}
\end{eqnarray}
This is the generic large--$R$ asymptotic potential including the full constant term and the first $1/R$ correction.

\medskip
\noindent
The aligned branch \(B=0\), \(A=R>0\) must be treated separately, since the factor \(1-c\) vanishes there. In this case
\begin{eqnarray}
\Phi(R,0)=I_0(R)-I_1(R).
\end{eqnarray}
Subtracting the asymptotic series for \(I_0\) and \(I_1\), one obtains
\begin{eqnarray}
I_0(R)-I_1(R)
&\sim&
\frac{e^R}{\sqrt{2\pi R}}
\left(
\frac{1}{2R}
+\frac{3}{16R^2}
+\frac{3}{16R^3}
+O(R^{-4})
\right).
\end{eqnarray}
Factoring out \(1/(2R)\),
\begin{eqnarray}
I_0(R)-I_1(R)
&\sim&
\frac{e^R}{\sqrt{2\pi R}}\frac{1}{2R}
\left(
1+\frac{3}{8R}+\frac{3}{8R^2}+O(R^{-3})
\right).
\end{eqnarray}
Hence the aligned large--$R$ potential is
\begin{eqnarray}
V_{\rm hol}^{\rm aligned}(R,0)
&=&
-\log\!\big(I_0(R)-I_1(R)\big)
\nonumber\\
&=&
-R+\frac12\log(2\pi R)+\log(2R)
-\log\!\left(
1+\frac{3}{8R}+\frac{3}{8R^2}+O(R^{-3})
\right)
\nonumber\\
&=&
-R+\frac32\log R+\frac12\log(8\pi)-\frac{3}{8R}+O(R^{-2}).
\label{Vhol_largeR_aligned_precise}
\end{eqnarray}
Since \(A=R\) on this branch, this may be rewritten as
\begin{eqnarray}
V_{\rm hol}^{\rm aligned}(A,0)
&=&
-A+\frac32\log A+\frac12\log(8\pi)-\frac{3}{8A}+O(A^{-2}),
\qquad A>0.
\label{Vhol_largeA_aligned_precise}
\end{eqnarray}

\medskip
\noindent
We now determine the stationary point of the aligned approximate potential
\begin{eqnarray}
V_{\rm hol}^{\rm aligned}(A)
:=
-A+\frac32\log A+\frac12\log(8\pi)-\frac{3}{8A}.
\label{Vapp_aligned}
\end{eqnarray}
Its derivative is
\begin{eqnarray}
\frac{dV_{\rm hol}^{\rm aligned}}{dA}
=
-1+\frac{3}{2A}+\frac{3}{8A^2}.
\end{eqnarray}
Setting this to zero gives
\begin{eqnarray}
8A^2-12A-3=0,
\end{eqnarray}
so the positive stationary point is
\begin{eqnarray}
A_*=\frac{3+\sqrt{15}}{4}\simeq 1.718.
\label{Astar_aligned}
\end{eqnarray}
The second derivative is

\begin{eqnarray}
\frac12\frac{d^2V_{\rm hol}^{\rm aligned}}{dA^2}=\frac12
\left(
-\frac{3}{2A_*^2}-\frac{3}{4A_*^3}
\right)
&=&
-\frac{78\sqrt{15}-270}{135}
\simeq -0.328<0,
\end{eqnarray}
so \(A_*\) is a local maximum.

\subsection{The low-temperature effective planar action}

\paragraph{Set up.}

\bigskip
\noindent
We now return to the exact planar Gaussian with the \(\beta_\Lambda\) coupling kept throughout:
\begin{eqnarray}
S_{\perp,0}
=
2\frac{N}{a}\sum_{a=1}^{d}
\left[
\alpha_\Lambda\big((V_a)^2+(W_a)^2\big)
-2\beta_\Lambda\,V_a\!\cdot\!W_a
\right].
\label{Sperp0_exact_beta}
\end{eqnarray}
Recall
\begin{eqnarray}
A:=\frac{2N}{a}\sum_{a=1}^{d}(W_a\!\cdot\!V_a),
\qquad
B:=\frac{2N}{a}\sum_{a=1}^{d}(W_a\!\times\!V_a),
\qquad
R^2:=A^2+B^2.
\end{eqnarray}

\medskip
\noindent
For simplicity, we focus on the aligned saddle branch \eqref{Astar_aligned}, for which the large--\(R\) holonomy potential takes the form \eqref{Vapp_aligned},
\begin{eqnarray}
V_{\rm hol}^{\rm aligned}(A)
:=
-A+\frac32\log A+\frac12\log(8\pi)-\frac{3}{8A}.
\label{perturbed1}
\end{eqnarray}
Thus it is natural to absorb the \(-A\) term into the unperturbed planar Gaussian,
\begin{eqnarray}
S_{\perp}^{(0)}
=
\lambda\sum_{a=1}^d
\left[
\alpha_\Lambda\big((V_a)^2+(W_a)^2\big)
-2\Big(\beta_\Lambda+\frac12\Big)V_a\!\cdot W_a
\right],\qquad \lambda:=2\frac{N}{a}.\label{unperturbed}
\end{eqnarray}
The potential left after extracting \(-A\) is
\begin{eqnarray}
\widetilde{V}_{\rm hol}^{\rm aligned}(R)
=\frac32\log A+\frac12\log(8\pi)-\frac{3}{8A}.\label{perturbed}
\end{eqnarray}
Defining
\begin{eqnarray}
\widehat\beta_\Lambda:=\beta_\Lambda+\frac12,
\qquad
\widehat D:=\alpha_\Lambda^2-\widehat\beta_\Lambda^{\,2},
\end{eqnarray}
the exact Gaussian moments are then given by (see appendix \eqref{appendix1})
\begin{eqnarray}
\langle A\rangle_0=
\frac{d\,\widehat\beta_\Lambda}{\widehat D},\qquad 
\langle R^2\rangle_0=
\frac{d}{2\widehat D^{\,2}}
\Big(3\alpha_\Lambda^2+2d\widehat\beta_\Lambda^{\,2}\Big).
\end{eqnarray}

\paragraph{Radial size.}

\bigskip
\noindent
In the continuum limit, we have 
\begin{eqnarray}
\alpha_\Lambda
&=&
\frac12+\frac{\mu}{2}\frac{1+x^2}{1-x^2}+O(\mu^2),
\qquad
\beta_\Lambda
=
\mu\frac{x}{1-x^2}+O(\mu^2),
\qquad
\mu:=as\ll1,
\qquad
x=e^{-\beta s},\nonumber\\
\end{eqnarray}
so that
\begin{eqnarray}
\widehat D
=
\alpha_\Lambda^2-\widehat\beta_\Lambda^{\,2}
=
\frac{\mu}{2}\frac{1-x}{1+x}+O(\mu^2).
\end{eqnarray}
Hence
\begin{eqnarray}
\langle A\rangle_0
&=&
\frac{d}{\mu}\frac{1+x}{1-x}
+O(1)
\nonumber\\
&=&
\frac{d}{\mu}\Big(1+2x+2x^2+O(x^3)\Big)+O(1),\label{A0}
\end{eqnarray}
while
\begin{eqnarray}
\langle R^2\rangle_0
&=&
\frac{d(2d+3)}{2\mu^2}\left(\frac{1+x}{1-x}\right)^2
+O(\mu^{-1})
\nonumber\\
&=&
\frac{d(2d+3)}{2\mu^2}\Big(1+4x+8x^2+O(x^3)\Big)+O(\mu^{-1}).
\label{R2_shifted_gaussian_final}
\end{eqnarray}
\medskip
\noindent
Thus the typical radial size in the low--temperature regime is

\begin{eqnarray}
R_{\rm typ}^2
\sim
\langle R^2\rangle_0
=
\frac{d(2d+3)}{2\mu^2},\label{so_basic}
\end{eqnarray}
so the continuum limit \(\mu\to0\) drives the system into the large--\(R\) regime.

\medskip
\noindent
It is important to distinguish here the rescaled variables entering the holonomy
asymptotics from the underlying endpoint bilinears. Writing
\begin{eqnarray}
A=\frac{2N}{a}\,\widehat A,
\qquad
B=\frac{2N}{a}\,\widehat B,
\qquad
R^2=\left(\frac{2N}{a}\right)^2\widehat R^{\,2},
\qquad
\widehat R^{\,2}=\widehat A^{\,2}+\widehat B^{\,2},\label{rescaled}
\end{eqnarray}
one finds that the continuum estimates
\begin{eqnarray}
A\sim \frac{d}{\mu},
\qquad
R^2\sim \frac{d(2d+3)}{2\mu^2},
\qquad
\mu=as,
\end{eqnarray}
translate into
\begin{eqnarray}
\widehat A\sim \frac{d}{2N\,s},
\qquad
\widehat B\sim \frac{d}{2N\,s},
\qquad
\widehat R^{\,2}\sim \frac{d(2d+3)}{2(2N)^2\,s^2}.
\end{eqnarray}
Thus the lattice spacing cancels from the underlying observables, as it must in
the continuum limit, and their physical scale is set by \(s\), not by \(a\).

\medskip
\noindent
Nevertheless, the requirement of a large--\(R\) saddle does not disappear after this
cancellation. Indeed, the large--\(R\) regime is precisely what produces the linear
term \(-R\) needed in the large--holonomy analysis, so the corresponding unhatted
radial variable must itself be parametrically large:

\begin{eqnarray}
\widehat R_{\rm typ}^2
\sim
\frac{d(2d+3)}{2(2N)^2s^2}\gg 1.
\end{eqnarray}
For large \(d\), this gives

\begin{eqnarray}
\widehat R_{\rm typ}\sim \frac{d}{2N\,s}.
\end{eqnarray}
Thus, up to the fixed numerical factor $2N\equiv 4$, the physical self-consistency
condition for the large--$R$ expansion is
\begin{eqnarray}
s\ll d.
\end{eqnarray}

\paragraph{The planar effective action.}

\bigskip
\noindent
Finally, the Gaussian integration over the planar modes gives in an obvious way the action
\begin{eqnarray}
S_{\perp,0}^{\rm eff}
=
-2d\log\!\left(\frac{\pi a}{2N}\right)
+d\log\!\left(\alpha_\Lambda^2-\widehat\beta_\Lambda^{\,2}\right)
=
-2d\log\!\left(\frac{\pi a}{2N}\right)
+d\log \widehat D.
\end{eqnarray}
Using
\begin{eqnarray}
\widehat D
=
\alpha_\Lambda^2-\widehat\beta_\Lambda^{\,2}
=
\frac{\mu}{2}\frac{1-x}{1+x}+O(\mu^2),
\end{eqnarray}
one finds
\begin{eqnarray}
S_{\perp,0}^{\rm eff}
&=&
-2d\log\!\left(\frac{\pi a}{2N}\right)
+d\log\!\left(\frac{\mu}{2}\right)
+d\log\!\left(\frac{1-x}{1+x}\right)
+O(\mu)
\nonumber\\
&=&
-2d\log\!\left(\frac{\pi a}{2N}\right)
+d\log\!\left(\frac{\mu}{2}\right)
-2d\,x-\frac{2d}{3}x^3+O(x^5)+O(\mu).
\end{eqnarray}

\medskip
\noindent
Thus, after separating the regulator-dependent pieces, the leading thermal
contribution is
\begin{eqnarray}
d\log\!\left(\frac{1-x}{1+x}\right)
=
-2d\,x-\frac{2d}{3}x^3+O(x^5),\label{Z_long_factor2}
\end{eqnarray}
which is twice the corresponding longitudinal contribution in
\eqref{Z_long_factor}, as expected.

\medskip
\noindent
At first sight, the remaining holonomy logarithmic contribution \eqref{perturbed} may appear as a perturbation around the present aligned Gaussian saddle. However, a more complete treatment of the holonomy-induced potential \eqref{exact}, developed in the subsequent discussion, will show that its continuum large-\(d\) physics is ultimately far richer than that of a simple aligned Gaussian saddle supplemented by a logarithmic correction.

\section{Relation to the Molien--Weyl partition function}
\subsection{The Molien--Weyl integral for \(SU(2)\)}
\medskip
\noindent
Let us now evaluate the Molien--Weyl integral explicitly for \(SU(2)\), keeping track of all factors carefully.

\medskip
\noindent
We start from
\begin{eqnarray}
Z_{\mathrm{MW}}^{SU(2)}(x)
&=&
(1-x)^d
\int d\mu(g)\,
\prod_{i,j=1}^{2}
\left(1-x\,z_i z_j^{-1}\right)^{-d}.
\label{MW_SU2_start}
\end{eqnarray}
For \(SU(2)\) we may write
\begin{eqnarray}
z_1=z=e^{i\theta},
\qquad
z_2=z^{-1}=e^{-i\theta},
\end{eqnarray}
so that
\begin{eqnarray}
\prod_{i,j=1}^{2}\left(1-x\,z_i z_j^{-1}\right)
=
(1-x)^2(1-xz^2)(1-xz^{-2}).
\end{eqnarray}
Hence
\begin{eqnarray}
Z_{\mathrm{MW}}^{SU(2)}(x)
=
\int d\mu(g)\,
\frac{1}{(1-x)^d(1-xz^2)^d(1-xz^{-2})^d}.
\label{MW_SU2_product}
\end{eqnarray}

\medskip
\noindent
Using the normalized Haar measure
\begin{eqnarray}
d\mu(g)=\frac{2}{\pi}\sin^2\theta\,d\theta,
\qquad
0\le \theta\le \pi,
\end{eqnarray}
we obtain
\begin{eqnarray}
Z_{\mathrm{MW}}^{SU(2)}(x)
=
\frac{2}{\pi}(1-x)^{-d}
\int_0^\pi d\theta\,
\sin^2\theta\,
\left(1-2x\cos 2\theta+x^2\right)^{-d}.
\label{MW_SU2_theta}
\end{eqnarray}

\medskip
\noindent
Now use the identity
\begin{eqnarray}
1-2x\cos 2\theta+x^2
=
(1-x)^2+4x\sin^2\theta
=
(1-x)^2\left(1+\frac{4x}{(1-x)^2}\sin^2\theta\right).
\end{eqnarray}
Therefore
\begin{eqnarray}
\left(1-2x\cos 2\theta+x^2\right)^{-d}
=
(1-x)^{-2d}
\left(1+\frac{4x}{(1-x)^2}\sin^2\theta\right)^{-d}.
\end{eqnarray}
Substituting into \eqref{MW_SU2_theta} gives
\begin{eqnarray}
Z_{\mathrm{MW}}^{SU(2)}(x)
=
\frac{2}{\pi}(1-x)^{-3d}
\int_0^\pi d\theta\,
\sin^2\theta\,
\left(1+\frac{4x}{(1-x)^2}\sin^2\theta\right)^{-d}.
\end{eqnarray}
By symmetry under \(\theta\mapsto \pi-\theta\), this becomes
\begin{eqnarray}
Z_{\mathrm{MW}}^{SU(2)}(x)
=
\frac{4}{\pi}(1-x)^{-3d}
\int_0^{\pi/2} d\theta\,
\sin^2\theta\,
\left(1+\frac{4x}{(1-x)^2}\sin^2\theta\right)^{-d}.
\label{MW_SU2_half}
\end{eqnarray}

\medskip
\noindent
Introduce
\begin{eqnarray}
\lambda:=\frac{4x}{(1-x)^2}.
\end{eqnarray}
Then \eqref{MW_SU2_half} becomes
\begin{eqnarray}
Z_{\mathrm{MW}}^{SU(2)}(x)
=
\frac{4}{\pi}(1-x)^{-3d}
\int_0^{\pi/2} d\theta\,
\sin^2\theta\,
(1+\lambda\sin^2\theta)^{-d}.
\label{MW_SU2_lambda}
\end{eqnarray}

\medskip
\noindent
We now use the Euler-type integral representation
\begin{eqnarray}
\int_0^{\pi/2}
\sin^{2b-1}\theta\,
\cos^{2c-2b-1}\theta\,
(1+\lambda\sin^2\theta)^{-a}\,d\theta
=
\frac12 B(b,c-b)\,
{}_2F_1(a,b;c;-\lambda),
\end{eqnarray}
valid for \(\Re(c)>\Re(b)>0\). Choosing
\begin{eqnarray}
a=d,
\qquad
b=\frac32,
\qquad
c=2,
\end{eqnarray}
we have
\begin{eqnarray}
2b-1=2,
\qquad
2c-2b-1=0,
\end{eqnarray}
so the integral matches exactly. Moreover,
\begin{eqnarray}
B\!\left(\frac32,\frac12\right)
=
\frac{\Gamma(3/2)\Gamma(1/2)}{\Gamma(2)}
=
\frac{(\sqrt{\pi}/2)\sqrt{\pi}}{1}
=
\frac{\pi}{2}.
\end{eqnarray}
Hence
\begin{eqnarray}
\int_0^{\pi/2} d\theta\,
\sin^2\theta\,
(1+\lambda\sin^2\theta)^{-d}
=
\frac{\pi}{4}\,
{}_2F_1\!\left(d,\frac32;2;-\lambda\right).
\end{eqnarray}
Substituting into \eqref{MW_SU2_lambda}, the factors \(\frac{4}{\pi}\) and \(\frac{\pi}{4}\) cancel, giving the exact closed form
\begin{eqnarray}
\boxed{
Z_{\mathrm{MW}}^{SU(2)}(x)
=
(1-x)^{-3d}\,
{}_2F_1\!\left(d,\frac32;2;-\frac{4x}{(1-x)^2}\right).
}
\label{MW_SU2_hyper}
\end{eqnarray}

\medskip
\noindent
We now expand the hypergeometric factor for small \(x\). Let
\begin{eqnarray}
u:=-\frac{4x}{(1-x)^2}.
\end{eqnarray}
Since
\begin{eqnarray}
\frac{1}{(1-x)^2}
=
1+2x+3x^2+4x^3+O(x^4),
\end{eqnarray}
we have
\begin{eqnarray}
u=-4x-8x^2-12x^3+O(x^4),
\qquad
u^2=16x^2+64x^3+O(x^4),
\qquad
u^3=-64x^3+O(x^4).\nonumber\\
\end{eqnarray}

\medskip
\noindent
The Gauss hypergeometric series is
\begin{eqnarray}
{}_2F_1(a,b;c;u)
=
1+\frac{ab}{c}u
+\frac{a(a+1)b(b+1)}{2c(c+1)}u^2
+\frac{a(a+1)(a+2)b(b+1)(b+2)}{6c(c+1)(c+2)}u^3
+O(u^4).\nonumber\\
\end{eqnarray}
With
\begin{eqnarray}
a=d,
\qquad
b=\frac32,
\qquad
c=2,
\end{eqnarray}
this gives
\begin{eqnarray}
{}_2F_1\!\left(d,\frac32;2;u\right)
=
1+\frac{3d}{4}u
+\frac{5d(d+1)}{16}u^2
+\frac{35d(d+1)(d+2)}{384}u^3
+O(u^4).
\end{eqnarray}
Substituting the small-\(x\) expansion of \(u\), one finds
\begin{eqnarray}
{}_2F_1\!\left(d,\frac32;2;-\frac{4x}{(1-x)^2}\right)
&=&
1+\frac{3d}{4}(-4x-8x^2-12x^3)
+\frac{5d(d+1)}{16}(16x^2+64x^3)
\nonumber\\
&&
+\frac{35d(d+1)(d+2)}{384}(-64x^3)
+O(x^4)
\nonumber\\
&=&
1-3dx+(5d^2-d)x^2
+\left(-\frac{35}{6}d^3+\frac{5}{2}d^2-\frac{2}{3}d\right)x^3
+O(x^4).\nonumber\\
\label{MW_SU2_hyper_expanded}
\end{eqnarray}

\medskip
\noindent
Next we expand the prefactor:
\begin{eqnarray}
(1-x)^{-3d}
=
1+3dx+\frac{3d(3d+1)}{2}x^2+\frac{3d(3d+1)(3d+2)}{6}x^3+O(x^4),
\end{eqnarray}
i.e.
\begin{eqnarray}
(1-x)^{-3d}
=
1+3dx+\frac{9d^2+3d}{2}x^2+\left(\frac{9}{2}d^3+\frac{9}{2}d^2+d\right)x^3+O(x^4).
\label{MW_SU2_pref_expanded}
\end{eqnarray}

\medskip
\noindent
We now multiply \eqref{MW_SU2_hyper_expanded} and \eqref{MW_SU2_pref_expanded}. Writing
\begin{eqnarray}
(1-x)^{-3d}
&=&
1+A_1x+A_2x^2+A_3x^3+O(x^4),\\
{}_2F_1
&=&
1+B_1x+B_2x^2+B_3x^3+O(x^4),
\end{eqnarray}
with
\begin{eqnarray}
A_1=3d,
\qquad
A_2=\frac{9d^2+3d}{2},
\qquad
A_3=\frac{9}{2}d^3+\frac{9}{2}d^2+d,
\end{eqnarray}
and
\begin{eqnarray}
B_1=-3d,
\qquad
B_2=5d^2-d,
\qquad
B_3=-\frac{35}{6}d^3+\frac{5}{2}d^2-\frac{2}{3}d,
\end{eqnarray}
the product is
\begin{eqnarray}
Z_{\mathrm{MW}}^{SU(2)}(x)
&=&
1+(A_1+B_1)x
+\left(A_2+A_1B_1+B_2\right)x^2
\nonumber\\
&&
+\left(A_3+A_2B_1+A_1B_2+B_3\right)x^3
+O(x^4).
\end{eqnarray}

\medskip
\noindent
The linear coefficient vanishes:
\begin{eqnarray}
A_1+B_1=3d-3d=0.
\end{eqnarray}

\medskip
\noindent
For the quadratic term,
\begin{eqnarray}
A_2+A_1B_1+B_2
&=&
\frac{9d^2+3d}{2}-9d^2+(5d^2-d)
\nonumber\\
&=&
\frac{d^2+d}{2}
=
\frac{d(d+1)}{2}.
\end{eqnarray}

\medskip
\noindent
For the cubic term,
\begin{eqnarray}
A_3+A_2B_1+A_1B_2+B_3
&=&
\left(\frac{9}{2}d^3+\frac{9}{2}d^2+d\right)
+\left(-\frac{27}{2}d^3-\frac{9}{2}d^2\right)
+(15d^3-3d^2)
\nonumber\\
&&
+\left(-\frac{35}{6}d^3+\frac{5}{2}d^2-\frac{2}{3}d\right)
\nonumber\\
&=&
\frac{1}{6}d^3-\frac{1}{2}d^2+\frac{1}{3}d
=
\frac{d(d-1)(d-2)}{6}.
\end{eqnarray}

\medskip
\noindent
Therefore the exact \(SU(2)\) Molien--Weyl series is
\begin{eqnarray}
\boxed{
Z_{\mathrm{MW}}^{SU(2)}(x)
=
1+\frac{d(d+1)}{2}x^2+\frac{d(d-1)(d-2)}{6}x^3+O(x^4).
}
\label{MW_SU2_final_series}
\end{eqnarray}

\medskip
\noindent
Thus the exact closed form \eqref{MW_SU2_hyper} is fully consistent with the expected Molien--Weyl expansion: the linear term cancels exactly between the prefactor and the hypergeometric factor, and the first nontrivial terms are precisely the quadratic and cubic invariant multiplicities.

\subsection{Molien--Weyl vs. planar endpoint descriptions}
\medskip
\noindent
We have in fact two equivalent ways of organizing the planar sector:

\begin{itemize}
\item \textbf{The Molien--Weyl route:} This is the \emph{holonomy-last} scheme, in which one first integrates the planar endpoint variables at fixed holonomy angle \(\Delta\), and only afterwards performs the Haar integral over \(\Delta\).
\item \textbf{The planar endpoint route:} This is the \emph{holonomy-first} scheme adopted here, in which one first performs the \(\Delta\)-integral and obtains the Bessel kernel, and only afterwards integrates over the planar endpoint variables.
\end{itemize}

\medskip
\noindent
The purpose of this subsection is to show that the two schemes differ, at the planar level, by a simple multiplicative factor
\begin{eqnarray}
\mathcal{X}(x)=(1-x)^d(1-x^2)^d,
\label{X_def_final}
\end{eqnarray}
so that
\begin{eqnarray}
\widetilde Z_\perp(x)=\mathcal{X}(x)\,Z_{\rm MW}(x).
\label{dictionary_main_final}
\end{eqnarray}
Equivalently,
\begin{eqnarray}
Z_{\rm MW}(x)=\frac{\widetilde Z_\perp(x)}{(1-x)^d(1-x^2)^d}.
\label{dictionary_inverse_final}
\end{eqnarray}

\subsubsection{The fixed-\(\Delta\) planar endpoint integral}

\medskip
\noindent
After integrating out the bulk fields, the exact planar Gaussian boundary action is
\begin{eqnarray}
S_{\perp,0}
=
\lambda\sum_{a=1}^{d}
\left[
\alpha_\Lambda\Big((V_a)^2+(W_a)^2\Big)
-2\beta_\Lambda\,V_a\!\cdot\!W_a
\right],
\qquad
\lambda:=2\frac{N}{a}.
\label{Sperp0_recall_factor}
\end{eqnarray}
At fixed holonomy angle \(\Delta\), the holonomy coupling adds
\begin{eqnarray}
-\lambda\sum_{a=1}^{d}
\Big[
\cos\Delta\,(W_a\!\cdot\!V_a)
+\sin\Delta\,(W_a\!\times\!V_a)
\Big].
\label{fixedDelta_coupling_factor}
\end{eqnarray}
Thus, for one planar species \(a\), the fixed-\(\Delta\) quadratic form is
\begin{eqnarray}
S_{\perp,a}(\Delta)
=
\lambda
\left[
\alpha_\Lambda\Big((V_a)^2+(W_a)^2\Big)
-
2\beta_\Lambda\,V_a\!\cdot\!W_a
-
\cos\Delta\,(W_a\!\cdot\!V_a)
-
\sin\Delta\,(W_a\!\times\!V_a)
\right].
\label{S_one_species_fixedDelta}
\end{eqnarray}

\medskip
\noindent
Introduce the complex planar variables
\begin{eqnarray}
v_a:=V_a^1+iV_a^2,
\qquad
w_a:=W_a^1+iW_a^2.
\end{eqnarray}
Then
\begin{eqnarray}
|v_a|^2=(V_a)^2,
\qquad
|w_a|^2=(W_a)^2,
\end{eqnarray}
and
\begin{eqnarray}
W_a\!\cdot\!V_a+i(W_a\!\times\!V_a)=w_a^*v_a.
\end{eqnarray}
Hence
\begin{eqnarray}
S_{\perp,a}(\Delta)
=
\lambda
\left[
\alpha_\Lambda\Big(|v_a|^2+|w_a|^2\Big)
-2\,\Re\!\left(
\beta_\Lambda+\frac{e^{-i\Delta}}{2}
\right)w_a^*v_a
\right].
\end{eqnarray}
Equivalently,
\begin{eqnarray}
S_{\perp,a}(\Delta)
=
\lambda
\begin{pmatrix}
v_a^* & w_a^*
\end{pmatrix}
\begin{pmatrix}
\alpha_\Lambda &
-\Big(\beta_\Lambda+\frac{e^{-i\Delta}}{2}\Big)
\\
-\Big(\beta_\Lambda+\frac{e^{i\Delta}}{2}\Big) &
\alpha_\Lambda
\end{pmatrix}
\begin{pmatrix}
v_a \\ w_a
\end{pmatrix}.
\end{eqnarray}
Therefore the Gaussian integral over \((v_a,w_a)\) gives, up to a \(\Delta\)-independent normalization,
\begin{eqnarray}
Z_{\perp,a}(\Delta)
\propto
\frac{1}{
\alpha_\Lambda^2-
\left|\beta_\Lambda+\frac{e^{i\Delta}}{2}\right|^2
}.
\end{eqnarray}
Since
\begin{eqnarray}
\left|\beta_\Lambda+\frac{e^{i\Delta}}{2}\right|^2
=
\beta_\Lambda^2+\beta_\Lambda\cos\Delta+\frac14,
\end{eqnarray}
we find
\begin{eqnarray}
Z_{\perp,a}(\Delta)
\propto
\frac{1}{
\alpha_\Lambda^2-\beta_\Lambda^2-\beta_\Lambda\cos\Delta-\frac14
}.
\end{eqnarray}
Because the \(d\) species factorize,
\begin{eqnarray}
Z_{\perp}^{\rm fixed\text{-}\Delta}
\propto
\Big(
\alpha_\Lambda^2-\beta_\Lambda^2-\beta_\Lambda\cos\Delta-\frac14
\Big)^{-d}.
\label{fixedDelta_planar_raw}
\end{eqnarray}

\medskip
\noindent
Now take the continuum low-temperature limit. With
\begin{eqnarray}
x=e^{-\beta s},
\end{eqnarray}
we may define the continuum coefficients
\begin{eqnarray}
\alpha_c(x):=\frac{1+x^2}{1-x^2}=\coth(\beta s),
\qquad
\beta_c(x):=\frac{2x}{1-x^2}=\operatorname{csch}(\beta s),
\label{alphac_betac_def}
\end{eqnarray}
so that
\begin{eqnarray}
\alpha_c^2-\beta_c^2=1.
\end{eqnarray}
After removing the overall \(x\)-independent normalization, \eqref{fixedDelta_planar_raw} becomes
\begin{eqnarray}
\widetilde Z_\perp(\Delta)
=
\Big(\alpha_c-\beta_c\cos\Delta\Big)^{-d}
=
\left(
\frac{1-x^2}{1-2x\cos\Delta+x^2}
\right)^d.
\label{Ztilde_fixedDelta_final}
\end{eqnarray}
This is the fixed-\(\Delta\) planar integrand obtained in the holonomy-last route.

\subsubsection{Identification of the planar Molien--Weyl integrand}

\medskip
\noindent
For \(SU(2)\), the planar part of the Molien--Weyl integrand is given by 
\begin{eqnarray}
Z_{\rm MW}^{(\perp)}(\Delta)
\propto
\frac{1}{(1-x)^d(1-xe^{i\Delta})^d(1-xe^{-i\Delta})^d}.
\end{eqnarray}
Using
\begin{eqnarray}
(1-xe^{i\Delta})(1-xe^{-i\Delta})
=
1-2x\cos\Delta+x^2,
\end{eqnarray}
this may be rewritten as
\begin{eqnarray}
Z_{\rm MW}^{(\perp)}(\Delta)
\propto
\frac{1}{(1-x)^d(1-2x\cos\Delta+x^2)^d}.
\label{MW_fixedDelta_planar}
\end{eqnarray}

\medskip
\noindent
Comparing \eqref{Ztilde_fixedDelta_final} and \eqref{MW_fixedDelta_planar}, we immediately obtain
\begin{eqnarray}
\widetilde Z_\perp(\Delta)
=
(1-x)^d(1-x^2)^d\,
Z_{\rm MW}^{(\perp)}(\Delta).
\label{dictionary_fixedDelta}
\end{eqnarray}
Thus the two fixed-holonomy integrands differ only by the \(\Delta\)-independent factor
\begin{eqnarray}
\mathcal{X}(x):=(1-x)^d(1-x^2)^d.
\end{eqnarray}

\medskip
\noindent
Indeed, the reduced planar partition function is obtained by integrating the fixed-\(\Delta\) expression \eqref{Ztilde_fixedDelta_final} against the normalized \(SU(2)\) Haar measure
\begin{eqnarray}
d\mu(\Delta)=\frac{1}{\pi}\sin^2\frac{\Delta}{2}\,d\Delta.
\end{eqnarray}
Since \(\mathcal{X}(x)\) depends only on \(x\), and not on \(\Delta\), it can be pulled out of the holonomy integral:
\begin{eqnarray}
\widetilde Z_\perp(x)
&=&
\int d\mu(\Delta)\,\widetilde Z_\perp(\Delta)
\nonumber\\
&=&
\mathcal{X}(x)\int d\mu(\Delta)\,Z_{\rm MW}^{(\perp)}(\Delta)
\nonumber\\
&=&
\mathcal{X}(x)\,Z_{\rm MW}(x).
\label{dictionary_integrated}
\end{eqnarray}

\medskip
\noindent
The derivation above shows that the reduced partition function obtained from the planar endpoint route and the  Molien--Weyl partition function are related by a simple universal factor:
\begin{eqnarray}
\boxed{
\widetilde Z_\perp(\Delta)
=
(1-x)^d(1-x^2)^d\,Z_{\rm MW}^{(\perp)}(\Delta)
}
\label{dictionary_box_fixedDelta}
\end{eqnarray}
at fixed holonomy, and therefore
\begin{eqnarray}
\boxed{
\widetilde Z_\perp(x)
=
(1-x)^d(1-x^2)^d\,Z_{\rm MW}(x)
}
\label{dictionary_box_integrated}
\end{eqnarray}
after the Haar integral. Equivalently,
\begin{eqnarray}
\boxed{
Z_{\rm MW}(x)
=
\frac{\widetilde Z_\perp(x)}{(1-x)^d(1-x^2)^d}.
}
\end{eqnarray}
Thus the factor \(\mathcal{X}\) is a genuine spectator factor: it is independent of the holonomy angle, independent of the endpoint variables, and is simply carried through both integration schemes unchanged.

\subsection{Explicit cubic expansion of the planar partition function from Molien--Weyl}

\medskip
\noindent
We now evaluate explicitly the low-\(x\) expansion of the planar endpoint reduced partition function and compare it with the exact Molien--Weyl series. The exact planar dictionary is
\begin{eqnarray}
\widetilde Z_\perp(x)
=
{\cal X}(x)\,Z_{\rm MW}(x),
\qquad
{\cal X}(x):=(1-x)^d(1-x^2)^d.
\label{dictionary_planar_again}
\end{eqnarray}
Thus the problem reduces simply to expanding the spectator factor \({\cal X}(x)\) and multiplying it by the Molien--Weyl series.

\medskip
\noindent
For \(SU(2)\), the exact Molien--Weyl partition function has the small-\(x\) expansion
\begin{eqnarray}
Z_{\rm MW}(x)
=
1+\frac{d(d+1)}{2}x^2+\frac{d(d-1)(d-2)}{6}x^3+O(x^4).
\label{MW_series_explicit}
\end{eqnarray}
In particular, there is no linear term.

\medskip
\noindent Expansion of the spectator factor \(\mathcal{X}(x)\) yields

\begin{eqnarray}
{\cal X}(x)
&=&
1-dx+\frac{d(d-3)}{2}x^2+\frac{d(-d^2+9d-2)}{6}x^3+O(x^4).
\label{X_series_explicit}
\end{eqnarray}

\medskip
\noindent
Substituting \eqref{MW_series_explicit} and \eqref{X_series_explicit} into \eqref{dictionary_planar_again}, we get
\begin{eqnarray}
\widetilde Z_\perp(x)
&=&
\left(
1-dx+\frac{d(d-3)}{2}x^2+\frac{d(-d^2+9d-2)}{6}x^3
\right)
\left(
1+\frac{d(d+1)}{2}x^2+\frac{d(d-1)(d-2)}{6}x^3
\right)
\nonumber\\
&&
+O(x^4)\nonumber\\
\end{eqnarray}
\medskip
\noindent Equivalently, 
\begin{eqnarray}
\boxed{
\widetilde Z_\perp(x)
=
1-dx+d(d-1)x^2-\frac12 d^2(d-1)x^3+O(x^4).
}
\label{Ztilde_perp_final_explicit}
\end{eqnarray}

\section{Low-\texorpdfstring{$x$}{x} expansion as a resummed moment expansion}

\medskip
\noindent
We now explain systematically how the low-temperature expansion in \(x\) should be organized in the \(\Delta\)-first scheme, i.e. in our planar endpoint route. The basic point is that the exact holonomy kernel is first integrated over \(\Delta\), producing a Bessel potential which is a function in the collective planar invariants $A$ and $R$, and only afterwards averaged with the Gaussian endpoint measure. The resulting small-\(x\) expansion is therefore naturally a \emph{resummed moment expansion}.

\subsection{Resummed moment expansion of the reduced planar partition function}

\medskip
\noindent
For the planar sector, define the collective invariants
\begin{eqnarray}
A:=\lambda \sum_{a=1}^{d}W_a\!\cdot\!V_a,
\qquad
B:=\lambda \sum_{a=1}^{d}W_a\!\times\!V_a,
\qquad
R^2:=A^2+B^2.
\label{ABR_defs_resummed}
\end{eqnarray}
After performing the holonomy integral first, one obtains the exact kernel
\begin{eqnarray}
\Phi(A,B)
=
I_0(R)-\frac{A}{R}I_1(R).
\label{Phi_AB_resummed}
\end{eqnarray}
Hence the reduced planar partition function is

\begin{eqnarray}
\widetilde Z_\perp
=Z_{\perp,0}
\langle \Phi(A,B)\rangle_0.
\label{ZtildePhiavg}
\end{eqnarray}
\medskip
\noindent
Here \(\langle\cdots\rangle_0\) denotes expectation value with respect to the planar Gaussian endpoint measure
\begin{eqnarray}
\langle {\cal O}(V,W)\rangle_0
&:=&
\frac{1}{Z_{\perp,0}}
\int \prod_{a=1}^{d} d^2V_a\,d^2W_a\;
\exp\!\left[
-\lambda\sum_{a=1}^{d}
\left(
\alpha_\Lambda\big((V_a)^2+(W_a)^2\big)
-2\beta_\Lambda\,V_a\!\cdot\!W_a
\right)
\right]
\,{\cal O}(V,W),\nonumber\\
\label{planar_measure_expectation}
\end{eqnarray}
with normalization
\begin{eqnarray}
Z_{\perp,0}
&=&
\int \prod_{a=1}^{d} d^2V_a\,d^2W_a\;
\exp\!\left[
-\lambda\sum_{a=1}^{d}
\left(
\alpha_\Lambda\big((V_a)^2+(W_a)^2\big)
-2\beta_\Lambda\,V_a\!\cdot\!W_a
\right)
\right]
\nonumber\\
&=&
\left(\frac{\pi}{\lambda}\right)^{2d}
\left(\alpha_\Lambda^2-\beta_\Lambda^2\right)^{-d},
\qquad
\lambda:=2\frac{N}{a}.
\label{planar_measure_normalization}
\end{eqnarray}

\medskip
\noindent
Expanding the Bessel functions gives
\begin{eqnarray}
I_0(R)
=
\sum_{n=0}^{\infty}\frac{1}{(n!)^2}\left(\frac{R^2}{4}\right)^n,
\qquad
\frac{A}{R}I_1(R)
=
A\sum_{n=0}^{\infty}\frac{1}{2\cdot 4^n\,n!(n+1)!}R^{2n},
\end{eqnarray}
so that
\begin{eqnarray}
\Phi(A,B)
=
\sum_{n=0}^{\infty}c_nR^{2n}
-
A\sum_{n=0}^{\infty}d_nR^{2n},
\label{Phi_cd_series}
\end{eqnarray}
with
\begin{eqnarray}
c_n=\frac{1}{4^n(n!)^2},
\qquad
d_n=\frac{1}{2\cdot 4^n\,n!(n+1)!}.
\label{cd_coeffs_resummed}
\end{eqnarray}
The first few terms are
\begin{eqnarray}
\Phi(A,B)
=
1-\frac{A}{2}
+\frac{R^2}{4}
-\frac{AR^2}{16}
+\frac{R^4}{64}
-\frac{AR^4}{384}
+\frac{R^6}{2304}
-\frac{AR^6}{18432}
+\cdots.
\label{Phi_first_terms_resummed}
\end{eqnarray}

\medskip
\noindent
The key observation is that \eqref{Phi_first_terms_resummed} is an expansion in the collective variables \(A\) and \(R\), \emph{not} an expansion in the thermal parameter \(x\). At \(x=0\), the Gaussian measure is invariant under
\begin{eqnarray}
W_a\mapsto -W_a,
\end{eqnarray}
under which
\begin{eqnarray}
A\mapsto -A,
\qquad
B\mapsto -B,
\qquad
R^2\mapsto R^2.
\end{eqnarray}
Therefore
\begin{eqnarray}
\langle R^{2n}\rangle_0
=
E_{n,0}+E_{n,2}x^2+E_{n,4}x^4+\cdots,
\label{even_moment_structure}
\end{eqnarray}
and
\begin{eqnarray}
\langle A\,R^{2n}\rangle_0
=
O_{n,1}x+O_{n,3}x^3+O_{n,5}x^5+\cdots.
\label{odd_moment_structure}
\end{eqnarray}
Thus every even tower contributes already to the constant and quadratic terms in \(x\), while every odd tower contributes already to the linear and cubic terms. This shows that a finite truncation of \eqref{Phi_first_terms_resummed} cannot, in general, produce a controlled expansion up to a fixed power of \(x\).

\medskip
\noindent
Using now \eqref{Phi_cd_series}, the reduced planar partition function becomes
\begin{eqnarray}
\frac{\widetilde Z_\perp}{Z_{\perp,0}}
=
\sum_{n=0}^{\infty}c_n\langle R^{2n}\rangle_0
-
\sum_{n=0}^{\infty}d_n\langle A\,R^{2n}\rangle_0.
\label{Ztilde_moment_sum}
\end{eqnarray}
Substituting \eqref{even_moment_structure} and \eqref{odd_moment_structure}, one finds the low-\(x\) expansion
\begin{eqnarray}
\frac{\widetilde Z_\perp}{Z_{\perp,0}}
&=&
\sum_{n=0}^{\infty}c_nE_{n,0}
-\left(\sum_{n=0}^{\infty}d_nO_{n,1}\right)x
+\left(\sum_{n=0}^{\infty}c_nE_{n,2}\right)x^2
\nonumber\\
&&
-\left(\sum_{n=0}^{\infty}d_nO_{n,3}\right)x^3
+O(x^4).
\label{Ztilde_cubic_general}
\end{eqnarray}
Therefore the cubic expansion is controlled by the four infinite families
\begin{eqnarray}
E_{n,0},\qquad E_{n,2},\qquad O_{n,1},\qquad O_{n,3},
\qquad n=0,1,2,\dots.
\label{four_families}
\end{eqnarray}

\subsection{Generating function for Gaussian moments}

\medskip
\noindent
The planar Gaussian moments are generated by
\begin{eqnarray}
{\cal M}(t,u)
:=
\big\langle e^{\,tA+uB}\big\rangle_0.
\label{Mtu_def_full}
\end{eqnarray}

\medskip
\noindent
Since the planar Gaussian action is a sum over \(a=1,\dots,d\), and \(A\) and \(B\) are also sums over \(a\), this generating function factorizes:
\begin{eqnarray}
{\cal M}(t,u)
=
\prod_{a=1}^{d}
\frac{1}{z_1(t,u)}
\int d^2V\,d^2W\;
\exp\!\left[
-\lambda
\left(
\alpha_\Lambda(V^2+W^2)-2\beta_\Lambda V\!\cdot\!W
\right)
+t\lambda\,W\!\cdot\!V
+u\lambda\,W\!\times\!V
\right].\nonumber\\
\end{eqnarray}
Here \(z_1(t,u)\) denotes the one-species normalization factor at \(t=u=0\). Thus it is enough to compute the one-species integral and then raise the result to the \(d\)-th power.

\medskip
\noindent
For one planar species, introduce
\begin{eqnarray}
V=(V^1,V^2),
\qquad
W=(W^1,W^2).
\end{eqnarray}
Then
\begin{eqnarray}
W\!\cdot\!V
=
W^1V^1+W^2V^2,
\qquad
W\!\times\!V
=
W^1V^2-W^2V^1.
\end{eqnarray}
Hence the exponent may be written as
\begin{eqnarray}
-\lambda\alpha_\Lambda(V^2+W^2)
+
\lambda(2\beta_\Lambda+t)\,W\!\cdot\!V
+
\lambda u\,W\!\times\!V.
\label{one_species_exponent_real}
\end{eqnarray}

\medskip
\noindent
It is convenient to pass to complex variables
\begin{eqnarray}
v:=V^1+iV^2,
\qquad
w:=W^1+iW^2.
\end{eqnarray}
Then
\begin{eqnarray}
|v|^2=V^2,
\qquad
|w|^2=W^2,
\end{eqnarray}
and
\begin{eqnarray}
W\!\cdot\!V+i\,W\!\times\!V
=
w^*v.
\end{eqnarray}
Therefore
\begin{eqnarray}
(2\beta_\Lambda+t)\,W\!\cdot\!V+u\,W\!\times\!V
=
\Re\!\Big((2\beta_\Lambda+t-iu)\,w^*v\Big).
\end{eqnarray}
The one-species exponent becomes
\begin{eqnarray}
-\lambda
\left[
\alpha_\Lambda(|v|^2+|w|^2)
-\Re\!\Big((2\beta_\Lambda+t-iu)\,w^*v\Big)
\right].
\end{eqnarray}

\medskip
\noindent
This is a two-complex-variable Gaussian, equivalently a four-real-variable Gaussian. Its quadratic matrix is
\begin{eqnarray}
Q(t,u)
=
\begin{pmatrix}
\alpha_\Lambda &
-\left(\beta_\Lambda+\frac{t-iu}{2}\right)
\\
-\left(\beta_\Lambda+\frac{t+iu}{2}\right) &
\alpha_\Lambda
\end{pmatrix}.
\end{eqnarray}
Hence the one-species Gaussian integral is proportional to
\begin{eqnarray}
\det Q(t,u)^{-1}.
\end{eqnarray}
A direct computation gives
\begin{eqnarray}
\det Q(t,u)
=
\alpha_\Lambda^2-
\left(\beta_\Lambda+\frac{t-iu}{2}\right)
\left(\beta_\Lambda+\frac{t+iu}{2}\right)
=
D_\Lambda-\beta_\Lambda t-\frac{t^2+u^2}{4}.
\label{detQtuexplicit}
\end{eqnarray}
At \(t=u=0\), this reduces to
\begin{eqnarray}
\det Q(0,0)=\alpha_\Lambda^2-\beta_\Lambda^2=D_\Lambda.
\end{eqnarray}
Therefore the normalized one-species generating function is
\begin{eqnarray}
{\cal M}_1(t,u)
=
\frac{\det Q(0,0)}{\det Q(t,u)}
=
\frac{D_\Lambda}{D_\Lambda-\beta_\Lambda t-\frac{t^2+u^2}{4}}.
\label{M1tu_final}
\end{eqnarray}

\medskip
\noindent
Since the \(d\) species are independent and identical, the full generating function is just
\begin{eqnarray}
{\cal M}(t,u)
=
\bigl({\cal M}_1(t,u)\bigr)^d
=
\left[
\frac{D_\Lambda}
{D_\Lambda-\beta_\Lambda t-\frac{t^2+u^2}{4}}
\right]^d.
\label{Mtu_exact_full_derived}
\end{eqnarray}
This generating function contains all mixed moments of \(A\) and \(B\).

\medskip
\noindent
The Gaussian moments entering \eqref{Ztilde_cubic_general} are extracted from \({\cal M}(t,u)\) by the differential rules
\begin{eqnarray}
\langle R^{2n}\rangle_0
=
\left.
(\partial_t^2+\partial_u^2)^n {\cal M}(t,u)
\right|_{t=u=0},
\label{extract_even}
\end{eqnarray}
and
\begin{eqnarray}
\langle A\,R^{2n}\rangle_0
=
\left.
\partial_t(\partial_t^2+\partial_u^2)^n {\cal M}(t,u)
\right|_{t=u=0}.
\label{extract_odd}
\end{eqnarray}
Thus the full problem is reduced to the systematic expansion of \({\cal M}(t,u)\) in powers of \(x\).

\subsection{Expansion of the generating function in \texorpdfstring{$x$}{x}}

\medskip
\noindent
Write
\begin{eqnarray}
\beta_\Lambda=\beta_1x+\beta_3x^3+O(x^5),
\qquad
D_\Lambda=D_0+D_2x^2+O(x^4).
\label{betaD_expand_full}
\end{eqnarray}
Then
\begin{eqnarray}
{\cal M}(t,u)
=
{\cal M}_0(t,u)
+x\,{\cal M}_1(t,u)
+x^2\,{\cal M}_2(t,u)
+x^3\,{\cal M}_3(t,u)
+O(x^4),
\label{M_expand_x_full}
\end{eqnarray}
with
\begin{eqnarray}
{\cal M}_0(t,u)
&=&
\left(
1-\frac{t^2+u^2}{4D_0}
\right)^{-d},
\label{M0_full}
\\
{\cal M}_1(t,u)
&=&
\frac{d\beta_1}{D_0}\,
t\,
\left(
1-\frac{t^2+u^2}{4D_0}
\right)^{-d-1},
\label{M1_full}
\\
{\cal M}_2(t,u)
&=&
-\frac{dD_2}{D_0}\,
\frac{t^2+u^2}{4D_0}\,
\left(
1-\frac{t^2+u^2}{4D_0}
\right)^{-d-1}
+\frac{d(d+1)\beta_1^2}{2D_0^2}\,
t^2\,
\left(
1-\frac{t^2+u^2}{4D_0}
\right)^{-d-2},
\label{M2_full}
\\
{\cal M}_3(t,u)
&=&
\frac{d\beta_3}{D_0}\,
t\,
\left(
1-\frac{t^2+u^2}{4D_0}
\right)^{-d-1}
-
\frac{d(d+1)D_2\beta_1}{D_0^2}\,
t\,
\left(
1-\frac{t^2+u^2}{4D_0}
\right)^{-d-2}
\nonumber\\
&&
+\frac{d^2D_2\beta_1}{D_0^2}\,
t\,
\left(
1-\frac{t^2+u^2}{4D_0}
\right)^{-d-1}
+
\frac{d(d+1)(d+2)\beta_1^3}{6D_0^3}\,
t^3\,
\left(
1-\frac{t^2+u^2}{4D_0}
\right)^{-d-3}.
\label{M3_full}
\end{eqnarray}

\medskip
\noindent
Applying \eqref{extract_even} and \eqref{extract_odd} to \eqref{M0_full}--\eqref{M3_full}, one obtains the four families needed in \eqref{Ztilde_cubic_general}.

\smallskip
\noindent
For the constant part of the even tower,
\begin{eqnarray}
E_{n,0}
=
\frac{n!\,(d)_n}{D_0^n}.
\label{En0_full}
\end{eqnarray}

\smallskip
\noindent
For the linear part of the odd tower,
\begin{eqnarray}
O_{n,1}
=
\frac{d\,\beta_1\,(n+1)!\,(d+1)_n}{D_0^{n+1}}.
\label{On1_full}
\end{eqnarray}

\smallskip
\noindent
For the quadratic part of the even tower,
\begin{eqnarray}
E_{0,2}=0,
\qquad
E_{n,2}
=
-\frac{d\,D_2\,n\,n!\,(d+1)_{n-1}}{D_0^{n+1}}
+
\frac{d\,\beta_1^2\,n\,n!\,(d+1)_n}{D_0^{n+1}},
\qquad n\ge1.
\label{En2_full}
\end{eqnarray}

\smallskip
\noindent
For the cubic part of the odd tower,
\begin{eqnarray}
O_{n,3}
&=&
\frac{d\,\beta_3\,(n+1)!\,(d+1)_n}{D_0^{n+1}}
-
\frac{d\,D_2\beta_1\,(n+1)(n+1)!\,(d+1)_n}{D_0^{n+2}}
\nonumber\\
&&
+
\frac{d\,\beta_1^3\,n\,(n+1)!\,(d+1)_{n+1}}{2D_0^{n+2}}.
\label{On3_full}
\end{eqnarray}

\medskip
\noindent
Combining \eqref{Ztilde_cubic_general} with \eqref{En0_full}--\eqref{On3_full}, the reduced planar partition function up to cubic order in \(x\) is
\begin{eqnarray}
\frac{\widetilde Z_\perp}{Z_{\perp,0}}
&=&
\sum_{n=0}^{\infty}\frac{E_{n,0}}{4^n(n!)^2}
-
\left[
\sum_{n=0}^{\infty}\frac{O_{n,1}}{2\cdot 4^n\,n!(n+1)!}
\right]x
+
\left[
\sum_{n=0}^{\infty}\frac{E_{n,2}}{4^n(n!)^2}
\right]x^2\nonumber\\
&-&
\left[
\sum_{n=0}^{\infty}\frac{O_{n,3}}{2\cdot 4^n\,n!(n+1)!}
\right]x^3
+O(x^4).
\label{Ztilde_resummed_cubic_final}
\end{eqnarray}
This is the desired resummed moment expansion. It makes explicit both the principle of the construction and the precise coefficient families entering the cubic low-\(x\) expansion.

\subsection{Explicit evaluation of the resummed planar cubic expansion}

\medskip
\noindent
Introduce
\begin{eqnarray}
z:=\frac{1}{4D_0}.
\label{z_def_last}
\end{eqnarray}

\medskip
\noindent
Then the constant term is given by 
\begin{eqnarray}
\sum_{n=0}^{\infty}\frac{E_{n,0}}{4^n(n!)^2}
&=&
\sum_{n=0}^{\infty}\frac{(d)_n}{n!}z^n
=
(1-z)^{-d}.
\label{sum_E0}
\end{eqnarray}

\medskip
\noindent
The coefficient of \(x\) is
\begin{eqnarray}
\sum_{n=0}^{\infty}\frac{O_{n,1}}{2\cdot 4^n\,n!(n+1)!}
&=&
\frac{d\beta_1}{2D_0}
\sum_{n=0}^{\infty}\frac{(d+1)_n}{n!}z^n
=
\frac{d\beta_1}{2D_0}(1-z)^{-d-1}.
\label{sum_O1}
\end{eqnarray}

\medskip
\noindent
The coefficient of \(x^2\) is
\begin{eqnarray}
\sum_{n=0}^{\infty}\frac{E_{n,2}}{4^n(n!)^2}
&=&
-\frac{dD_2}{4D_0^2}
\sum_{n=1}^{\infty}\frac{(d+1)_{n-1}}{(n-1)!}z^{\,n-1}
+
\frac{d(d+1)\beta_1^2}{4D_0^2}
\sum_{n=1}^{\infty}\frac{(d+2)_{n-1}}{(n-1)!}z^{\,n-1}
\nonumber\\
&=&
-\frac{dD_2}{4D_0^2}(1-z)^{-d-1}
+
\frac{d(d+1)\beta_1^2}{4D_0^2}(1-z)^{-d-2}.
\label{sum_E2}
\end{eqnarray}

\medskip
\noindent
The coefficient of \(x^3\) is
\begin{eqnarray}
\sum_{n=0}^{\infty}\frac{O_{n,3}}{2\cdot 4^n\,n!(n+1)!}
&=&
\frac{d\beta_3}{2D_0}
\sum_{n=0}^{\infty}\frac{(d+1)_n}{n!}z^n
-
\frac{dD_2\beta_1}{2D_0^2}
\sum_{n=0}^{\infty}\frac{(n+1)(d+1)_n}{n!}z^n\nonumber\\
&+&
\frac{d(d+1)(d+2)\beta_1^3}{16D_0^3}
\sum_{n=1}^{\infty}\frac{(d+3)_{n-1}}{(n-1)!}z^{\,n-1}
\nonumber\\
&=&
\frac{d\beta_3}{2D_0}(1-z)^{-d-1}
-
\frac{dD_2\beta_1}{2D_0^2}(1+dz)(1-z)^{-d-2}\nonumber\\
&+&
\frac{d(d+1)(d+2)\beta_1^3}{16D_0^3}(1-z)^{-d-3}.
\label{sum_O3}
\end{eqnarray}
In the second line we used
\begin{eqnarray}
\sum_{n=0}^{\infty}\frac{(n+1)(d+1)_n}{n!}z^n
=
(1+dz)(1-z)^{-d-2}.
\label{aux_sum}
\end{eqnarray}

\medskip
\noindent
Substituting \eqref{sum_E0}--\eqref{sum_O3} into \eqref{Ztilde_resummed_cubic_final}, we obtain
\begin{eqnarray}
\frac{\widetilde Z_\perp}{Z_{\perp,0}}&=&
(1-z)^{-d}
-
\frac{d\beta_1}{2D_0}(1-z)^{-d-1}x
\nonumber\\
&&
+
\left[
-\frac{dD_2}{4D_0^2}(1-z)^{-d-1}
+\frac{d(d+1)\beta_1^2}{4D_0^2}(1-z)^{-d-2}
\right]x^2
\nonumber\\
&&
-
\left[
\frac{d\beta_3}{2D_0}(1-z)^{-d-1}
-\frac{dD_2\beta_1}{2D_0^2}(1+dz)(1-z)^{-d-2}
+\frac{d(d+1)(d+2)\beta_1^3}{16D_0^3}(1-z)^{-d-3}
\right]x^3
\nonumber\\
&&
+O(x^4).
\label{Ztilde_before_cont}
\end{eqnarray}

\medskip
\noindent
At low temperature in the continuum limit one has
\begin{eqnarray}
\beta_\Lambda=\beta_1x+\beta_3x^3+O(x^5),
\qquad
D_\Lambda=D_0+D_2x^2+O(x^4),
\end{eqnarray}
with
\begin{eqnarray}
\beta_1=\beta_3=\mu,
\qquad
D_0=\frac{1+2\mu}{4},
\qquad
D_2=\mu,
\qquad
\mu:=as.
\label{continuum_data_last}
\end{eqnarray}
Hence
\begin{eqnarray}
z=\frac{1}{4D_0}=\frac{1}{1+2\mu},
\qquad
1-z=\frac{2\mu}{1+2\mu}.
\end{eqnarray}

\medskip
\noindent
The overall factor \((1-z)^{-d}\), arising in the resummed moment expansion, is the same massless normalization artifact that also appears in the raw fixed-\(\Delta\) planar partition function \eqref{fixedDelta_planar_raw}. Indeed, in the continuum low-temperature regime one has
\begin{eqnarray}
(1-z)^{-d}
=
\left(\frac{1+2\mu}{2\mu}\right)^d.
\end{eqnarray}
Thus it diverges as \(\mu^{-d}\) in the strict limit \(\mu\to0\). It should therefore not be regarded as part of the normalized thermal answer, but rather as the raw \(x=0\) normalization of the massless planar Gaussian.

\medskip
\noindent Factoring out this \(x\)-independent normalization \((1-z)^{-d}\), we define the normalized planar partition function by
\begin{eqnarray}
\widehat Z_\perp
:=
(1-z)^d\frac{\widetilde Z_\perp}{Z_{\perp,0}}.
\label{Zhat_def}
\end{eqnarray}

\medskip
\noindent The partition function \eqref{Ztilde_before_cont}, after projecting out this factor, becomes
\begin{eqnarray}
\widehat Z_\perp
&=&
1
-\frac{d\beta_1}{2D_0(1-z)}x
+
\left[
-\frac{dD_2}{4D_0^2(1-z)}
+\frac{d(d+1)\beta_1^2}{4D_0^2(1-z)^2}
\right]x^2
\nonumber\\
&&
-
\left[
\frac{d\beta_3}{2D_0(1-z)}
-\frac{dD_2\beta_1}{2D_0^2}\frac{1+dz}{(1-z)^2}
+\frac{d(d+1)(d+2)\beta_1^3}{16D_0^3(1-z)^3}
\right]x^3
+O(x^4).
\label{Zhat_before_subst}
\end{eqnarray}
Substituting \eqref{continuum_data_last}, one finds
\begin{eqnarray}
\frac{\beta_1}{2D_0(1-z)}=1,
\qquad
\frac{D_2}{4D_0^2(1-z)}=\frac{2}{1+2\mu},
\qquad
\frac{\beta_1^2}{4D_0^2(1-z)^2}=1,
\end{eqnarray}
and
\begin{eqnarray}
\frac{\beta_3}{2D_0(1-z)}=1,
\qquad
\frac{D_2\beta_1}{2D_0^2}\frac{1+dz}{(1-z)^2}
=
\frac{2(1+2\mu+d)}{1+2\mu},
\qquad
\frac{\beta_1^3}{16D_0^3(1-z)^3}=1.
\end{eqnarray}
Therefore
\begin{eqnarray}
\widehat Z_\perp
&=&
1-dx
+
\left[
-\frac{2d}{1+2\mu}+d(d+1)
\right]x^2
\nonumber\\
&&
-
\left[
d-\frac{2d(1+2\mu+d)}{1+2\mu}+d(d+1)(d+2)
\right]x^3
+O(x^4).
\label{Zhat_mu_finite}
\end{eqnarray}
Finally, in the strict continuum limit \(\mu\to0\),
\begin{eqnarray}
\boxed{
\widehat Z_\perp(x)
=
1-dx+d(d-1)x^2-\frac12 d^2(d-1)x^3+O(x^4).
}
\label{Zhat_final_explicit}
\end{eqnarray}
\medskip
\noindent
This is precisely the expansion \eqref{Ztilde_perp_final_explicit}.

\section{Physics and geometry of the holonomy potential}\label{section 4}

\subsection{Exact saddle structure of the full holonomy potential}

\medskip
\noindent
Consider the exact holonomy potential
\begin{eqnarray}
V(A,R)
:=V_{\rm hol}(A,R)=
-\log\!\left(
I_0(R)-\frac{A}{R}I_1(R)
\right),
\qquad
-R\le A\le R,
\qquad
R\ge 0.
\label{V_exact_full}
\end{eqnarray}
Thus we treat \(A\) and \(R\) as independent variables on the physical domain
\begin{eqnarray}
|A|\le R.
\end{eqnarray}

\paragraph{The \texorpdfstring{$(A,R^2)$}{(A,R²)} formulation.}

\medskip
\noindent
It is useful to reformulate the exact holonomy potential in terms of the variables
\begin{eqnarray}
A,
\qquad
X:=R^2.
\end{eqnarray}
The reason is that both
\begin{eqnarray}
I_0(R)=I_0(\sqrt{X})
\quad\text{and}\quad
\frac{I_1(R)}{R}=\frac{I_1(\sqrt{X})}{\sqrt{X}},
\end{eqnarray}
are functions of \(X\). Thus the exact effective potential can be written as
\begin{eqnarray}
V(A,X)
=
-\log\!\left(
I_0(\sqrt{X})-A\,\frac{I_1(\sqrt{X})}{\sqrt{X}}
\right).
\label{Veff_AX_exact}
\end{eqnarray}
Equivalently, introducing
\begin{eqnarray}
f(X):=I_0(\sqrt{X}),
\qquad
g(X):=\frac{I_1(\sqrt{X})}{\sqrt{X}},
\label{fg_def}
\end{eqnarray}
one has
\begin{eqnarray}
V(A,X)
=
-\log\!\Big(f(X)-A\,g(X)\Big).
\label{Veff_AX_fg}
\end{eqnarray}

\medskip
\noindent
The physical domain is not the whole \((A,X)\)-plane. Since
\begin{eqnarray}
R^2=A^2+B^2,
\end{eqnarray}
one must impose
\begin{eqnarray}
A^2\le X.
\label{physical_domain_AX}
\end{eqnarray}
The boundary of the physical domain is therefore
\begin{eqnarray}
X=A^2,
\label{boundary_AX}
\end{eqnarray}
and for the aligned branch one takes \(A\ge0\), so that \(A=R\).

\paragraph{Absence of an unconstrained critical point.}

\medskip
\noindent
The derivatives of \eqref{Veff_AX_fg} are
\begin{eqnarray}
\partial_A V(A,X)
=
\frac{g(X)}{f(X)-A\,g(X)},
\label{dVdA_AX}
\end{eqnarray}
and
\begin{eqnarray}
\partial_X V(A,X)
=
-\frac{f'(X)-A\,g'(X)}{f(X)-A\,g(X)}.
\label{dVdX_AX}
\end{eqnarray}
In particular, \(\partial_A V>0\) throughout the physical domain, since
\begin{eqnarray}
g(X)>0,
\qquad
f(X)-A\,g(X)>0.
\end{eqnarray}
Hence there is no interior critical point obtained by setting \(\partial_A V=0\).

\medskip
\noindent
Similarly, if one asks for an unconstrained critical point in the ambient \((A,X)\)-plane, one would have to solve
\begin{eqnarray}
\partial_A V(A,X)=0,
\qquad
\partial_X V(A,X)=0.
\label{unconstrained_critical_eqs}
\end{eqnarray}
But the first equation in \eqref{unconstrained_critical_eqs} has no solution. Therefore:

\begin{eqnarray}
\boxed{
\text{there is no unconstrained critical point of }V(A,X)\text{ in the physical domain.}
}
\end{eqnarray}

\paragraph{Restriction to the physical boundary.}

\medskip
\noindent
The relevant stationary structure therefore lives on the boundary
\begin{eqnarray}
X=A^2,
\qquad
A\ge0.
\end{eqnarray}
Restricting \eqref{Veff_AX_exact} to this branch gives
\begin{eqnarray}
V_{\rm b}(A)
:=
V(A,A^2)
=
-\log\!\Big(I_0(A)-I_1(A)\Big).
\label{Vb_def}
\end{eqnarray}
The boundary stationary point \(A=A_*\) is determined by
\begin{eqnarray}
\frac{d}{dA}V_{\rm b}(A_*)=0.
\label{boundary_saddle_eq}
\end{eqnarray}
Using
\begin{eqnarray}
\frac{d}{dA}\Big(I_0(A)-I_1(A)\Big)
=
I_1(A)-I_1'(A),
\end{eqnarray}
together with
\begin{eqnarray}
I_1'(A)=I_0(A)-\frac{1}{A}I_1(A),
\end{eqnarray}
one obtains
\begin{eqnarray}
I_0(A_*)=\left(1+\frac{1}{A_*}\right)I_1(A_*).
\label{Astar_eq}
\end{eqnarray}
Numerically,
\begin{eqnarray}
A_*\approx 1.545.
\label{Astar_num}
\end{eqnarray}
Since on the aligned branch \(A=R\), one also has
\begin{eqnarray}
R_*=A_*,
\qquad
X_*=R_*^2=A_*^2\approx 2.387.
\label{Rstar_Xstar_num}
\end{eqnarray}

\paragraph{Why this is not an unconstrained saddle.}

\medskip
\noindent
The point \((A_*,X_*)\) is a constrained saddle because the effective potential is defined on the restricted domain \(X\ge A^2\), rather than on the full \((A,X)\)-plane. The relevant stationary point is realized on the boundary \(X=A^2\), and more specifically on its aligned component \(A=R\). Equivalently, \((A_*,X_*)\) becomes stationary only after restricting to the boundary

\begin{eqnarray}
X=A^2.
\end{eqnarray}
Indeed, by the chain rule,
\begin{eqnarray}
\frac{d}{dA}V(A,A^2)
=
\partial_A V(A,X)\Big|_{X=A^2}
+
2A\,\partial_X V(A,X)\Big|_{X=A^2}.
\label{chain_rule_boundary}
\end{eqnarray}

\medskip
\noindent Therefore the boundary saddle condition \eqref{boundary_saddle_eq} means
\begin{eqnarray}
\partial_A V(A_*,X_*)
+
2A_*\,\partial_X V(A_*,X_*)
=
0.
\label{boundary_stationarity_vector}
\end{eqnarray}
This is the precise sense in which \((A_*,X_*)\) is a \emph{constrained} saddle: the full ambient gradient
\begin{eqnarray}
\nabla V(A_*,X_*)
=
\left(\partial_A V(A_*,X_*),\,\partial_X V(A_*,X_*)\right)
\end{eqnarray}
does not vanish, i.e.
\begin{eqnarray}
\partial_A V(A_*,X_*)\neq 0,
\qquad
\partial_X V(A_*,X_*)\neq 0,
\end{eqnarray}
but its contraction with the tangent vector to the boundary \(X=A^2\),
\begin{eqnarray}
(1,2A_*),
\end{eqnarray}
does vanish, as expressed in equation \eqref{boundary_stationarity_vector}.

\medskip
\noindent
Indeed, an \emph{unconstrained} saddle would have required the full gradient to vanish:
\begin{eqnarray}
\partial_A V(A_*,X_*)=0,
\qquad
\partial_X V(A_*,X_*)=0.
\end{eqnarray}
This does not happen here. Hence the surviving linear terms in the ambient Taylor expansion are not a contradiction: they encode the fact that the saddle is a boundary saddle, not an interior one.

\medskip
\noindent 
Let us compute explicitly the ambient gradient at the saddle. At the boundary saddle one has the condition
\begin{eqnarray}
I_0(R_*)-I_1(R_*)
=
\frac{I_1(R_*)}{R_*},
\label{F0_identity}
\end{eqnarray}
by rearranging \eqref{Astar_eq}. Using this in \eqref{dVdA_AX}, one finds
\begin{eqnarray}
\partial_A V(A_*,X_*)
=
1.
\label{VAstar}
\end{eqnarray}
Similarly, one finds
\begin{eqnarray}
\partial_X V(A_*,X_*)
=
-\frac{1}{2R_*}.
\label{VXstar}
\end{eqnarray}
Thus
\begin{eqnarray}
\partial_A V(A_*,X_*)
+
2A_*\,\partial_X V(A_*,X_*)
=
1-\frac{A_*}{R_*}
=
0,
\end{eqnarray}
since \(A_*=R_*\). This verifies \eqref{boundary_stationarity_vector} explicitly.

\medskip
\noindent
This constrained saddle is actually a maximum along the aligned branch. The second derivative of the boundary potential \eqref{Vb_def} is
\begin{eqnarray}
V_{\rm b}''(A)
=
-\frac{F''(A)}{F(A)}
+
\frac{F'(A)^2}{F(A)^2},
\qquad
F(A):=I_0(A)-I_1(A).
\end{eqnarray}
At the saddle \(A=A_*\), since \(F'(A_*)=0\), this simplifies to
\begin{eqnarray}
V_{\rm b}''(A_*)
=
-\frac{F''(A_*)}{F(A_*)}.
\label{Vb_second}
\end{eqnarray}
Numerically one finds
\begin{eqnarray}
V_{\rm b}''(A_*)<0,
\end{eqnarray}
hence the point is a local maximum along the physical aligned branch.

\medskip
\noindent
This is the correct physical picture: \((A_*,X_*)\) is not a genuine ambient extremum, but it is the maximizer of the exact effective potential along the boundary branch \(X=A^2\), \(A\ge0\), which is the relevant branch selected by the holonomy counting problem.

\subsection{Quadratic expansion in the ambient variables \texorpdfstring{$(A,X)$}{(A,X)}}

\paragraph{Expansion around \texorpdfstring{$(A-A_*,X-X_*)=(0,0)$}.}

\medskip
\noindent
We now expand \(V(A,X)\) around the saddle point
\begin{eqnarray}
(A_*,X_*)=(R_*,R_*^2).
\end{eqnarray}
Writing
\begin{eqnarray}
\delta A:=A-A_*,
\qquad
\delta X:=X-X_*,
\end{eqnarray}
the Taylor expansion to second order is
\begin{eqnarray}
V(A,X)
&\simeq&
V_*
+
V_A^*\,\delta A
+
V_X^*\,\delta X
+
\frac12 V_{AA}^*\,\delta A^2
+
V_{AX}^*\,\delta A\,\delta X
+
\frac12 V_{XX}^*\,\delta X^2,
\label{ambient_taylor}
\end{eqnarray}
where all derivatives are evaluated at \((A_*,X_*)\), and
\begin{eqnarray}
V_*:=V(A_*,X_*)
=
-\log\!\Big(I_0(R_*)-I_1(R_*)\Big).
\end{eqnarray}

\medskip
\noindent
The derivatives at the saddle are
\begin{eqnarray}
V_A^*=1,
\qquad
V_X^*=-\frac{1}{2R_*},
\label{first_derivs_star}
\end{eqnarray}
\begin{eqnarray}
V_{AA}^*=1,
\qquad
V_{AX}^*=-\frac{1}{2R_*^2},
\qquad
V_{XX}^*=\frac{2-R_*}{2R_*^3}.
\label{second_derivs_star}
\end{eqnarray}
Therefore the quadratic approximation is
\begin{eqnarray}
V(A,X)
&=&
V_*
+
(A-A_*)
-\frac{1}{2R_*}(X-X_*)
+\frac12 (A-A_*)^2\nonumber\\
&-&\frac{1}{2R_*^2}(A-A_*)(X-X_*)
+\frac{2-R_*}{4R_*^3}(X-X_*)^2.
\label{quadratic_AX_final}
\end{eqnarray}

\medskip
\noindent
Restoring \(X=R^2\), this becomes
\begin{eqnarray}
V(A,R^2)
&=&
V_*
+
(A-A_*)
-\frac{1}{2R_*}(R^2-R_*^2)
+\frac12 (A-A_*)^2\nonumber\\
&-&\frac{1}{2R_*^2}(A-A_*)(R^2-R_*^2)
+\frac{2-R_*}{4R_*^3}(R^2-R_*^2)^2.
\label{quadratic_AR2_final}
\end{eqnarray}

\medskip
\noindent
The linear terms in \eqref{quadratic_AX_final} or \eqref{quadratic_AR2_final} survive because the expansion is taken in the ambient variables \((A,X)\), not along the constrained boundary alone. Writing
\begin{eqnarray}
\delta A:=A-A_*,
\qquad
\delta X:=X-X_*,
\end{eqnarray}
the linear part of the Taylor expansion is
\begin{eqnarray}
V_{\rm lin}
=
V_A^*\,\delta A+V_X^*\,\delta X.
\end{eqnarray}
Now, along the boundary \(X=A^2\), one has to first order around \((A_*,X_*)\), \(\delta X=2A_*\,\delta A\). Hence, for displacements tangent to the boundary, the linear term becomes
\begin{eqnarray}
V_{\rm lin}
=
\left(V_A^*+2A_*V_X^*\right)\delta A.
\end{eqnarray}
This vanishes by the boundary saddle condition
\begin{eqnarray}
V_A^*+2A_*V_X^*=0,
\end{eqnarray}
which is precisely the statement that the linear variation vanishes along the tangent direction \((1,2A_*)\). Thus the linear part in \eqref{quadratic_AX_final} or \eqref{quadratic_AR2_final} is precisely measuring the displacement away from the boundary-stationary direction.

\medskip
\noindent
This is the main conceptual point: the point \((A_*,R_*^2)\) is a maximum only after imposing the physical constraint \(R^2\ge A^2\), and more specifically after restricting to the aligned boundary branch \(R^2=A^2\), \(A\ge0\). In the ambient \((A,R^2)\)-plane it is not an unconstrained extremum, and the Taylor expansion correctly remembers this through the non-vanishing linear terms.

\paragraph{Expansion around \texorpdfstring{$(R^2-R_*^2,R^2-A^2)=(0,0)$}.}

\medskip
\noindent
It is useful to reorganize the local expansion in terms of the two variables
\begin{eqnarray}
\Sigma:=R^2-R_*^2,
\qquad
\Delta:=R^2-A^2.
\label{SigmaDelta_def}
\end{eqnarray}
Here \(\Sigma\) measures the displacement along the boundary direction, while \(\Delta\) measures the departure from the boundary \(A^2=R^2\).  The Taylor expansion of the exact holonomy potential around the constrained saddle \((\Sigma,\Delta)=(0,0)\) then reads
\begin{eqnarray}
V(\Delta,\Sigma)
&=&
V_*
+
V_\Delta^*\,\Delta
+
V_\Sigma^*\,\Sigma
+
V_{\Delta\Sigma}^*\,\Delta\,\Sigma
+\frac12 V_{\Sigma\Sigma}^*\,\Sigma^2
+\frac12 V_{\Delta\Delta}^*\,\Delta^2
+O(3),
\end{eqnarray}
with
\begin{eqnarray}
V_\Delta^*
=
-\frac{1}{2R_*},
\qquad
V_\Sigma^*
=
0,
\qquad
V_{\Delta\Sigma}^*
=
\frac{2-R_*}{4R_*^3},
\qquad
V_{\Sigma\Sigma}^*
=
-\frac{R_*-1}{4R_*^3},
\qquad
V_{\Delta\Delta}^*
=
\frac{1}{4R_*^2}.
\end{eqnarray}
Hence
\begin{eqnarray}
V(A,R)
&=&
V_*
-\frac{1}{2R_*}\,\Delta
+\frac{2-R_*}{4R_*^3}\,\Delta\,\Sigma
-\frac{R_*-1}{8R_*^3}\,\Sigma^2
+\frac{1}{8R_*^2}\,\Delta^2
+O(3),
\label{Vhol_SigmaDelta_quad}
\end{eqnarray}
where
\begin{eqnarray}
V_*
=
-\log\!\Big(I_0(R_*)-I_1(R_*)\Big).
\end{eqnarray}
The absence of a linear term in \(\Sigma\) expresses stationarity along the boundary direction, while the linear term in \(\Delta\) reflects the constrained nature of the saddle.

\begin{figure}[htbp]
  \centering
  \includegraphics[width=0.75\textwidth]{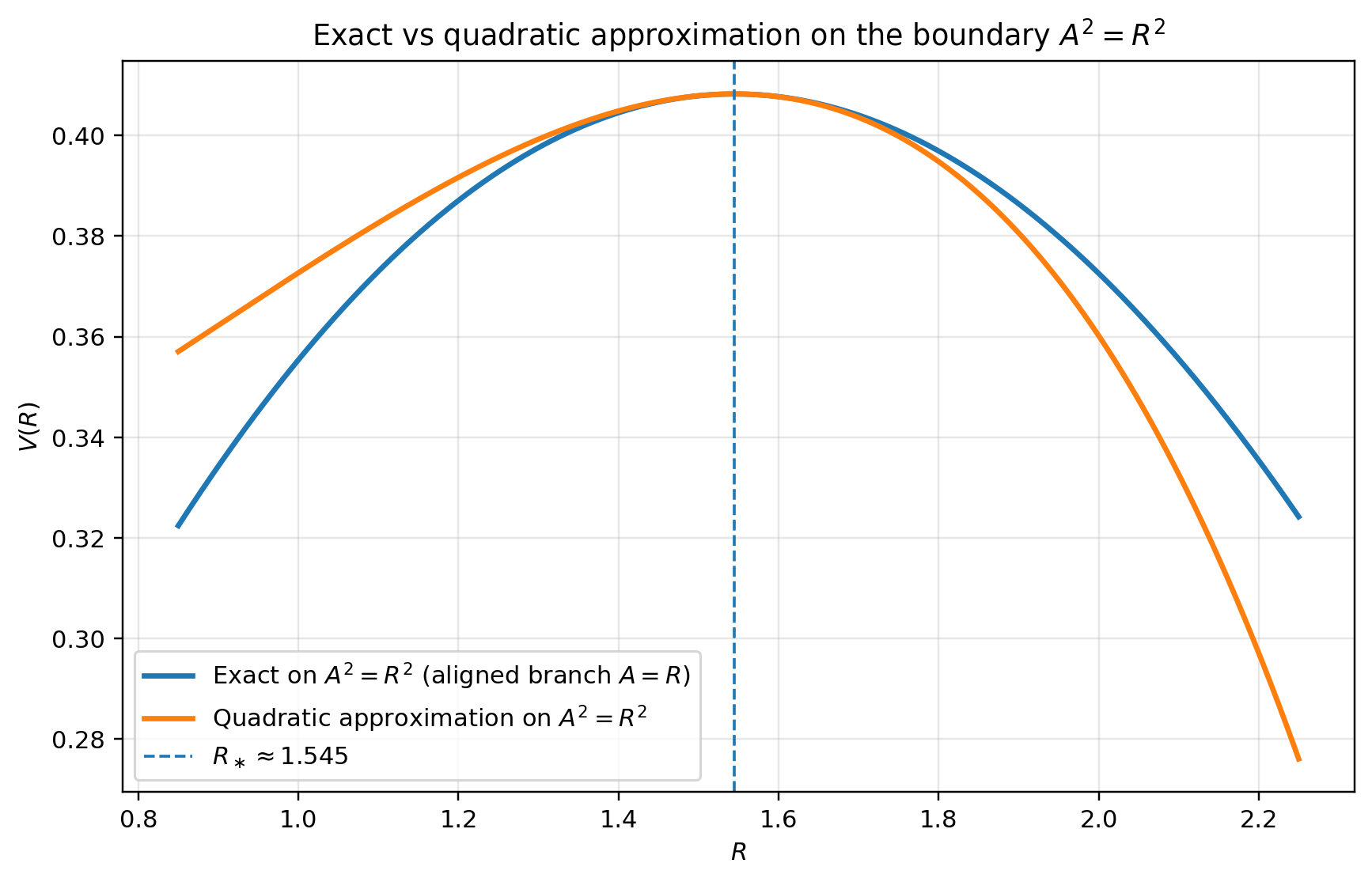}
  \caption{The exact versus the quadratic holonomy potential on the aligned component \(A=R\) showing a visible maximum near $A_*=R_* = 1.545$.}
  \label{fig:energy-comparison0}
\end{figure}

\medskip
\noindent
Restricting now to the boundary
\begin{eqnarray}
A^2=R^2,
\qquad\Longleftrightarrow\qquad
\Delta=0,
\end{eqnarray}
one obtains immediately

\begin{eqnarray}
V(R)\Big|_{A=R}
=
V_*
-\frac{R_*-1}{8R_*^3}\,
\left(R^2-R_*^2\right)^2
+O\!\left((R^2-R_*^2)^3\right).
\label{boundary_R2_expansion}
\end{eqnarray}
This is precisely the quadratic approximation in the variable \(R^2\) that was used in the comparison plot \eqref{fig:energy-comparison0}. More explicitly, in that figure we compare the exact holonomy potential restricted to the aligned boundary branch,
\begin{eqnarray}
V(R)\Big|_{A=R}
=
-\log\!\Big(I_0(R)-I_1(R)\Big),
\end{eqnarray}
with its local quadratic approximation \eqref{boundary_R2_expansion} obtained by restricting \eqref{Vhol_SigmaDelta_quad} to \(\Delta=0\). Thus the plot tests how well the exact potential along the physical boundary branch is reproduced by the leading quadratic expansion in the natural variable \(R^2-R_*^2\) around the maximum at \(R=R_*\).

\subsection{Geometry of the constrained boundary saddle}

\medskip
\noindent
The expansion \eqref{Vhol_SigmaDelta_quad} is simply a reorganization of the expansion \eqref{quadratic_AR2_final} around the constrained saddle
\begin{eqnarray}
(A_*,X_*)=(R_*,R_*^2),
\qquad
X:=R^2.
\end{eqnarray}
To see this explicitly, introduce
\begin{eqnarray}
\delta A:=A-A_*,
\qquad
\delta X:=X-X_*,
\qquad
A_*=R_*,
\qquad
X_*=R_*^2.
\end{eqnarray}
In these variables, \eqref{quadratic_AR2_final} reads
\begin{eqnarray}
V(A,X)
&\simeq&
V_*
+
V_A^*\,\delta A
+
V_X^*\,\delta X
+
\frac12 V_{AA}^*\,\delta A^2
+
V_{AX}^*\,\delta A\,\delta X
+
\frac12 V_{XX}^*\,\delta X^2,
\label{Taylor_deltaA_deltaX}
\end{eqnarray}
with
\begin{eqnarray}
V_A^*=1,
\qquad
V_X^*=-\frac{1}{2R_*},
\qquad
V_{AA}^*=1,
\qquad
V_{AX}^*=-\frac{1}{2R_*^2},
\qquad
V_{XX}^*=\frac{2-R_*}{2R_*^3}.
\label{derivatives_star_recall}
\end{eqnarray}

\medskip
\noindent
We now pass to the boundary-adapted variables
\begin{eqnarray}
\Sigma:=X-X_*=\delta X,
\qquad
\Delta:=X-A^2.
\label{SigmaDelta_def_recall}
\end{eqnarray}
Since
\begin{eqnarray}
A=A_*+\delta A,
\end{eqnarray}
one has
\begin{eqnarray}
\Delta
=
(X_*+\delta X)-(A_*+\delta A)^2
=
\delta X-2A_*\delta A-\delta A^2.
\end{eqnarray}
Using \(A_*=R_*\) and \(\delta X=\Sigma\), this becomes
\begin{eqnarray}
\Delta
=
\Sigma-2R_*\delta A-\delta A^2.
\label{Delta_forward_map}
\end{eqnarray}
To the order needed here, this can be inverted as
\begin{eqnarray}
\delta A
=
\frac{\Sigma-\Delta}{2R_*}
+O(2),
\label{deltaA_linear_inverse}
\end{eqnarray}
and, one order more accurately,
\begin{eqnarray}
\delta A
=
\frac{\Sigma-\Delta}{2R_*}
-\frac{(\Sigma-\Delta)^2}{8R_*^3}
+O(3).
\label{deltaA_quadratic_inverse}
\end{eqnarray}

\medskip
\noindent
Substituting \eqref{deltaA_linear_inverse} into the linear part of \eqref{Taylor_deltaA_deltaX}, one finds
\begin{eqnarray}
V_A^*\,\delta A+V_X^*\,\delta X
=
\frac{\Sigma-\Delta}{2R_*}
-\frac{\Sigma}{2R_*}
+O(2)
=
-\frac{\Delta}{2R_*}
+O(2).
\end{eqnarray}
Thus the linear terms in the ambient variables \((\delta A,\delta X)\) reorganize into a single linear term in the boundary-distance variable \(\Delta\).

\medskip
\noindent
For the quadratic terms, it is enough to use \eqref{deltaA_linear_inverse}. One gets
\begin{eqnarray}
\frac12 V_{AA}^*\,\delta A^2
&=&
\frac12
\left(
\frac{\Sigma-\Delta}{2R_*}
\right)^2
+O(3)
=
\frac{1}{8R_*^2}
\left(
\Sigma^2-2\Sigma\Delta+\Delta^2
\right)
+O(3),
\\
V_{AX}^*\,\delta A\,\delta X
&=&
-\frac{1}{2R_*^2}
\left(
\frac{\Sigma-\Delta}{2R_*}
\right)\Sigma
+O(3)
=
-\frac{1}{4R_*^3}
\left(
\Sigma^2-\Sigma\Delta
\right)
+O(3),
\\
\frac12 V_{XX}^*\,\delta X^2
&=&
\frac{2-R_*}{4R_*^3}\,\Sigma^2
+O(3).
\end{eqnarray}
Adding these contributions, and also keeping the quadratic correction coming from substituting \eqref{deltaA_quadratic_inverse} into the linear term \(V_A^*\,\delta A\), one obtains after simplification
\begin{eqnarray}
V(A,R)
&=&
V_*
-\frac{1}{2R_*}\,\Delta
+\frac{2-R_*}{4R_*^3}\,\Delta\,\Sigma
-\frac{R_*-1}{8R_*^3}\,\Sigma^2
+\frac{1}{8R_*^2}\,\Delta^2
+O(3),
\end{eqnarray}
which is precisely \eqref{Vhol_SigmaDelta_quad}.

\medskip
\noindent
Hence \eqref{Vhol_SigmaDelta_quad} is not a new expansion, but the same quadratic Taylor series \eqref{quadratic_AR2_final} rewritten in variables adapted to the boundary geometry. In particular, the geometry of the saddle in these variables becomes transparent:

\begin{itemize}
\item \(\Sigma=X-X_*=R^2-R_*^2\) measures the displacement along the boundary direction; and there is no term linear in \(\Sigma\) because the potential is stationary along the boundary direction;
\item \(\Delta=X-A^2=R^2-A^2\) measures the distance from the boundary itself; and the term linear in \(\Delta\) reflects the fact that the saddle is constrained rather than unconstrained;
  \item the linear terms in \((\delta A,\delta X)\) reorganize into a single linear term in \(\Delta\);
\item the first nontrivial variation along the boundary is the quadratic term
\begin{eqnarray}
-\frac{R_*-1}{8R_*^3}\,\Sigma^2,
\end{eqnarray}
which shows that the saddle is a local maximum along the aligned branch.
\end{itemize}

\subsection{Transverse expansion at fixed \texorpdfstring{$A=R_*$}{A=R*}}

\medskip
\noindent
Substituting \(\Delta=R^2-A^2\) and \(\Sigma=R^2-R_*^2\) into \eqref{Vhol_SigmaDelta_quad} gives
\begin{eqnarray}
V(A,R)
&=&
V_*
-\frac{1}{2R_*}\,\Big(R^2-A^2\Big)
+\frac{2-R_*}{4R_*^3}\,\Big(R^2-A^2\Big)\Big(R^2-R_*^2\Big)
\nonumber\\
&&
-\frac{R_*-1}{8R_*^3}\,\Big(R^2-R_*^2\Big)^2
+\frac{1}{8R_*^2}\,\Big(R^2-A^2\Big)^2
+O(3).
\label{Vhol_grouped_AR}
\end{eqnarray}

\medskip
\noindent
It is convenient to trade the variables \((A,R)\) for \((A,B)\), where
\begin{eqnarray}
B^2:=R^2-A^2,
\qquad\text{so that}\qquad
R^2=A^2+B^2.
\end{eqnarray}
In particular,
\begin{eqnarray}
R^2-A^2=B^2,
\qquad
R^2-R_*^2=A^2+B^2-R_*^2.
\end{eqnarray}
Substituting these relations into \eqref{Vhol_grouped_AR}, one obtains
\begin{eqnarray}
V(A,B)
&=&
V_*
-\frac{1}{2R_*}\,B^2
+\frac{2-R_*}{4R_*^3}\,B^2\Big(A^2+B^2-R_*^2\Big)
\nonumber\\
&&
-\frac{R_*-1}{8R_*^3}\,\Big(A^2+B^2-R_*^2\Big)^2
+\frac{1}{8R_*^2}\,B^4
+O(3).
\label{Vhol_grouped_AB}
\end{eqnarray}
Expanding and collecting powers of \(B^2\), this takes the form
\begin{eqnarray}
V(A,B)
&=&
V(A)+c_2(A)\,B^2+c_4\,B^4+O_3\!\left(A^2-R_*^2,\;B^2\right),
\label{Vhol_Vc2c4_AB}
\end{eqnarray}
with
\begin{eqnarray}
V(A)
&=&
V_*
-\frac{R_*-1}{8R_*^3}\,\Big(A^2-R_*^2\Big)^2,
\\
c_2(A)
&=&
-\frac{1}{2R_*}
+\frac{3-2R_*}{8R_*^3}\,\Big(A^2-R_*^2\Big),
\\
c_4
&=&
\frac{3-R_*}{8R_*^3}.
\end{eqnarray}
Thus the local expansion is naturally organized in terms of \(A^2\), which parametrizes motion along the aligned branch, and \(B^2\), which measures the departure from it.

\medskip
\noindent
A particularly natural probe of the constrained saddle is obtained by fixing the longitudinal variable at its saddle value,
\begin{eqnarray}
A=R_*,
\end{eqnarray}
while allowing a nonzero transverse component \(B\). Since \(R^2=A^2+B^2\), this implies
\begin{eqnarray}
R^2=R_*^2+B^2.
\end{eqnarray}
Equivalently, in the variables
\begin{eqnarray}
\Sigma:=R^2-R_*^2,
\qquad
\Delta:=R^2-A^2,
\end{eqnarray}
one has along this slice
\begin{eqnarray}
\Sigma=\Delta=B^2.
\end{eqnarray}
Substituting into \eqref{Vhol_SigmaDelta_quad}, one finds
\begin{eqnarray}
V(A=R_*,B)
&=&
V_*
-\frac{1}{2R_*}\,B^2
+\frac{3-R_*}{8R_*^3}\,B^4
+O(B^6).
\label{Vhol_fixedAstar_B0}
\end{eqnarray}
Thus, at fixed \(A=R_*\), the holonomy potential takes the standard Landau form in the transverse variable \(B\): the quadratic term is negative, so the potential immediately decreases as one moves away from the aligned branch, while the quartic term is positive and provides the leading stabilizing correction.

\medskip
\noindent
Here \(O(3)\) in \eqref{Vhol_grouped_AB} denotes terms of total degree at least three in the variables \((\Delta,\Sigma)\), namely terms such as
\begin{eqnarray}
\Delta^3,\qquad \Delta^2\Sigma,\qquad \Delta\Sigma^2,\qquad \Sigma^3,
\end{eqnarray}
and higher. Accordingly, in \eqref{Vhol_Vc2c4_AB} the remainder \(O_3\!\left(A^2-R_*^2,\;B^2\right)\) is understood in the sense of total degree in the two small quantities \(A^2-R_*^2\) and \(B^2\). Thus, after rewriting in the variables \((A,B)\), the omitted terms include not only \(B^6\) but also mixed contributions such as \((A^2-R_*^2)^2B^2\) and \((A^2-R_*^2)B^4\).

\medskip
\noindent
Along the slice \(A=R_*\), however, one has
\begin{eqnarray}
\Delta=\Sigma=B^2,
\end{eqnarray}
so that a term of total degree \(n\) in \((\Delta,\Sigma)\) becomes of order \(B^{2n}\). In particular,
\begin{eqnarray}
O(3)=O(B^6).
\end{eqnarray}
Hence no omitted term contributes to the coefficients of \(B^2\) or \(B^4\). Moreover, only even powers of \(B\) can appear, since the potential depends on \(B\) only through \(B^2\).

\section{Continuum limit and a non-polynomial toy model}
\subsection{The low-\texorpdfstring{$x$}{x} expansion in the transverse expansion}

\medskip
\noindent
We now examine in detail the quartic transverse approximation \eqref{Vhol_fixedAstar_B0} to the exact holonomy potential \eqref{V_exact_full} in the low-temperature regime \(x\ll 1\). Instead of the exact holonomy kernel \(\Phi(A,B)\), we therefore consider the approximation \(\exp\!\big(-V_{\rm hol}(A=R_*,B)\big)\), where

\begin{eqnarray}
V_{\rm hol}(A=R_*,B)=-aB^2+bB^4.
\label{quartic_potential_ab}
\end{eqnarray}
Here
\begin{eqnarray}
a:=\frac{1}{2R_*},
\qquad
b:=\frac{3-R_*}{8R_*^3},
\end{eqnarray}
Then the planar partition function is
\begin{eqnarray}
\widetilde Z_{\perp}^{}(x)
=
Z_{\perp,0}(x)\,
\Big\langle e^{-V_{\rm hol}(A=R_*,B)}\Big\rangle_0.
\label{Zperp_app_def}
\end{eqnarray}
\medskip
\noindent
Equivalently, defining
\begin{eqnarray}
G(D):=
\Big\langle e^{\,aB^2-bB^4}\Big\rangle_D,\label{G_of_D_def}
\end{eqnarray}
one has
\begin{eqnarray}
\widetilde Z_{\perp}^{}(x)
=
Z_{\perp,0}(x)\,G(D_\Lambda(x)).
\end{eqnarray}
\medskip
\noindent Here \(\langle\cdots\rangle_0\equiv \langle\cdots\rangle_D\) denotes Gaussian averaging with respect to the planar endpoint measure, and
\begin{eqnarray}
Z_{\perp,0}(x)
=\left(\frac{\pi a}{2N}\right)^{2d}
D_\Lambda^{-d},
\label{Zperp0_again}
\end{eqnarray}
with
\begin{eqnarray}
D=D_\Lambda=\alpha_\Lambda^2-\beta_\Lambda^2.\label{Zperp0_again0}
\end{eqnarray}

\medskip
\noindent
Since
\begin{eqnarray}
D_\Lambda=D_0+D_2x^2+O(x^4),
\label{D_expand_x}
\end{eqnarray}
the normalized partition function
\begin{eqnarray}
\widehat Z_{\perp}^{}(x)
:=
\frac{\widetilde Z_{\perp}^{}(x)}{\widetilde Z_{\perp}^{}(0)}
\label{Zhat_app_def}
\end{eqnarray}
has only even powers in \(x\):
\begin{eqnarray}
\widehat Z_{\perp}^{}(x)
=
1+{\cal C}_2x^2+O(x^4).
\label{Zhat_C2_form}
\end{eqnarray}
Indeed, one finds
\begin{eqnarray}
\widehat Z_{\perp}^{}(x)
=
\left(\frac{D_0}{D_\Lambda(x)}\right)^d
\frac{G(D_\Lambda(x))}{G(D_0)}.
\label{Zhat_ratio_form}
\end{eqnarray}
Expanding in \(x^2\),
\begin{eqnarray}
\left(\frac{D_0}{D_\Lambda(x)}\right)^d
=
1-\frac{dD_2}{D_0}x^2+O(x^4),
\label{gaussian_pref_exp}
\end{eqnarray}
and
\begin{eqnarray}
\frac{G(D_\Lambda(x))}{G(D_0)}
=
1+D_2\frac{G'(D_0)}{G(D_0)}x^2+O(x^4).
\label{G_ratio_exp}
\end{eqnarray}
Thus
\begin{eqnarray}
{\cal C}_2
=
-\frac{dD_2}{D_0}
+
D_2\frac{G'(D_0)}{G(D_0)}.
\label{C2_basic_formula}
\end{eqnarray}
\medskip
\noindent
The coefficient \({\cal C}_2\) therefore has a trivial continuum limit. In the quartic transverse approximation, the only \(x\)-dependence enters through \(D_\Lambda(x)\), while \(G(D)\) remains regular at \(D=D_0\). Thus the ratio in \eqref{G_ratio_exp} is an ordinary Taylor expansion, controlled by finite derivatives of \(G\) at \(D_0\). This regularity prevents the approximation from reproducing the singular continuum contribution of the exact holonomy kernel. It therefore yields only a Gaussian correction, and the corresponding \(D\)-channel contribution is trivial.

\subsection{Low-\texorpdfstring{$x$}{x} expansion of the exact theory revisited}

\medskip
\noindent
In the exact theory we define
\begin{eqnarray}
G_{\rm ex}(D_\Lambda,\beta_\Lambda)
:=
\frac{\widetilde Z_\perp(x)}{Z_{\perp,0}(x)}
=
\langle \Phi(A,B)\rangle_0,
\qquad
\Phi(A,B):=
I_0(R)-\frac{A}{R}I_1(R).
\label{Gex_def_diff}
\end{eqnarray}
Since the Gaussian measure contains the anisotropic coupling \(-2\beta_\Lambda A\), and the exact kernel depends explicitly on \(A\), the function \(G_{\rm ex}\) depends on both \(D_\Lambda\) and \(\beta_\Lambda\).

\medskip
\noindent
The normalized exact partition function is then
\begin{eqnarray}
\widehat Z_\perp(x)
=
\frac{G_{\rm ex}(D_\Lambda(x),\beta_\Lambda(x))}
{G_{\rm ex}(D_0,0)}.
\label{Zhat_Gex_ratio_diff}
\end{eqnarray}
Now expand
\begin{eqnarray}
D_\Lambda(x)=D_0+D_2x^2+O(x^4),
\qquad
\beta_\Lambda(x)=\beta_1x+\beta_3x^3+O(x^5).
\label{Dbeta_expand_diff}
\end{eqnarray}
Since \(D_\Lambda\) has no linear term in \(x\), while \(\beta_\Lambda\) starts at order \(x\), the Taylor expansion of \(G_{\rm ex}(D_\Lambda,\beta_\Lambda)\) around \((D_0,0)\) gives
\begin{eqnarray}
G_{\rm ex}(D_\Lambda(x),\beta_\Lambda(x))
&=&
G_{\rm ex}(D_0,0)
+
\beta_1x\,\partial_\beta G_{\rm ex}(D_0,0)
\nonumber\\
&&
+
\left[
D_2\,\partial_D G_{\rm ex}(D_0,0)
+\frac{\beta_1^2}{2}\,\partial_\beta^2 G_{\rm ex}(D_0,0)
\right]x^2
+O(x^3).
\label{Taylor_Gex_diff}
\end{eqnarray}
Dividing by \(G_{\rm ex}(D_0,0)\), one obtains
\begin{eqnarray}
\widehat Z_\perp(x)
=
1
+
\beta_1\frac{\partial_\beta G_{\rm ex}(D_0,0)}{G_{\rm ex}(D_0,0)}\,x
+
{\cal C}_2^{\rm ex}\,x^2
+O(x^3),
\label{Zhat_Gex_diff_expand}
\end{eqnarray}
with
\begin{eqnarray}
\boxed{
{\cal C}_2^{\rm ex}
=
D_2\frac{\partial_D G_{\rm ex}(D_0,0)}{G_{\rm ex}(D_0,0)}
+
\frac{\beta_1^2}{2}\frac{\partial_\beta^2 G_{\rm ex}(D_0,0)}{G_{\rm ex}(D_0,0)}.
}
\label{C2ex_diff_form}
\end{eqnarray}

\medskip
\noindent
This is the clean differential form of the exact answer after normalization by the free Gaussian partition function. If one now wishes to restore the universal Gaussian contribution explicitly, one simply recalls that the full partition function is
\begin{eqnarray}
\widetilde Z_\perp(x)
=
Z_{\perp,0}(x)\,G_{\rm ex}(D_\Lambda(x),\beta_\Lambda(x)),
\qquad
Z_{\perp,0}(x)\propto D_\Lambda(x)^{-d}.
\end{eqnarray}
Thus the normalized full endpoint ratio, in which the free Gaussian factor is restored, is
\begin{eqnarray}
\frac{\widetilde Z_\perp(x)}{\widetilde Z_\perp(0)}
=
\frac{Z_{\perp,0}(x)}{Z_{\perp,0}(0)}
\widehat Z_\perp(x).
\end{eqnarray}
This contains the additional Gaussian contribution
\begin{eqnarray}
-\frac{dD_2}{D_0}
\end{eqnarray}
from the prefactor \(Z_{\perp,0}(x)/Z_{\perp,0}(0)\), in addition to the two exact holonomy contributions in \eqref{C2ex_diff_form}. In other words, for the full endpoint ratio one obtains
\begin{eqnarray}
\boxed{
{\cal C}_{2,\rm full}^{\rm ex}
=
-\frac{dD_2}{D_0}
+
D_2\frac{\partial_D G_{\rm ex}(D_0,0)}{G_{\rm ex}(D_0,0)}
+
\frac{\beta_1^2}{2}\frac{\partial_\beta^2 G_{\rm ex}(D_0,0)}{G_{\rm ex}(D_0,0)}.
}
\label{C2ex_full_diff_form}
\end{eqnarray}

\medskip
\noindent
In the continuum low-temperature regime one has
\begin{eqnarray}
D_0=\frac{1+2\mu}{4},
\qquad
D_2=\mu,
\qquad
\beta_1=\mu,
\qquad
1-z=\frac{2\mu}{1+2\mu},
\qquad
\mu=as.
\label{continuum_data_Glanguage_clean}
\end{eqnarray}

\medskip
\noindent
We now compare the Taylor expansion of the exact kernel average around \((D_0,0)\), given in \eqref{Taylor_Gex_diff}, with the explicit resummed expression \eqref{Ztilde_before_cont}, namely

\begin{eqnarray}
G_{\rm ex}(x)
&=&
(1-z)^{-d}
-\frac{d\beta_1}{2D_0}(1-z)^{-d-1}x+
\left[
-\frac{dD_2}{4D_0^2}(1-z)^{-d-1}
+\frac{d(d+1)\beta_1^2}{4D_0^2}(1-z)^{-d-2}
\right]x^2\nonumber\\
&+&O(x^3),
\qquad
z=\frac{1}{4D_0}.
\label{Gex_expansion_raw_clean}
\end{eqnarray}

\medskip
\noindent
Comparing \eqref{Taylor_Gex_diff} with \eqref{Gex_expansion_raw_clean}, one then reads off the derivatives directly from the coefficients of \(x\), \(D_2x^2\), and \(\beta_1^2x^2\):
\begin{eqnarray}
G_{\rm ex}(D_0,0)
&=&
(1-z)^{-d},
\\
\partial_\beta G_{\rm ex}(D_0,0)
&=&
-\frac{d}{2D_0}(1-z)^{-d-1},
\\
\partial_D G_{\rm ex}(D_0,0)
&=&
-\frac{d}{4D_0^2}(1-z)^{-d-1},
\\
\frac{1}{2}\partial_\beta^2 G_{\rm ex}(D_0,0)
&=&
\frac{d(d+1)}{4D_0^2}(1-z)^{-d-2}.
\label{derivs_Gex_eval_clean}
\end{eqnarray}
Dividing the last two expressions by \(G_{\rm ex}(D_0,0)=(1-z)^{-d}\), one finds
\begin{eqnarray}
D_2\frac{\partial_D G_{\rm ex}(D_0,0)}{G_{\rm ex}(D_0,0)}
&=&
-\frac{dD_2}{4D_0^2(1-z)},
\\
\frac{\beta_1^2}{2}\frac{\partial_\beta^2 G_{\rm ex}(D_0,0)}{G_{\rm ex}(D_0,0)}
&=&
\frac{d(d+1)\beta_1^2}{4D_0^2(1-z)^2}.
\label{D_beta_terms_clean}
\end{eqnarray}
Substituting these into \eqref{C2ex_full_diff_form} one obtains
\begin{eqnarray}
\boxed{
{\cal C}_{2,\rm full}^{\rm ex}
=
-\frac{dD_2}{D_0}
-\frac{dD_2}{4D_0^2(1-z)}
+\frac{d(d+1)\beta_1^2}{4D_0^2(1-z)^2}.
}
\label{C2ex_threepiece_clean}
\end{eqnarray}
Thus the exact theory contains, in addition to the universal Gaussian term \(-dD_2/D_0\), two extra contributions coming from the \(D\)- and \(\beta\)-dependence of the exact holonomy kernel.

\medskip
\noindent
In more detail, the Gaussian piece \(-{dD_2}/{D_0}\) vanishes in the continuum limit, whereas the exact holonomy pieces contain inverse powers of \(\mu\):
\begin{eqnarray}
-\frac{dD_2}{4D_0^2(1-z)}
\sim
-2d,
\qquad
\frac{d(d+1)\beta_1^2}{4D_0^2(1-z)^2}
\sim
d(d+1).
\end{eqnarray}
Hence the exact kernel generates nontrivial finite contributions in the continuum limit, and these reorganize the Gaussian endpoint result into the exact Molien--Weyl answer.

\subsection{Continuum limit of \(B\)-theories}

\medskip
\noindent
For the transverse \(B\)-theory, obtained by freezing the longitudinal direction at \(A=R_*\), the truncated kernel becomes a function of \(B\) alone:
\begin{eqnarray}
e^{-V_{\rm hol}(B)}=e^{aB^2-bB^4}.
\end{eqnarray}
The corresponding kernel average is
\begin{eqnarray}
G_B(D_\Lambda)
:=
\frac{\widetilde Z_\perp^{(B)}(x)}{Z_{\perp,0}(x)}=\langle e^{aB^2-bB^4} \rangle_0,
\end{eqnarray}
so that the normalized \(B\)-theory partition function is
\begin{eqnarray}
\widehat Z_\perp^{(B)}(x)
=
\frac{G_B(D_\Lambda(x))}{G_B(D_0)}.\label{ssi}
\end{eqnarray}

\medskip
\noindent
The crucial point is that \(G_B\) depends only on \(D_\Lambda\), and not separately on \(\beta_\Lambda\), because the explicit \(A\)-dependence has been removed from the kernel. Once the kernel depends only on \(B\), the only Gaussian datum relevant for its distribution is the variance parameter \(D_\Lambda\). Hence
\begin{eqnarray}
\partial_\beta G_B(D_0)=0.
\end{eqnarray}
It follows that the quadratic coefficient of the normalized \(B\)-theory partition function \(\widehat Z_\perp^{(B)}\) takes the form
\begin{eqnarray}
{\cal C}_2^{(B)}
=
D_2\frac{G_B'(D_0)}{G_B(D_0)}.
\label{C2_B_general}
\end{eqnarray}

\medskip
\noindent
Thus the \(B\)-theory contains only the Gaussian contribution together with a single \(D\)-derivative correction. The \(\beta\)-derivative term is absent precisely because the pure \(B\)-kernel has lost all explicit dependence on \(A\).

\medskip
\noindent
The crucial difference between the exact theory and the pure \(B\)-theory is that the \(D\)-derivative channel is singular in the former but regular in the latter. Indeed, for the normalized exact quantity \eqref{Zhat_Gex_ratio_diff}, the quadratic coefficient contains
\begin{eqnarray}
D_2\frac{\partial_D G_{\rm ex}(D_0,0)}{G_{\rm ex}(D_0,0)}
=
-\frac{dD_2}{4D_0^2(1-z)}\longrightarrow -\frac{d(\mu)}{4(\tfrac{1}{4})^2(2\mu)}=-2d.
\end{eqnarray}
Thus the \(D\)-derivative contribution survives the continuum limit. The reason is that

\begin{eqnarray}
G_{\rm ex}(D,0)
&=&
\sum_{n=0}^{\infty}
\frac{(d)_n}{n!}
\left(\frac{1}{4D}\right)^n=
\left(1-\frac{1}{4D}\right)^{-d}
\end{eqnarray}
has a singularity at \(D=1/4\), and the continuum limit drives \(D_0\) precisely to this point.

\medskip
\noindent
By contrast, in the pure \(B\)-theory one has the normalized partition function \eqref{ssi}. For fixed \(a,b\), this function is regular at \(D=1/4\), so that
\begin{eqnarray}
\frac{G_B'(D_0)}{G_B(D_0)}=O(1)
\qquad
(\mu\to0).
\end{eqnarray}
Hence
\begin{eqnarray}
D_2\frac{G_B'(D_0)}{G_B(D_0)}
=
\mu\cdot O(1)\to0.
\end{eqnarray}
Therefore the \(D\)-derivative channel survives in the exact theory but disappears in the pure \(B\)-theory.

\subsection{A non-polynomial \(B\)-toy model}

\medskip
\noindent
The analysis of the pure quartic \(B\)-theory showed that its failure is not merely numerical but structural. If one keeps only the transverse quartic approximation
\begin{eqnarray}
V_{\rm hol}(B)
=
V_*-\frac{1}{2R_*}B^2+\frac{3-R_*}{8R_*^3}B^4+O(B^6),
\label{VlocB_recall}
\end{eqnarray}
and defines
\begin{eqnarray}
G_B(D_\Lambda)
:=
\frac{\widetilde Z_\perp^{(B)}(x)}{Z_{\perp,0}(x)}
=
\Big\langle e^{-V_{\rm hol}(B)}\Big\rangle_0,
\end{eqnarray}
then the normalized partition function
\begin{eqnarray}
\widehat Z_\perp^{(B)}(x)
=
\frac{G_B(D_\Lambda(x))}{G_B(D_0)}
\end{eqnarray}
has quadratic coefficient
\begin{eqnarray}
\widehat{\mathcal C}_2^{(B)}
=
D_2\frac{G_B'(D_0)}{G_B(D_0)}.
\label{C2B_recall_toy}
\end{eqnarray}
For any finite polynomial truncation of the \(B\)-potential, the corresponding function \(G_B(D)\) is regular at the continuum point
\begin{eqnarray}
D_{\Lambda}=\frac14.
\end{eqnarray}
Hence
\begin{eqnarray}
\frac{G_B'(D_0)}{G_B(D_0)}=O(1),
\qquad
D_2=\mu\to0,
\end{eqnarray}
so that
\begin{eqnarray}
\widehat{\mathcal C}_2^{(B)}=O(\mu)\to0.
\end{eqnarray}
Thus the quartic \(B\)-theory becomes trivial in the ordinary continuum limit.

\medskip
\noindent
This conclusion is not specific to the quartic \(B\)-theory. Rather, it is conjectured to hold for \emph{any finite polynomial truncation} of the holonomy potential, whether it involves only the \(B\)-field, only the \(A\)-field, or any finite mixture of the two. The essential point is that a finite polynomial truncation leads, after Gaussian averaging, to a function
\begin{eqnarray}
G(D_\Lambda,\beta_\Lambda)
\end{eqnarray}
which is regular at the continuum point
\begin{eqnarray}
(D_\Lambda,\beta_\Lambda)=\left(\frac14,0\right).
\end{eqnarray}
As a result, the corresponding derivative channels remain finite,
\begin{eqnarray}
\partial_{D_\Lambda}\log G=O(1),
\qquad
\partial_{\beta_\Lambda}^2\log G=O(1),
\end{eqnarray}
so that the explicit factors
\begin{eqnarray}
D_2=\mu,
\qquad
\beta_1^2=\mu^2
\end{eqnarray}
force all such contributions to vanish in the ordinary continuum limit \(\mu\to0\). In this sense, the quartic \(B\)-theory is merely the simplest representative of a much broader class: all finite polynomial truncations are expected to become trivial in the ordinary continuum limit.

\medskip
\noindent
This suggests a sharper question. Is there a simple one-variable \(B\)-kernel which, while still depending only on \(B\), reproduces the singular \(D\)-dependence needed for the \(D\)-derivative channel to survive? The answer is yes, and the simplest choice is provided by the non-polynomial potential
\begin{eqnarray}
V_{\rm toy}(B):=-\log\cosh B.
\label{Vtoy_def_full}
\end{eqnarray}
Equivalently,
\begin{eqnarray}
e^{-V_{\rm toy}(B)}=\cosh B.
\label{toy_kernel_full}
\end{eqnarray}

\medskip
\noindent
The reason for choosing \(\cosh B\) is that its Gaussian average can be computed exactly from the \(B\)-moment generating function. Recall that
\begin{eqnarray}
{\cal M}_B(u):=\langle e^{uB}\rangle_0
=
\left[
\frac{D_\Lambda}{D_\Lambda-\frac{u^2}{4}}
\right]^d.
\label{MB_recall_full}
\end{eqnarray}
Since
\begin{eqnarray}
\cosh B=\frac{e^B+e^{-B}}{2},
\end{eqnarray}
and \({\cal M}_B(u)\) is even in \(u\), one obtains
\begin{eqnarray}
\big\langle \cosh B\big\rangle_0
=
{\cal M}_B(1)
=
\left(
1-\frac{1}{4D_\Lambda}
\right)^{-d}.
\label{Gtoy_closed_full}
\end{eqnarray}
This is precisely the singular factor that was missing from the finite quartic \(B\)-theory. In this sense, the toy model is not chosen to fit the local saddle expansion coefficient by coefficient, but rather to reproduce the \emph{global singular \(D\)-structure} responsible for the survival of the \(D\)-channel in the exact theory.

\medskip
\noindent
Thus, let us define
\begin{eqnarray}
G_{\rm toy}(D_\Lambda)
:=
\frac{\widetilde Z_\perp^{\rm toy}(x)}{Z_{\perp,0}(x)}
=
\big\langle \cosh B\big\rangle_0
=
\left(
1-\frac{1}{4D_\Lambda}
\right)^{-d},
\label{Gtoy_def_full}
\end{eqnarray}
and hence
\begin{eqnarray}
\widehat Z_\perp^{\rm toy}(x)
:=
\frac{G_{\rm toy}(D_\Lambda(x))}{G_{\rm toy}(D_0)}.
\label{Zhat_toy_def_full}
\end{eqnarray}
Its quadratic coefficient is
\begin{eqnarray}
\widehat{\mathcal C}_2^{\rm toy}
=
D_2\frac{G_{\rm toy}'(D_0)}{G_{\rm toy}(D_0)}.
\end{eqnarray}
But
\begin{eqnarray}
\log G_{\rm toy}(D)
=
-d\log\!\left(1-\frac{1}{4D}\right)\qquad \Rightarrow \qquad \frac{G_{\rm toy}'(D)}{G_{\rm toy}(D)}
=
-\frac{d}{4D^2\left(1-\frac{1}{4D}\right)}.
\end{eqnarray}
Therefore
\begin{eqnarray}
\widehat{\mathcal C}_2^{\rm toy}
=
-\frac{dD_2}{4D^2(1-z)}.
\label{C2_toy_general_full}
\end{eqnarray}
In the continuum regime, this becomes
\begin{eqnarray}
\boxed{
\widehat{\mathcal C}_2^{\rm toy}
=
-\frac{2d}{1+2\mu}
\longrightarrow -2d
\qquad
(\mu\to0).
}
\label{C2_toy_cont_full}
\end{eqnarray}
Thus the toy model does exactly what the quartic \(B\)-theory could not do: it produces a nontrivial continuum limit for the \(D\)-channel.

\medskip
\noindent
The toy potential has the small-\(B\) expansion
\begin{eqnarray}
V_{\rm toy}(B)
=
-\log\cosh B
=
-\frac12 B^2+\frac1{12}B^4-\frac1{45}B^6+O(B^8).
\label{Vtoy_smallB_full}
\end{eqnarray}
\medskip
\noindent
For the physical value \(R_*\approx 1.545\), the corresponding coefficients of the exact local potential \eqref{VlocB_recall} are

\begin{eqnarray}
-\frac{1}{2R_*}\approx -0.324,
\qquad
\frac{3-R_*}{8R_*^3}\approx 0.049,
\end{eqnarray}
whereas the small-\(B\) expansion \eqref{Vtoy_smallB_full} of the toy model gives
\begin{eqnarray}
-\frac12=-0.500,
\qquad
\frac1{12}\approx 0.083.
\end{eqnarray}
So the toy potential is qualitatively similar to the quartic saddle expansion: it has the same Landau shape, namely a negative quadratic term stabilized by a positive quartic term. However, it is quantitatively steeper than the exact local \(B\)-potential.

\medskip
\noindent
The quartic \(B\)-potential captures the local shape near \(B=0\), but its Gaussian average is regular at \(D=1/4\), and therefore its \(D\)-channel vanishes in the continuum limit. The toy potential, by contrast, is less faithful locally, but it is engineered so that its Gaussian average is exactly
\begin{eqnarray}
\left(1-\frac{1}{4D}\right)^{-d},
\end{eqnarray}
which is singular at the continuum point. This is why the toy model reproduces the entire \(D\)-derivative channel of the exact theory, namely the full \(-2d\) contribution.

\medskip
\noindent
What it does not capture is the second exact contribution,
\begin{eqnarray}
\frac{\beta_1^2}{2}\frac{\partial_\beta^2 G_{\rm ex}(D_0,0)}{G_{\rm ex}(D_0,0)}
=
\frac{d(d+1)\beta_1^2}{4D_0^2(1-z)^2},
\end{eqnarray}
because the toy kernel depends only on \(B\), not on \(A\). Thus \(G_{\rm toy}\) depends only on \(D_\Lambda\), whereas the exact function \(G_{\rm ex}\) depends on both \(D_\Lambda\) and \(\beta_\Lambda\).

\section{Conclusion}

\medskip
\noindent
In this paper we studied the \(N=2\), large--\(d\) sector of BFSS/BMN-type matrix quantum mechanics on the lattice, focusing on the Gaussian regime in which the theory reduces to a gauged matrix harmonic oscillator. The main objective was to clarify the relation between two complementary descriptions of the same gauge-projected dynamics: the angular Molien--Weyl formulation, in which the gauge-field holonomy is retained, and the radial endpoint formulation, in which the coordinate matrices are retained after integrating out the bulk, gauge, and longitudinal degrees of freedom.

\medskip
\noindent
The first result was to show that the planar endpoint formulation reproduces the Molien--Weyl structure of the \(N=2\) theory. The fixed-holonomy endpoint integrand and the Molien--Weyl integrand differ only by a universal spectator factor, independent of both the holonomy angle and the endpoint variables. This establishes that the radial endpoint formulation and the angular Molien--Weyl representation are not different theories, but two organizations of the same planar gauge-projected partition function.

\medskip
\noindent
This equivalence allowed us to derive the low-temperature expansion of the planar endpoint partition function directly from the Molien--Weyl result. In particular, the quadratic coefficient of the Molien--Weyl partition function, \(d(d+1)/2\), counts the Gaussian singlet states above the vacuum.

In the endpoint formulation, the same singlet-counting information is reorganized through the radial variables and the spectator factor. We also confirmed the expansion directly by integrating the exact Bessel holonomy kernel and resumming the resulting Gaussian moment towers. This direct computation shows explicitly how the radial endpoint variables reproduce the gauge-projected physics encoded more compactly by the Molien--Weyl integral.

\medskip
\noindent
A central part of the analysis concerned the continuum limit. We showed that the quadratic coefficient naturally decomposes into a Gaussian contribution, a \(D\)-channel, and a \(\beta\)-channel. The \(D\)-channel is controlled by the isotropic Gaussian width of the endpoint variables, whereas the \(\beta\)-channel is controlled by the direct anisotropic endpoint-to-endpoint coupling. In the strict continuum limit, the naive Gaussian contribution becomes trivial, while the exact holonomy kernel develops the singular dependence needed to produce finite \(D\)- and \(\beta\)-channel contributions. Together, these channels reconstruct the correct planar endpoint coefficient and hence the Molien--Weyl singlet-counting result.

\medskip
\noindent
We then studied the geometry of the holonomy potential itself. The exact holonomy potential has no unconstrained critical point in the ambient space of variables. Its relevant stationary point lies instead on the physical aligned boundary. This point is a constrained boundary saddle: the ambient gradient does not vanish, but its projection along the tangent direction to the boundary does. This explains the survival of linear terms in the ambient Taylor expansion and clarifies the geometric meaning of the saddle.

\medskip
\noindent
Expanding around this constrained saddle gives a natural separation between the aligned, longitudinal direction and the transverse variable \(B\). The transverse expansion captures the local instability away from the aligned branch and produces a quartic Landau-type approximation. However, we showed that any finite polynomial transverse truncation has a trivial continuum limit. The reason is structural: after Gaussian averaging, a finite polynomial transverse theory produces a regular function at the continuum point. Consequently, its \(D\)-channel vanishes, while its \(\beta\)-channel is absent because the longitudinal variable has been frozen.

\medskip
\noindent
This led us to introduce a non-polynomial toy model based on the kernel \(\cosh B\), or equivalently the potential \(-\log\cosh B\). Locally, this toy model has the same qualitative transverse geometry as the quartic expansion: a central maximum followed by transverse descent. Globally, however, it is closer to the exact holonomy potential than the finite quartic truncation, since it avoids the artificial off-center minima created by polynomial approximations. More importantly, its Gaussian average has precisely the singular \(D\)-dependence required for a nontrivial continuum limit. It therefore reproduces the full continuum \(D\)-channel exactly, namely the finite contribution \(-2d\).

\medskip
\noindent
In this precise sense, the toy model provides a completion of the transverse expansion. It does not reproduce the full exact holonomy kernel, and in particular it does not capture the \(\beta\)-channel, because it depends only on the transverse variable \(B\). Nevertheless, it restores exactly the part of the continuum limit that finite transverse truncations necessarily miss. This makes it a useful intermediate model between the local saddle expansion and the full non-polynomial holonomy kernel.

\medskip
\noindent
The broader significance of this result is that the \(D\)-channel is not merely a technical correction. In the next installments of this work, we will show that the same contribution has a natural geometric interpretation as the Wishart--Stiefel entropy of the endpoint formulation. Indeed, the transverse endpoint variables \(V_a^\mu\) and \(W_a^\mu\), with \(a=1,\ldots,d\) and \(\mu=1,2\), define two real \(2\times d\) matrices. Equivalently, the two vectors \(V^\mu\in\mathbb R^d\) span a two-plane in \(\mathbb R^d\), and similarly the two vectors \(W^\mu\in\mathbb R^d\) span a second two-plane. Thus the natural geometric data are two Stiefel frames, or after quotienting by internal \(O(2)\) rotations, two points in the Grassmannian \(G(2,d)\). This connects the continuum limit of the \(N=2\), large--\(d\) gauge theory to the geometry of Stiefel and Grassmann manifolds, and suggests a route by which emergent geometry, including an effective four-dimensional geometry, may arise dynamically from the coupled pair of gauge-projected endpoint planes.

\medskip
\noindent
The main lesson of this first paper is therefore that the correct continuum limit of the \(N=2\), large--\(d\) theory is not captured by a naive Gaussian approximation or by any finite transverse truncation. It requires the exact holonomy structure, or at least a non-polynomial completion capable of reproducing its singular \(D\)-dependence. The Molien--Weyl and endpoint formulations provide two complementary windows onto this structure: one organized by gauge symmetry and singlet counting, the other by radial endpoint spacetime geometry and its emergent entropic interpretation.

\section{Acknowledgments}

\medskip
\noindent
The author would like to acknowledge helpful discussions with Denjoe O'Connor from the Dublin Institute for Advanced Studies. The author is especially grateful for Denjoe O'Connor's continued institutional hosting and generous support over the years, including travel, accommodation, and living expenses.

\medskip
\noindent
The author also acknowledges the use of ChatGPT-5.5, as well as previous versions, in several auxiliary capacities: 
(1) as a language editor; 
(2) as a LaTeX generator; 
(3) as a Mathematica-like symbolic tool; 
(4) as an assistant in searching for and reviewing references; 
and, more importantly, 
(5) as an ``artificial'' sounding board for testing, organizing, and refining ideas, effectively replacing in this role the function often played by human collaborators. However, the scientific vision, concept, design, direction, final scientific and mathematical editing, and all intellectual responsibility for this work remain solely with the author.

\appendix 
\section{Covariance matrix}\label{appendix1}

\bigskip
\noindent
For a single planar component \((v,w)\), the quadratic form in the exponent is
\begin{eqnarray}
S_{(v,w)}
=
\lambda\Big(\alpha_\Lambda v^2+\alpha_\Lambda w^2-2\widehat\beta_\Lambda\,vw\Big)
=
\begin{pmatrix}
v & w
\end{pmatrix}
M
\begin{pmatrix}
v\\
w
\end{pmatrix},\qquad 
M
=
\lambda
\begin{pmatrix}
\alpha_\Lambda & -\widehat\beta_\Lambda\\
-\widehat\beta_\Lambda & \alpha_\Lambda
\end{pmatrix}.
\end{eqnarray}
The inverse matrix is
\begin{eqnarray}
M^{-1}
=
\frac{1}{\lambda(\alpha_\Lambda^2-\widehat\beta_\Lambda^{\,2})}
\begin{pmatrix}
\alpha_\Lambda & \widehat\beta_\Lambda\\
\widehat\beta_\Lambda & \alpha_\Lambda
\end{pmatrix}
=
\frac{1}{\lambda\widehat D}
\begin{pmatrix}
\alpha_\Lambda & \widehat\beta_\Lambda\\
\widehat\beta_\Lambda & \alpha_\Lambda
\end{pmatrix}.
\end{eqnarray}

\medskip
\noindent
The covariance matrix is of the form 
\begin{eqnarray}
\big\langle (v,w)^T(v,w)\big\rangle
=
\frac12\,M^{-1}.
\end{eqnarray}
Therefore
\begin{eqnarray}
\langle v^2\rangle
=
\frac{\alpha_\Lambda}{2\lambda\widehat D},
\qquad
\langle w^2\rangle
=
\frac{\alpha_\Lambda}{2\lambda\widehat D},
\qquad
\langle vw\rangle
=
\frac{\widehat\beta_\Lambda}{2\lambda\widehat D}.
\label{single_component_covariance}
\end{eqnarray}

\medskip
\noindent
Restoring the indices \(a=1,\dots,d\) and the two planar components \(\mu=1,2\), one obtains
\begin{eqnarray}
\langle V_{a\mu}V_{b\nu}\rangle_0
=
\delta_{ab}\delta_{\mu \nu}\,
\frac{\alpha_\Lambda}{2\lambda\widehat D},\qquad 
\langle W_{a\mu }W_{b\nu}\rangle_0=
\delta_{ab}\delta_{\mu \nu}\,
\frac{\alpha_\Lambda}{2\lambda\widehat D},\qquad 
\langle V_{a\mu }W_{b\nu}\rangle_0=
\delta_{ab}\delta_{\mu \nu}\,
\frac{\widehat\beta_\Lambda}{2\lambda\widehat D}.
\label{planar_covariance_full}
\end{eqnarray}
This is the covariance matrix of the shifted planar Gaussian.

\medskip
\noindent
Using the covariance matrix \eqref{planar_covariance_full}, we can now extract the moments of
\begin{eqnarray}
A:=\lambda\sum_{a=1}^{d}W_a\!\cdot\!V_a
=
\lambda\sum_{a=1}^{d}\sum_{\mu=1}^{2}W_{a\mu}V_{a\mu},
\qquad
B:=\lambda\sum_{a=1}^{d}W_a\!\times\!V_a
=
\lambda\sum_{a=1}^{d}\Big(W_{a1}V_{a2}-W_{a2}V_{a1}\Big).\nonumber\\
\end{eqnarray}

\medskip
\noindent
For the one-point function of \(A\), one finds
\begin{eqnarray}
\langle A\rangle_0
=
\lambda\sum_{a=1}^{d}\sum_{\mu=1}^{2}\langle W_{a\mu}V_{a\mu}\rangle_0
=
\lambda\sum_{a=1}^{d}\sum_{\mu=1}^{2}
\frac{\widehat\beta_\Lambda}{2\lambda\widehat D}
=
\frac{d\,\widehat\beta_\Lambda}{\widehat D}.
\label{A_mean_shifted}
\end{eqnarray}
By symmetry,
\begin{eqnarray}
\langle B\rangle_0=0.
\end{eqnarray}

\medskip
\noindent
To compute \(A^2\), write
\begin{eqnarray}
A^2
=
\lambda^2
\sum_{a,b=1}^{d}\sum_{\mu,\nu=1}^{2}
W_{a\mu}V_{a\mu}\,W_{b\nu}V_{b\nu}.
\end{eqnarray}
Applying Wick's theorem and using \eqref{planar_covariance_full}, one obtains
\begin{eqnarray}
\langle A^2\rangle_0
&=&
\lambda^2
\sum_{a,b=1}^{d}\sum_{\mu,\nu=1}^{2}
\Big(
\langle W_{a\mu}V_{a\mu}\rangle_0\langle W_{b\nu}V_{b\nu}\rangle_0
+
\langle W_{a\mu}W_{b\nu}\rangle_0\langle V_{a\mu}V_{b\nu}\rangle_0
+
\langle W_{a\mu}V_{b\nu}\rangle_0\langle V_{a\mu}W_{b\nu}\rangle_0
\Big)
\nonumber\\
&=&
\frac{d^2\widehat\beta_\Lambda^{\,2}}{\widehat D^{\,2}}
+
\frac{d\,\alpha_\Lambda^2}{\widehat D^{\,2}}
+
\frac{d\,\widehat\beta_\Lambda^{\,2}}{2\widehat D^{\,2}}
\nonumber\\
&=&
\frac{d}{2\widehat D^{\,2}}
\Big(2\alpha_\Lambda^2+(2d+1)\widehat\beta_\Lambda^{\,2}\Big).
\label{A2_shifted}
\end{eqnarray}

\medskip
\noindent
\medskip
\noindent
Similarly, for \(B^2\), we have
\begin{eqnarray}
B^2
=
\lambda^2
\sum_{a,b=1}^{d}
\Big(W_{a1}V_{a2}-W_{a2}V_{a1}\Big)
\Big(W_{b1}V_{b2}-W_{b2}V_{b1}\Big).
\end{eqnarray}
Using Wick's theorem gives
\begin{eqnarray}
\langle B^2\rangle_0
=
\frac{d}{2\widehat D}.
\label{B2_shifted}
\end{eqnarray}

\medskip
\noindent
Therefore
\begin{eqnarray}
\langle R^2\rangle_0
=
\langle A^2+B^2\rangle_0=
\frac{d}{2\widehat D^{\,2}}
\Big(3\alpha_\Lambda^2+2d\widehat\beta_\Lambda^{\,2}\Big).
\label{R2_shifted}
\end{eqnarray}
Equations \eqref{A_mean_shifted} and \eqref{R2_shifted} are the basic Gaussian moments needed in the shifted planar scheme.